\documentclass[12pt]{article}
\usepackage{latexsym}
\usepackage{amsmath}
\usepackage{amssymb}
\usepackage{epsf}
\usepackage{epsfig}
\usepackage{cite}
\usepackage{accents}
\usepackage{booktabs}
\usepackage{lineno}

\usepackage{epsfig}
\usepackage{epstopdf}
\usepackage {lineno} 
\usepackage{ifthen} 
\newboolean{pdflatex}
\setboolean{pdflatex}{true} 

\newboolean{uprightparticles}
\setboolean{uprightparticles}{false} 

\RequirePackage{xspace}

\newcommand {\Bd} {\ensuremath{B^0_d}}
\newcommand {\Bs} {\ensuremath{B^0_s}}

\newcommand {\Ds} {\ensuremath{D_s}}

\newcommand {\Dsmp} {\ensuremath{\Ds^{\mp}}}
\newcommand {\Bsbar} {\ensuremath{\bar B^0_s}}
\newcommand {\Bdbar} {\ensuremath{\bar B^0_d}}

\newcommand {\asld} {\ensuremath{a^d_{\mathrm{sl}}}}
\newcommand {\asls} {\ensuremath{a^s_{\mathrm{sl}}}}
\newcommand {\aslq} {\ensuremath{a^q_{\mathrm{sl}}}}
\newcommand {\aslb} {\ensuremath{A^b_{\mathrm{sl}}}}

\newcommand {\pt} {\ensuremath{p_T}}

\newcommand{\gevc}{\ensuremath{{\mathrm{\,Ge\kern -0.1em V\!/}c}}\xspace}
\def\mum  {\ensuremath{\mu\rm m}}
\newcommand{\chisq}{\ensuremath{\chi^2}}
\newcommand {\ra} {\ensuremath{\rightarrow}}
\newcommand {\BsToMuMu} {\ensuremath{\Bs \rightarrow \mu^+ \mu^-}}
\newcommand {\BsToJpsiPhi} {\ensuremath{\Bs \rightarrow J/\psi \phi}}

\newcommand {\BsToDsK} {\ensuremath{\Bs \rightarrow \Dsmp K^{\pm} }}
\def\invfb {\ensuremath{\mbox{\,fb}^{-1}}\xspace}
\def\invpb {\ensuremath{\mbox{\,pb}^{-1}}\xspace}
\def\cms {\ensuremath{{\rm \,cm}^{-2} {\rm s}^{-1}}\xspace}
\def\dms{\ensuremath{\Delta M_s}\xspace}
\def\dmd{\ensuremath{\Delta M_d}\xspace}
\newcommand{\lsim}{\mathrel{\hbox{\rlap{\hbox{\lower4pt\hbox{$\sim$}}}\hbox{$<$}}}}

\begin{document}

\input epsf.tex    

\input psfig.sty

\begin{titlepage}

\vspace*{-0.0truecm}

\begin{flushright}
Nikhef-2013-009\\
LAL 13-72 
\end{flushright}

\vspace*{1.3truecm}

\begin{center}
\boldmath
{\Large{\bf Rare Decays and CP Violation in the $B_s$ System}}
\unboldmath
\end{center}

\vspace{0.9truecm}

\begin{center}
{\bf Guennadi Borissov\,${}^a$, Robert Fleischer\,${}^{b,c}$ and Marie-H\'el\`ene Schune\,${}^d$}

\vspace{0.5truecm}

${}^a${\sl Department of Physics, Lancaster University, Lancaster LA1 4YB, England, UK}

${}^b${\sl Nikhef, Science Park 105, NL-1098 XG Amsterdam, Netherlands}

${}^c${\sl  Department of Physics and Astronomy, Vrije Universiteit Amsterdam,\\
NL-1081 HV Amsterdam, Netherlands}

${}^d${\sl LAL, Universit\'{e} Paris-Sud, CNRS/IN2P3, Orsay, France}
\end{center}

\vspace{1.4cm}
\begin{abstract}
\vspace{0.2cm}\noindent
CP violating phenomena and rare decays of $\Bs$ mesons offer interesting probes to test the 
quark-flavor sector of the Standard Model. In view of plenty of data reported in particular from
the Large Hadron Collider, this topic has received a lot of attention in 2012. 
We give an overview of the the most recent experimental results, new theoretical developments,
and discuss the prospects for the future exploration of the $\Bs$-meson system.
\end{abstract}

\vspace*{1.5truecm}

\begin{center}
Invited contribution to 

\vspace*{0.5truecm}

{\it Annual Review of Nuclear and Particle Science}, Vol. {\bf 63} (2013)
\end{center}

\vspace*{0.5truecm}
\vfill
\noindent
December 2012
\vspace*{0.5truecm}

\end{titlepage}

\thispagestyle{empty}
\vbox{}
\newpage

\setcounter{page}{1}

\section{Introduction}
\label{introduction}

\subsection{Setting the Stage}
Weak decays of $B$ mesons offer various strategies for the exploration of the quark-flavor sector
of the Standard Model (SM) of particle physics. In the previous decade, decays of $\Bd$ and
$B^+$ mesons have been the focus of the $e^+e^-$ $B$ factories at SLAC and KEK with
the BaBar and Belle detectors, respectively. Pioneering first results on the $\Bs$-meson system
were first obtained by the ALEPH~\cite{detector-aleph}, DELPHI~\cite{detector-delphi} and 
OPAL~\cite{detector-opal} experiments at LEP. Few years later  
the CDF \cite{detector-cdf,muon-cdf,trigger-cdf} 
and D\O\ \cite{detector-d0,muon-d0} experiments at Fermilab have greatly improved the knowledge of the 
$\Bs$-meson system. In this decade, the exploration of $\Bs$ decays
is one of the key topics of the $B$-physics program of the Large Hadron Collider (LHC) at CERN,
with its dedicated $B$-decay experiment LHCb \cite{LHCb-detector}. 

The valence quark content of a $\Bs$ meson is given by a strange quark $s$ and an
anti-bottom quark $\bar b$. The $\Bs$-meson system exhibits a fascinating 
quantum-mechanical phenomenon, 
$\Bs$--$\Bsbar$ mixing, and provides an interesting laboratory to explore CP violation. 
Key features of the $B_s$-meson system are the large mass difference $\Delta M_s$
and the expectation of a sizable width difference 
$\Delta\Gamma_s$ between the $B_s$ mass eigenstates, and smallish CP violation in the 
$\Bs \to J/\psi \phi$ decay in the SM. 

In the SM, all the flavour and CP violation is governed by the Cabibbo--Kobayashi--Maskawa 
(CKM) quark-mixing matrix \cite{cab,KM}, connecting the electroweak eigenstates of the 
down, strange and bottom quarks with their mass eigenstates through a unitary transformation.
In extensions of the SM, typically new sources of flavor and CP violation are present, with
experimental data putting severe constraints on them. A recent overview of this topic in
view of the recent LHC data was given in \cite{BuGi}.

The study of the CP asymmetry in mixing offers one of the promising possibilities 
to search for the new sources of CP violation because of a very small SM expectation. 
The results of this study using the semileptonic $\Bs$ decays obtained at 
the Tevatron \cite{asl-d0, asld-d0,asls-d0}
show an indication of the deviation from the SM prediction, although the recent measurement
by the LHCb experiment \cite{asls-lhcb} is consistent with the SM expectation.

The current results about CP violation in $\Bs \to J/\psi \phi$ \cite{cdf-jpsi,d0-jpsi,jpsi-lhcb,jpsi-atlas} 
and $\Bs \to J/\psi f_0(980)$ \cite{jpsif0-lhcb,jpsipipi-lhcb} are consistent with the SM corresponding 
to tiny CP violation. 
From the theoretical point of view, these measurements are affected by uncertainties 
from doubly Cabibbo-suppressed penguin contributions \cite{RF-BsJpsiK}--\cite{MJ}. 
Since these effects are of non-perturbative nature, they cannot be calculated in a
reliable way within QCD. However, the corresponding hadronic parameters can be
constrained and determined with the help of control channels.

The study of CP violation in $\Bs$ decays is also important for the determination of
the angle $\gamma$ of the unitarity triangle of the CKM matrix. On the one hand, this
angle can be determined by means of the pure tree decays $B^0_s\to D_s^\pm K^\mp$.
On the other hand, it can also be extracted from the $\Bs\to K^+K^-$ decay and its
partner channel $\Bd\to\pi^+\pi^-$, involving loop contributions. 

Complementing these studies of CP violation, the rare decay $\Bs\to\mu^+\mu^-$ plays
an outstanding role for the testing of the SM, where this transition emerges only from 
loop processes. The theoretical prediction of the branching ratio of this channel involves
only a single non-perturbative, hadronic parameter and is very clean, with an uncertainty 
limited by lattice QCD. The search for this decay started at the Tevatron 
\cite{Rare_BsToMuMu_D0,Rare_BsToMuMu_CDF} and continued
at LHC\cite{Rare_BsToMuMu_CMS,Rare_BsToMuMu_ATLAS,LHCb-Bsmumu}. 
In November 2012, the LHCb collaboration has eventually reported 
the first evidence of this channel at the $3.5\,\sigma$ level, with a branching ratio in agreement
with the SM picture although the experimental errors are still large~\cite{LHCb-Bsmumu}. 

A further highlight of the experimental $B_s$ results reported in 2012 is a sizable 
$\Delta\Gamma_s$, which has been established by the CDF, D\O\, LHCb and ATLAS
collaborations~\cite{cdf-jpsi,d0-jpsi,jpsi-lhcb,jpsi-atlas,
bs-kk-lhcb-1,bs-kk-lhcb-2,bs-f0-cdf,bs-f0-lhcb}. This quantity 
leads to subtleties in the conversion of experimental data into branching ratios of $\Bs$ 
decays \cite{BR-paper}, but offers also new observables that can be exploited in the search 
for New Physics (NP) with the $\Bs\to\mu^+\mu^-$ channel \cite{Bsmumu-paper}.

In this review, we shall give an overview of these topics. As a large part will deal with 
experimental data, let give a brief description of the main detectors in the next subsection.

\boldmath
\subsection{Experimental Detectors for $\Bs$ Physics}
\unboldmath
\subsubsection{Detectors at Tevatron}

The CDF and D\O\ experiments are general purpose collider detectors 
designed to maximally exploit the possibilities
provided by the $p \bar p$ collisions at $\sqrt{s} = 1.96$ TeV and operate at the instantaneous luminosity up to 
$5 \times 10^{32}$ cm$^{-2}$ s$^{-1}$. Although the main emphasis in their design is made on the detection
of events with the highest possible invariant mass, they also
contain the elements necessary to endeavour the $B$-physics research. 

Both of them have the tracking system \cite{detector-cdf,detector-d0} 
consisting of the solenoidal magnet, the silicon microstrip
detectors and the central tracker. The instrumented volume of the CDF tracking system extends up to the radius of 137 cm,
while the outer radius of the D\O\ tracking system is 53 cm.

The muon identification system \cite{muon-cdf,muon-d0} covers
the pseudorapidity range up to $|\eta| < 1$ in CDF and up to $|\eta| < 2$ in the D\O\ detector.
The muon system of the D\O\ experiment also includes the toroidal magnets. They allow an independent measurement
of the muon momentum. This property helps to improve the quality of the identified muons.

An important part of the CDF detector essential for the $B$-physics studies is its special
trigger \cite{trigger-cdf} to select events with displaced tracks. 
It is the basis for many CDF measurements with fully hadronic $B$ decays. 
Its another trigger configurations select
the events with one or two muons. The trigger system of the D\O\ detector does not provide 
a possibility to collect events with displaced tracks, although its muon and di-muon triggers
are very efficient and robust. Therefore the focus
of the $B$-physics measurements in D\O\ experiment is shifted towards the
semileptonic $B$ decays and decays with $J/\psi \to \mu^+ \mu^-$ in the final state.

The polarities of the toroidal and solenoidal magnetic fields of the D\O\ detector are regularly reversed.
This reversal helps to significantly reduce the systematic uncertainties of the measurements sensitive 
to the differences
in the reconstruction efficiency between the positive and negative particles, like the measurements of the
$CP$ violating charge asymmetries.

Thus, both the CDF and D\O\ experiments have sufficient 
and powerful  tools to fulfill their $B$-physics research program. 
They also contain several special features which make them different and complementary. The CDF detector has
a larger tracking volume. Therefore its charged particle momentum resolution is superior to that of the
D\O\ detector. It also has the possibility to select the hadronic $B$ decays.
The D\O\ detector includes a sophisticated muon identification system with local measurement 
of the muon momentum. The reversal of the magnet polarities
allows it to perform several measurements of the charge asymmetry in the semileptonic $B$ decays which are
at the world best level.

\begin{figure}[t]
\begin{center}
{\includegraphics[height=8cm]{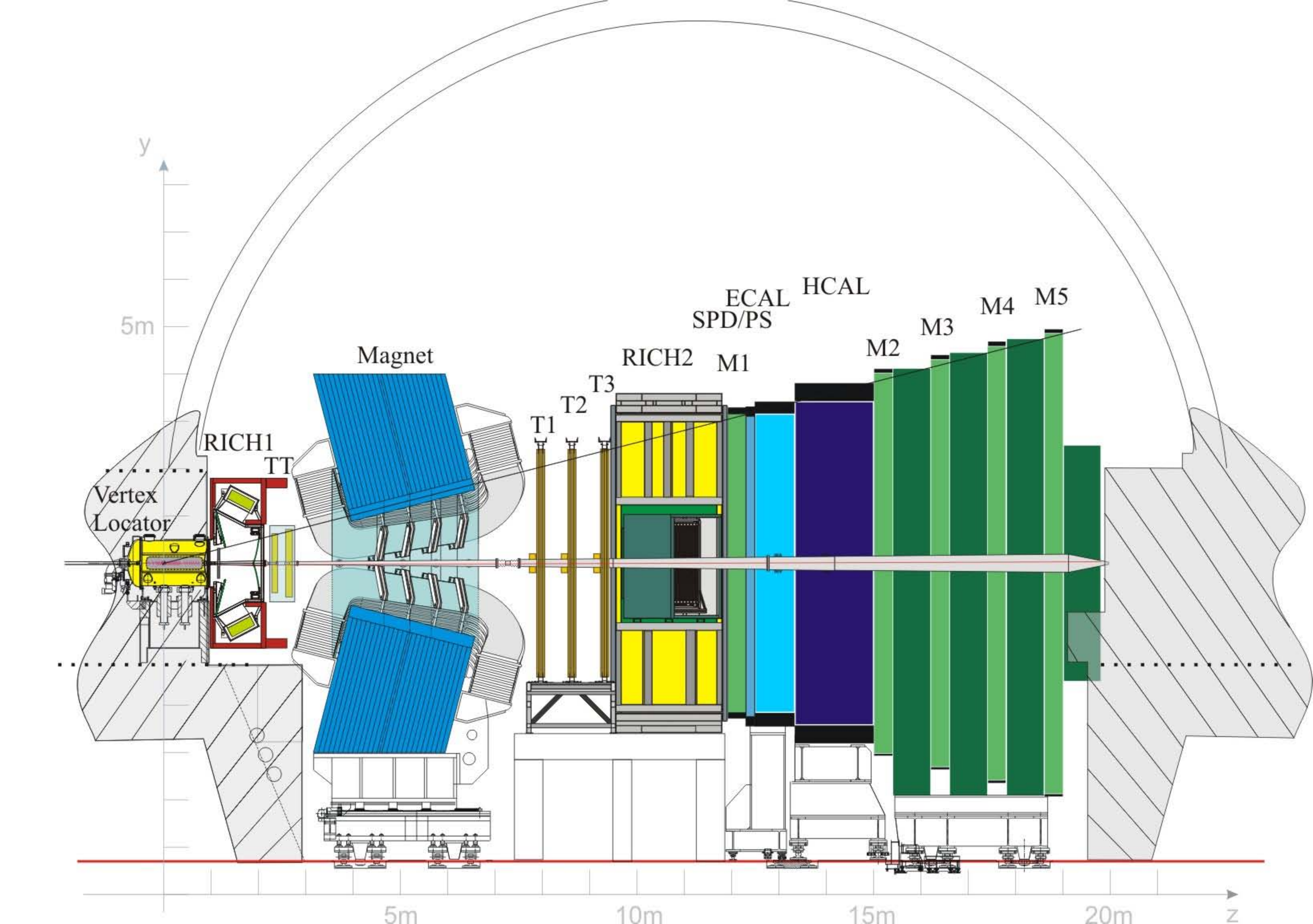}}
\caption{Vertical view of the LHCb detector (from Ref.~\cite{LHCb-detector}).}
\label{fig:LHCb}
\end{center}
\end{figure}

\subsubsection{The LHCb detector}
The LHCb detector~\cite{LHCb-detector}, shown in Fig.~\ref{fig:LHCb}, is a single-arm forward
spectrometer covering the \mbox{pseudorapidity} range $2<\eta <5$,
designed for the study of particles containing $b$-quark or $c$-quark. 
The detector includes a high precision tracking system
consisting of a silicon-strip vertex detector surrounding the $pp$
interaction region, a large-area silicon-strip detector located
upstream of a dipole magnet with a bending power of about
$4{\rm\,Tm}$, and three stations of silicon-strip detectors and straw
drift tubes placed downstream. The combined tracking system has a
momentum resolution $\Delta p/p$ that varies from 0.4\% at 5\gevc\ to
0.6\% at 100\gevc\, and an impact parameter resolution of 20\mum\ for
tracks with high transverse momentum. The dipole magnet can be operated in either polarity 
and this feature is used to reduce systematic effects due to detector asymmetries. 
Charged hadrons are identified using two ring-imaging Cherenkov detectors. Photon, electron and
hadron candidates are identified by a calorimeter system consisting of
scintillating-pad and preshower detectors, an electromagnetic
calorimeter and a hadronic calorimeter. Muons are identified by a
system composed of alternating layers of iron and multiwire
proportional chambers. 

A two-stage trigger is employed~\cite{bib:LHCbTrigger}. First a hardware-based decision is taken at a frequency 
up to 40 MHz. It accepts high transverse energy clusters in either the electromagnetic 
calorimeter or hadron calorimeter, or a muon of high \pt\ .
A second trigger level, implemented in software, receives 1 MHz of events and retains $\sim 0.3\%$
  of them. 
The software trigger requires a two-, three- or four-track
  secondary vertex with a high sum of the transverse momentum, \pt\, of
  the tracks and a significant displacement from the primary $pp$
  interaction vertices~(PVs). At least one track should have $\pt >
  1.7\gevc$ and impact parameter~(IP) \chisq\ with respect to the
  primary interaction greater than 16. The IP \chisq\ is defined as the
  difference between the \chisq\ of the PV reconstructed with and
  without the considered track. A multivariate algorithm is used for
  the identification of secondary vertices consistent with the decay
  of a $b$-hadron.
  
\subsubsection{The Atlas and CMS detectors}
The Atlas~\cite{bib:ATLAS} and CMS~\cite{bib:CMS} detectors are multi-purpose central detectors optimized for searches of heavy objects. At high luminosity, their potential for flavour physics is limited by their triggering capabilities and focus mainly on $b$ and charmonium decays involving dimuons.
Their impact on \Bs\ physics is currently mainly related to the search for the very rare \BsToMuMu\ decay and to the study of  $\Bs \to J/\psi \phi$.

\boldmath
\subsection{Production of $\Bs$ Mesons}\label{sec:Bsprod}
\unboldmath
At the LHC, the $b \bar {b}$ production cross-section is large: 
it is expected to be of the order of 500 $\mu$b at 14 TeV\cite{bib:Pythia}. 
The LHC is currently running at 
7 or 8 TeV and the cross-section has been measured to be of the order of 290 
$\mu$b \cite{LHCb-JpsiProd} for a center of mass energy of 7 TeV. The detector has taken data at an instantaneous luminosity of about 3.5 to 4 $\times 10^{32}$\cms , and with a number of pp interactions per crossing of $\sim 1.4$.  During the 2012 data taking, the center of mass energy has been increased to  $ \sqrt s$ = 8 TeV which corresponds to an increase of about  15 \% in the number of  $b \bar {b}$ events. The recorded data amounts to more than 3 \invfb , the world largest $b$-hadron sample. 

The knowledge of the production rate of \Bs\ mesons is required to determine any \Bs\ branching fraction. To be specific, the measurement of branching ratios of $\Bs\to f$ decays at hadron 
colliders relies
on certain normalization channels $B_q\to X$, where the $B_u^+\to J/\psi K^+$, 
$B^0_d\to K^+\pi^-$ and/or $B_d^0 \to J/\psi K^{*0}$ modes play key roles. 
The $\Bs$ decay branching ratio can then be extracted with the help of the relation
\begin{equation}
\mbox{BR}(B^0_s\to f)
=\mbox{BR}(B_q\to X)\frac{f_q}{f_s}
\frac{\epsilon_{X}}{\epsilon_{\mu\mu}}
\frac{N_{\mu\mu}}{N_{X}}, \label{BRmumu-exp}
\end{equation}
where the $\epsilon$ and $N$ factors denote the total detector efficiencies and the 
observed number of events, respectively. In practical terms, the ratio of the ``fragmentation
functions"  $f_q$ represents usually the major source of the systematic uncertainty, in particular
for the measurement of the branching ratio of the rare $\Bs\to\mu^+\mu^-$ decay \cite{FST}. 
The $f_q$
describe the probability that a $b$ quark will fragment in a $\bar B_q$ meson ($q\in\{u,d,s\}$),
and depend on the hadronic environment of the collider. 

A new method for determining $f_s/f_d$ using nonleptonic $\bar B^0_s\to D_s^+\pi^-$, 
$B^0_d\to D^+K^-$, $B^0_d\to D^+\pi^-$ decays \cite{FST,FST-fact} was  
implemented at LHCb \cite{LHCb-hadr}, with a result in good agreement with 
measurement using semileptonic decays \cite{LHCb-sl}. The $SU(3)$-breaking form-factor 
ratio entering this method has recently been calculated with lattice QCD \cite{FF-lat}.
An updated experimental result obtained with the nonleptonic decays and the data recorded 
in 2011 by the LHCb experiment~\cite{LHCb-hadr-update}, combined the measurement using semileptonic decays, leads to 
\begin{equation}
\frac{f_s}{f_d} = 0.256 \pm 0.020,
\end{equation}
where  the various sources of correlated systematics uncertainties, notably the $D$ branching fractions and $B$ lifetimes, are taken into account.

\subsection{Outline of the Review}
The remainder of this review is organized as follows: in Section~\ref{mixing}, we discuss
$\Bs$--$\Bsbar$ mixing and the current status of the measurements of the corresponding
mixing parameters. In view of the large value of $\Delta\Gamma_s$, we have a closer
look at subtleties in the extraction of $\Bs$ decay branching ratios and point out the usefulness 
of effective decay lifetimes. In Section~\ref{CPV}, we turn to CP violation in the \Bs\ system,
which is a central part of this review. After discussing first CP violation in 
$\Bs$--$\Bsbar$ oscillations that is probed through the semileptonic charge asymmetry, 
we review the extraction of the mixing phase $\phi_s$ from the $\Bs\to J/\psi \phi$ and 
$\Bs\to J/\psi f_0(980)$ decays and the associated theoretical uncertainties through
penguin topologies. In addition to these benchmark channels, we shall also address 
CP-violating phenomena in a variety of other $\Bs$ decays. In Section~\ref{rare_decays}, 
we discuss rare \Bs\ decays, with a focus on the most prominent $B^0_s\to \mu^+\mu^-$ 
channel, and brief discussions of $\Bs\to \phi\gamma$ and $\Bs\to\phi\mu^+\mu^-$. Finally, 
we summarize in Section~\ref{sec:concl} the main conclusions and give an outlook of \Bs\ physics. 

\boldmath
\section{Time Evolution of the $\Bs$ System}
\unboldmath
\label{mixing}

\begin{figure}[t] 
   \centering
    \includegraphics[width=5.0truecm]{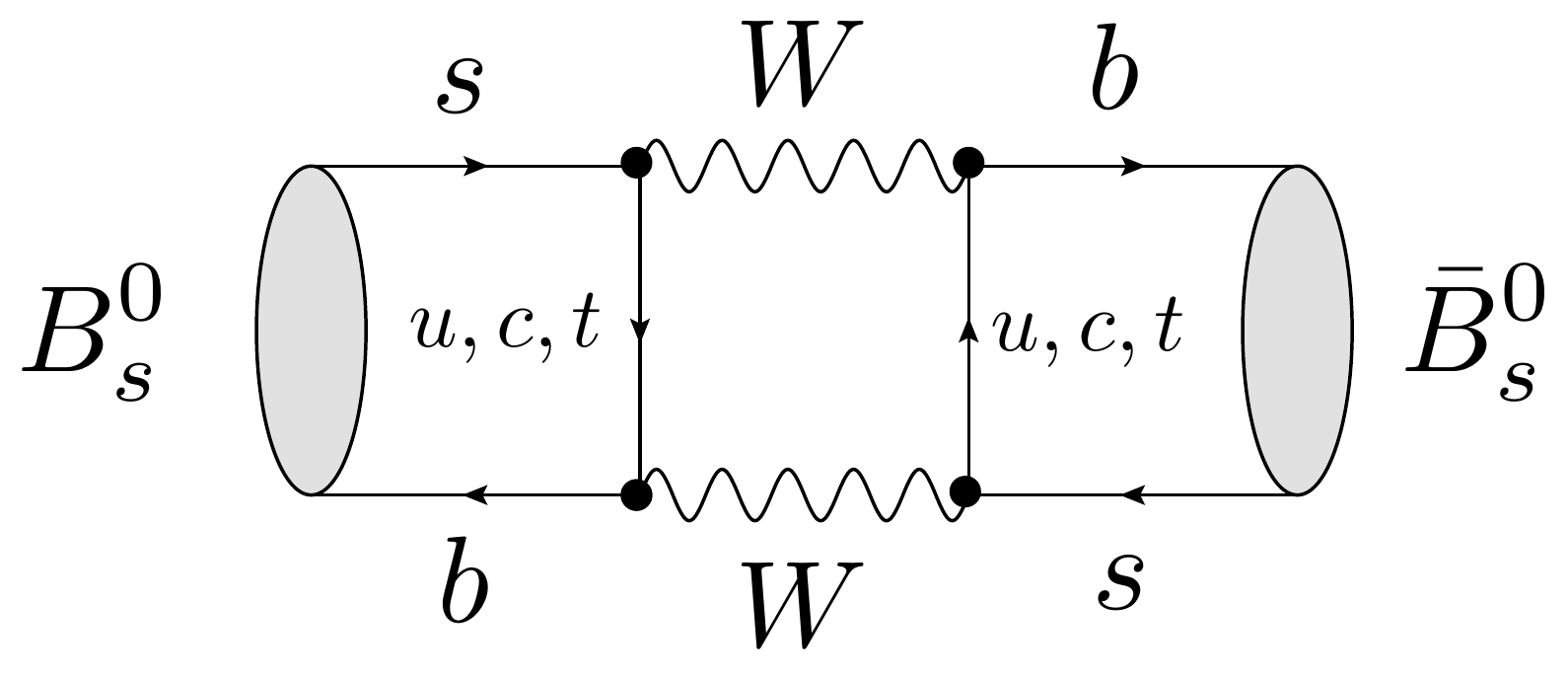} \hspace*{0.8truecm}
    \includegraphics[width=5.0truecm]{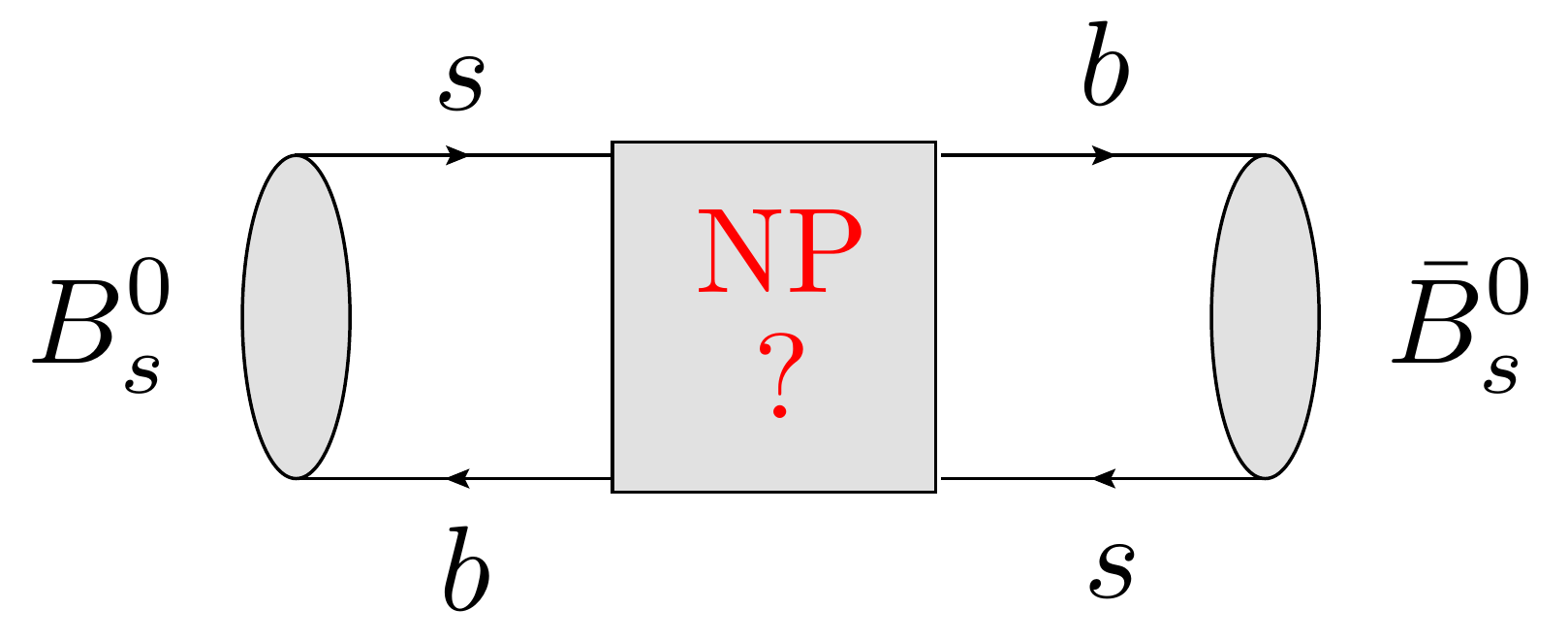} 
  \caption{Illustration of $\Bs$--$\Bsbar$ mixing in the SM (left panel) and in the presence of
  NP contributions (right panel).}\label{fig:Bs-mix}
\end{figure}

\subsection{General Features}
The neutral $B_s$ mesons show the quantum-mechanical phenomenon of 
$\Bs$--$\Bsbar$ mixing, which is caused in the SM by box-diagram topologies, as 
illustrated in the left panel of Fig.~\ref{fig:Bs-mix}. 
Consequently, an initially, i.e.\ at time 
$t=0$, produced $B^0_s$-meson state evolves into a time-dependent linear combination 
of $\Bs$ and $\Bsbar$ states:
\begin{equation}
|B_s(t)\rangle=a(t)|\Bs\rangle + b(t)|\Bsbar \rangle.
\end{equation}
The time-dependent functions $a(t)$ and $b(t)$ can be calculated in a straightforward way
by solving an appropriate Schr\"odinger equation, where ``heavy" and ``light" mass eigenstates
are introduced, which are characterized by the differences
\begin{equation}\label{DMDG-def}
\Delta M_s\equiv M_{\rm H}^{(s)}-M_{\rm L}^{(s)},\quad
\Delta\Gamma_s\equiv\Gamma_{\rm L}^{(s)}
-\Gamma_{\rm H}^{(s)}
\end{equation}
of their masses and decay widths. These quantities enter the analytic expressions for the
time-dependent decay rates $\Gamma(B^0_s(t)\to f)$ and $\Gamma(\bar B^0_s(t)\to f)$,
which correspond to decays of initially present $\Bs$ and $\Bsbar$ mesons, respectively. 
Moreover, $\Bs$--$\Bsbar$ mixing involves also a CP-violating  phase, which takes
the general form
\begin{equation}\label{phis-def}
\phi_s=\phi_s^{\rm SM} + \phi_s^{\rm NP}.
\end{equation}
Here the former piece is the SM contribution
\begin{equation}\label{phis-SM}
\phi_s^{\rm SM}=2\mbox{arg}(V_{ts}^*V_{tb})=-2\lambda^2\eta=-(2.08\pm0.09)^\circ,
\end{equation}
where $\lambda\equiv|V_{us}|$ and $\eta$ (measuring the height of the unitarity triangle)
are Wolfenstein parameters \cite{wolf}, and the numerical value refers to the analysis of the
CKM matrix performed in Refs.~\cite{CKM-fitter,bib:UTFit}.
For a detailed discussion of  the formalism of $\Bs$--$\Bsbar$ mixing, 
the reader is referred to \cite{RF-Phys-Rep}.

In the presence of NP,  new particles may enter the box diagram 
(as illustrated in the right panel of Fig.~\ref{fig:Bs-mix}), or may give 
rise to new contributions at the tree level, which 
is forbidden in the SM. Should new CP-violating phases be involved, they would manifest
themselves through the $\phi_s^{\rm NP}$ term in (\ref{phis-def}), which could then 
make $\phi_s$ differ sizably from its SM value (see, for instance, 
Refs.~\cite{BaFl}--\cite{bur-fla} and references therein). 

Another characteristic feature of the $B_s$-meson system
is the decay width difference $\Delta\Gamma_s$. Thanks to 
$b \to c \bar c s$ quark-level processes, it is expected to be sizable in the SM 
\cite{LN,CFLMT}, with $\Delta\Gamma_s/\Gamma_s|_{\rm SM}\sim 0.15$, while the counterpart
of this quantity for the $B_d$-meson system is expected at the $0.1\%$ level. The current 
situation has recently been summarized in Ref.~\cite{lenz}.

\boldmath
\subsection{$B^0_s$--$\bar B^0_s$ Oscillations}
\unboldmath
Since the first observation of particle--antiparticle transformations in neutral $B$ mesons in 1987 \cite{Dmd_Argus}, the determination of the  \Bs--\Bsbar\  oscillation frequency \dms from a time-dependent measurement of \Bs--\Bsbar\ oscillations has been a major objective of experimental particle physics. A long standing search was performed during more than 19 years, mainly due to the fact that the \Bs--\Bsbar\  oscillation frequency is 35 times larger than that for the \Bd--\Bdbar\ system, posing a considerable challenge for the decay time resolution of the detectors. The large statistics available at the Tevatron and the good proper time reconstruction allowed in 2006 the  D\O\  and CDF experiments to produce the first precise measurements \cite{Dms_D0,Dms_CDF}. More recently, with only 36 \invpb,  the LHCb experiment has confirmed the measurement with a similar statistical precision but a smaller systematical uncertainty \cite{Dms_LHCb}. The \Bs--\Bsbar\ mixing frequency is now known with a precision better than 0.5~\% \cite {pdg-2012} :
\begin{equation}
\dms = ( 17.69 \pm 0.08 ) \  {\rm ps}^{-1}
\end{equation}

Note that the ability to resolve these fast  \Bs--\Bsbar\ oscillations is a prerequisite for many physics analyses. In particular it is essential for the study of the time-dependent CP asymmetry of \BsToJpsiPhi.

The mass difference $\dms$ is proportional to the CKM coefficient $|V_{ts}|^2$. However, 
the value of $|V_{ts}|$ directly extracted from $\dms$ has large theoretical uncertainties related to the
contribution of non-perturbative QCD effects. Many such uncertainties cancel
in the ratio $\dms / \dmd$, which can be expressed as
\begin{equation}
\frac{\dms}{\dmd} =  \xi^2 \frac{M_{\Bs}}{M_{\Bd}}  \left | \frac{V_{ts}}{V_{td}} \right | ^2.
\end{equation}
Here $\xi$ is an $SU(3)$ flavor-symmetry-breaking factor, while the $M_{B^0_q}$ denote the
masses of the $B^0_q$ mesons. The former non-perturbative parameter can be determined with 
the help lattice QCD, where the most recent value reads as follows \cite{lattice-xi}:
\begin{equation}
\xi = 1.237 \pm 0.032.
\end{equation}
Using this result and the average of the $\dms$ measurements by the CDF and LHCb 
experiments, the value 
\begin{equation}
\left | \frac{V_{ts}}{V_{td}} \right | = 0.2111 \pm 0.0010 \mbox{(exp)} \pm 0.0055 \mbox{(lattice)}
\end{equation}
has been extracted \cite{pdg-2012}.
It can be seen that the theoretical uncertainties still dominate in this ratio.
They need to be improved in the future for a precise test of the unitarity relation of the CKM matrix.

\boldmath
\subsection{Untagged $B_s$ Decay Rates and Branching Ratios}\label{sec:untagged}
\unboldmath
A particularly interesting case arises if no distinction, i.e.\ ``tagging", is made between 
initially  present $B^0_s$ or $\bar B^0_s$ mesons. The corresponding ``untagged"
decay rate is a sum of two exponentials:
\begin{equation}\label{untag1}
	\langle \Gamma(B_s(t)\to f)\rangle
	\equiv\ \Gamma(B^0_s(t)\to f)+ \Gamma(\bar B^0_s(t)\to f) 
=R^f_{\rm H} e^{-\Gamma_{\rm H}^{(s)} t} + R^f_{\rm L} e^{-\Gamma_{\rm L}^{(s)} t}.
\end{equation}
This expression can be rewritten as
\begin{equation}
	\langle \Gamma(B_s(t)\to f)\rangle=
	\left(R^f_{\rm H} + R^f_{\rm L}\right) e^{-\Gamma_s\,t}\left[ \cosh\left(y_s\, t / \tau_{B_s}\right)+
	{\cal A}^f_{\rm \Delta\Gamma}\,\sinh\left(y_s\, t/\tau_{B_s}\right)\right],\label{untag2}
\end{equation}
where the observable 
\begin{equation}\label{ADG-def}
{\cal A}^f_{\rm \Delta\Gamma}\equiv\frac{R^f_{\rm H} - R^f_{\rm L}}{R^f_{\rm H} + R^f_{\rm L}}
\end{equation}
depends on the final state $f$, and 
\begin{equation}\label{ys-def}
	y_s \equiv \frac{\Delta\Gamma_s}{2\,\Gamma_s}\equiv
	\frac{\Gamma_{\rm L}^{(s)} - \Gamma_{\rm H}^{(s)}}{2\,\Gamma_s}= 0.088 \pm 0.014
\end{equation}
describes the impact of a non-vanishing decay width difference $\Delta\Gamma_s$; the
average decay width
\begin{equation}\label{Gams}
\Gamma_s \equiv 
	\frac{\Gamma_{\rm L}^{(s)} + \Gamma_{\rm H}^{(s)}}{2\,\Gamma_s} = \tau_{B_s}^{-1}=
	\left(0.6580 \pm 0.0085\right)\mbox{ps}^{-1}
\end{equation}
is given by the inverse of the $B_s$ lifetime $\tau_{B_s}$. The numerical values in (\ref{ys-def})
and (\ref{Gams}) correspond to the results reported in Ref.~\cite{LHCb-Mor-12}.

The untagged rates (\ref{untag2}) are used by experiments for the extraction of branching
ratios. However, usually no time information for the untagged data sample is taken into account, 
which corresponds to the following time-integrated, ``experimental" branching ratios 
\cite{DFN,BR-paper}:
\begin{equation}\label{BR-exp}
{\rm BR}\left(B_s \to f\right)_{\rm exp} 
	\equiv \frac{1}{2}\int_0^\infty \langle \Gamma(B_s(t)\to f)\rangle\, dt
\end{equation}
\begin{displaymath}	
= \frac{1}{2}\left[ \frac{R^f_{\rm H}}{\Gamma^{(s)}_{\rm H}} + 
	\frac{R^f_{\rm L}}{\Gamma^{(s)}_{\rm L}}\right]
	= \frac{\tau_{B_s}}{2}\left(R^f_{\rm H} + R^f_{\rm L}\right)
	\left[\frac{1 + {\cal A}^f_{\Delta\Gamma}\, y_s}{1-y_s^2} \right].
\end{displaymath}
On the other hand, in the theory community, the $B^0_s$--$\bar B^0_s$ oscillations are 
usually ``switched off" by choosing $t=0$, and the following CP-averaged branching ratios are
calculated:
\begin{equation}\label{BR-theo}
{\rm BR}\left(B_s \to f\right)_{\rm theo}\equiv 
	\frac{\tau_{B_s}}{2}\langle \Gamma(B^0_s(t)\to f)\rangle\Big|_{t=0}
= \frac{\tau_{B_s}}{2}\left(R^f_{\rm H} + R^f_{\rm L}\right).
\end{equation}
The advantage of this $B_s$ branching ratio concept is the possibility of 
comparing straightforwardly with branching ratios of decays of $B^0_d$ or $B^+_u$ mesons
that are related to one another by the $SU(3)_{\rm F}$ flavor symmetry of strong interactions. 

The conversion between the time-integrated and theoretical branching ratios defined in 
(\ref{BR-exp}) and (\ref{BR-theo}), respectively,  can be accomplished with the help of
the following relation \cite{BR-paper}: 
\begin{equation}\label{BRratio-1}
       {\rm BR}\left(B_s \to f\right)_{\rm theo}
 = \left[\frac{1-y_s^2}{1 + {\cal A}^f_{\Delta\Gamma}\, y_s}\right]
{\rm BR}\left(B_s \to f\right)_{\rm exp}.
\end{equation}
While the term in square brackets would be equal to one in the presence of a vanishing 
decay width difference $\Delta\Gamma_s$, the experimental value of $y_s$ in (\ref{ys-def})
can lead to a difference between theoretical $B_s\to f$ branching ratios and their experimental
counterparts as large as $10\%$, depending on the final state $f$. A compilation of these
effects, based on theoretical analyses which make in particular use of the $SU(3)$ flavor
symmetry, can be found in Ref.~\cite{BR-paper}. Subtleties related to $\Delta\Gamma_s$
in the experimental analysis of the $B_s\to K^{*0}\bar K^{*0}$ channel were also discussed 
in Refs.~\cite{LHCbKstarKstar,DGMV}.

\boldmath
\subsection{Effective $B_s$ Decay Lifetimes}\label{sec:Bslifetimes}
\unboldmath
The theoretical input for the conversion of the branching ratios into the corresponding theoretical
branching ratios can be avoided as soon as time information for the untagged $B_s$ data is
available. In this case, the effective $B_s$ lifetime of the decay at hand \cite{FK}, 
which is defined as
\begin{equation} \label{taueff}
       \tau_f \equiv \frac{\int_0^\infty t\,\langle \Gamma(B_s(t)\to f)\rangle\, dt}
        {\int_0^\infty \langle \Gamma(B_s(t)\to f)\rangle\, dt}
               \quad = \frac{\tau_{B_s}}{1-y_s^2}\left[\frac{1+2\,{\cal A}^f_{\Delta\Gamma}y_s + y_s^2}
        {1 + {\cal A}^f_{\Delta\Gamma} y_s}\right],
       \end{equation}
 can be extracted, thereby yielding the following expression  \cite{BR-paper}:
\begin{equation}\label{BRratioT}
    {\rm BR}\left(B_s \to f\right)_{\rm theo}
      = \left[2 - \left(1-y_s^2\right)\frac{\tau_f}{\tau_{B_s}}\right]{\rm BR}\left(B_s \to f\right)_{\rm exp};
\end{equation}
it should be emphasized that only measurable quantities appear on the right-hand side. 
The measurement of effective $B_s$ decay lifetimes is hence an integral part of the extraction 
of the theoretical branching ratios (\ref{BR-theo}) from the data. 

Another interesting application of effective lifetimes of $B_s$ decays is that they allow us
to probe the CP-violating mixing phase $\phi_s$ and the decay width difference
$\Delta\Gamma_s$. In particular, the lifetimes can be converted into contours in the
$\phi_s$--$\Delta\Gamma_s$ plane, where the intersection of the contours related to the effective 
lifetimes of $B_s$ decays into CP-even (such as $B_s\to K^+K^-$) and CP-odd 
(such as $B_s \to J/\psi f_0$ with $f_0\equiv f_0(980)$) final states allows  the extraction of
$\phi_s$ and $\Delta\Gamma_s$ \cite{FK,RK}. This determination is extremely robust with 
respect to the hadronic penguin uncertainties and complements nicely studies of CP violation
in $B^0_s\to J/\psi \phi$ and $\Bs\to J/\psi f_0$ decays to be discussed in Section~\ref{CPV}.

\begin{figure}
   \centering
   \includegraphics[width=8.0truecm]{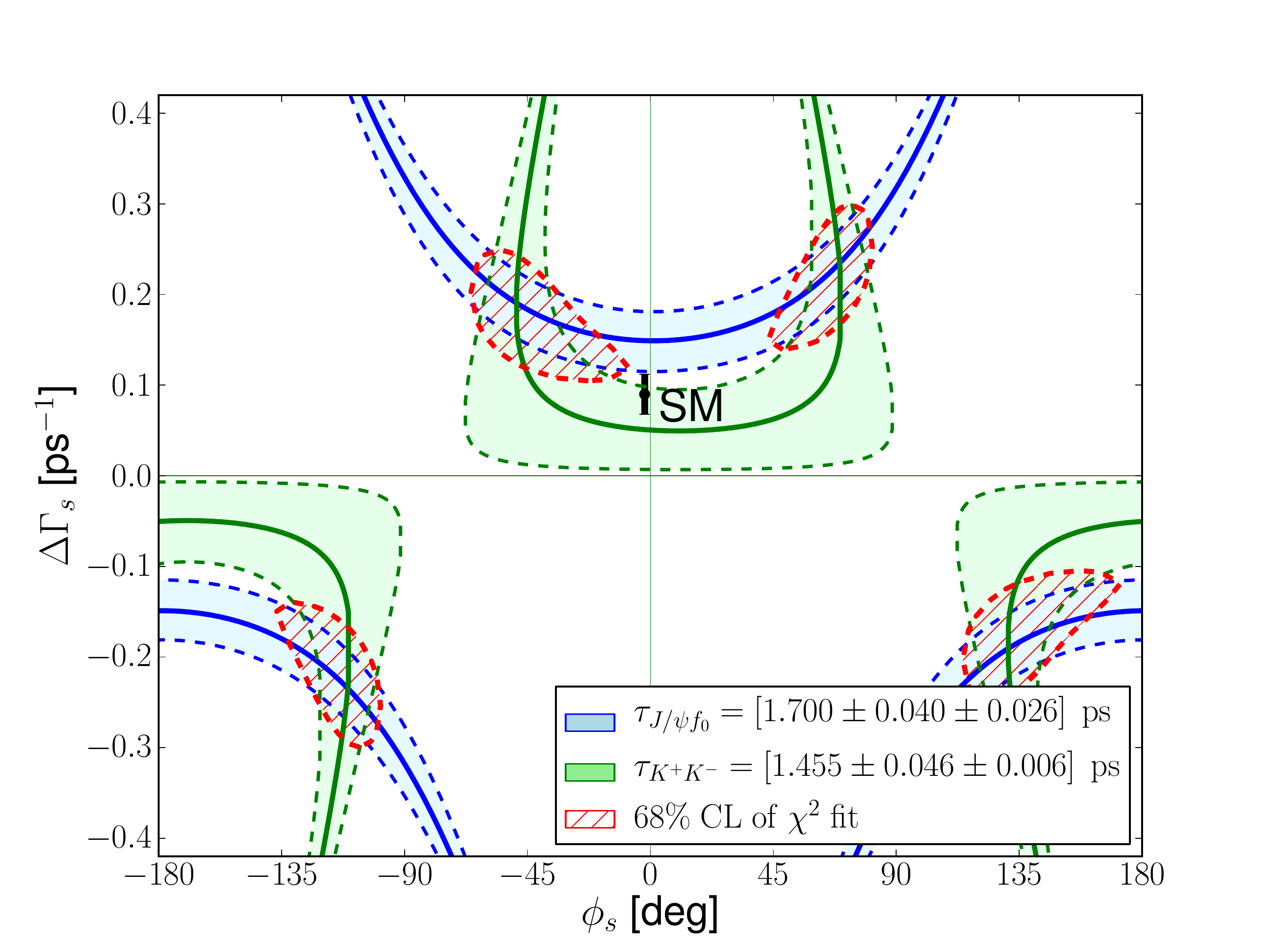} 
\caption{Constraints in the $\phi_s$--$\Delta\Gamma_s$ plane from measurements of 
the effective lifetimes of the $B^0_s\to K^+K^-$ and $B^0_s\to J/\psi f_0$ decays \cite{FK,RK}.}\label{fig:lifetimes}
 \end{figure}

The first measurements of effective lifetimes for $B_s$ decays into final CP eigenstates by 
the CDF and LHCb collaborations have recently become available.
The LHCb collaboration measured the effective lifetime of the $\Bs \to K^+ K^-$ decay 
and published two independent results \cite{bs-kk-lhcb-1,bs-kk-lhcb-2} using the statistics
collected in 2010 and 2011: 
\begin{eqnarray}
\tau_{K^+ K^-} & = & 
[1.440 \pm 0.096 \mbox{(stat)} \pm 0.009 \mbox{(syst)} ]~\mbox{ps (37 pb}^{-1}), \\
\tau_{K^+ K^-}& = & [1.455 \pm 0.046 \mbox{(stat)} \pm 0.006 \mbox{(syst)} ]~\mbox{ps (1 fb}^{-1}).
\label{kk2}
\end{eqnarray}

The $\Bs$ lifetime in the decay mode $\Bs \to J/\psi f_0(980)$ with $f_0(980) \to \pi^+ \pi^-$ 
was measured by the CDF and LHCb collaborations.
The CDF collaboration has analyzed 3.7 fb$^{-1}$ and reconstructed $502 \pm 37$ such decays and has found~\cite{bs-f0-cdf}
\begin{equation}
\label{f0-cdf}
\tau_{J/\psi f_0}  =  [1.70 ^{+0.12}_{-0.11}\mbox{(stat)} \pm 0.03 \mbox{(syst)} ]~\mbox{ps}.
\end{equation}
The LHCb collaboration has reconstructed $4040 \pm 75$ decays $\Bs \to J/\psi f_0$ using the statistics
corresponding to the integrated luminosity of 1 fb$^{-1}$ and has obtained~\cite{bs-f0-lhcb}
\begin{equation}
\label{f0-lhcb}
\tau_{J/\psi f_0}  =  [1.700 \pm 0.040\mbox{(stat)} \pm 0.026 \mbox{(syst)} ]~\mbox{ps}.
\end{equation}
The results (\ref{f0-cdf}) and (\ref{f0-lhcb}) are in very good agreement, although 
with a better precision for the LHCb result. 
It is important to note that the difference between the $\Bs$ lifetime in the decay modes
$\Bs \to K^+ K^-$ and $\Bs \to J/\psi f_0(980)$ 
exceeds three standard deviation, which is an independent evidence of the non-zero decay width 
difference $\Delta\Gamma_s$ of the $B_s$-meson system.

In Fig.~\ref{fig:lifetimes}, we show the constraints in the $\phi_s$--$\Delta\Gamma_s$ plane
that follow from the effective \Bs\ decay lifetime measurements discussed above. 
Future lifetime measurements with $1\%$ uncertainty would be most interesting.


\boldmath
\section{ CP Violation in $\Bs$ Decays}
\unboldmath
\label{CPV}

\subsection{Introduction}
Decays of \Bs\ mesons allow interesting studies of CP violation. 
In the analyses of the 
corresponding CP-violating rate asymmetries it is essential that ``tagging" information is 
available, allowing the distinction between initially present $B^0_s$ or $\bar B^0_s$ meson 
states. Let us, for simplicity, consider a decay into a CP eigenstate $f$, which will also be 
particularly relevant for the major part of the discussion in this section. 
The CP-violating rate asymmetry takes then the following form:
\begin{equation}\label{CP-asym}
\frac{\Gamma(B^0_s(t)\to f)-\Gamma(\bar B^0_s(t)\to f)}{\Gamma(B^0_s(t)\to f)
+\Gamma(\bar B^0_s(t)\to f)}=\frac{C(B_s\to f)\cos(\Delta M_st) -
S(B_s\to f) \sin(\Delta M_st)}{\cosh(\Delta\Gamma_st/2)+
{\cal A}_{\Delta\Gamma}(B_s\to f)\sinh(\Delta\Gamma_st/2)} ,
\end{equation}
where the the time-dependent rates refer to initially present $B^0_s$ or $\bar B^0_s$ states.

The observable $C(B_s\to f)$ describes  ``direct" CP violation,  which is caused by the 
interference between different amplitudes contributing to the decay at hand, with non-trivial 
CP-conserving strong and CP-violating weak phase differences. 
On the other hand, the observable $S(B_s\to f)$ originates from interference between 
$B^0_s$--$\bar B^0_s$ mixing and $B^0_s, \bar B^0_s\to f$ decay processes and is 
referred to as ``mixing-induced" CP violation. The observable 
${\cal A}_{\Delta\Gamma}(B_s\to f)$ arises in  the untagged rate, as we have already 
discussed in (\ref{untag2}). It should be noted that these observables are not independent 
from one another, satisfying the following relation:
\begin{equation}
[C(B_s\to f)]^2+[S(B_s\to f)]^2+[{\cal A}_{\Delta\Gamma}(B_s\to f)]^2=1.
\end{equation}
For a detailed discussion of the calculation of these observables, the reader is referred 
to \cite{RF-Phys-Rep}.

In the rate asymmetry in Eq.~(\ref{CP-asym}), CP violation in $B^0_s$--$\bar B^0_s$ 
oscillations has been neglected as this phenomenon has here a tiny impact. Before 
turning to $B^0_s\to J/\psi \phi$, which is one of the most prominent $\Bs$-meson
decays to explore CP violation, let us first have a closer look at CP violation in 
$\Bs$--$\Bsbar$ mixing. 

\boldmath
\subsection{CP Violation in $\Bs$--$\Bsbar$ Mixing}\label{CPV-mixing}
\unboldmath
CP violation in the mixing of the neutral $B^0_q$ mesons ($q=d,s$) is described by the 
phase $\phi_{12}^q$, which is defined as
\begin{equation}
\phi_{12}^q \equiv \arg \left( - \frac {M_{12}^q}{\Gamma^q_{12}} \right).
\end{equation}
The phase $\phi_{12}^s$ should not be mixed up with the $\phi_s$ introduced in (\ref{phis-def}). 
In the presence of NP contributions to $B^0_s$--$\bar B^0_s$ mixing, it takes the form
\begin{equation}
\phi_{12}^s = \phi_{12}^s|_{\rm SM} + \phi_s^{\rm NP}, 
\end{equation}
where the SM piece takes the following numerical value \cite{LN}:
\begin{equation}
\phi_{12}^s|_{\rm SM}=(0.22 \pm 0.06)^\circ,
\end{equation}
and $\phi_s^{\rm NP}$ is the same NP phase entering also (\ref{phis-def}). The notation
agrees with that of Ref.~\cite{LHCb-Strat}.

The parameters $M^q_{12}$ and $\Gamma^q_{12}$ are the complex non-diagonal elements of
the mass mixing matrix. They are related to the observable quantities $\Delta M_q$ and 
$\Delta \Gamma_q$ introduced in (\ref{DMDG-def}) as 
\begin{equation}
\Delta M_q = 2 \left| M^q_{12} \right|, \qquad
\Delta \Gamma_q = 2 \left| \Gamma^q_{12} \right| \cos \phi_{12}^q,
\end{equation}
where it should be emphasized that $\phi_{12}^q$ enters the decay width difference 
\cite{grossman}.
 The CP-violating phase $\phi_{12}^q$ can be extracted from the charge asymmetry 
$\aslq$ for ``wrong-charge"
semileptonic $B^0_q$-meson decays induced by oscillations, which  is defined as
\begin{equation}
\aslq = \frac{\Gamma(\bar{B}^0_q(t)\rightarrow \ell^+ X) -
              \Gamma(    {B}^0_q(t)\rightarrow \ell^- X)}
             {\Gamma(\bar{B}^0_q(t)\rightarrow \ell^+ X) +
              \Gamma(    {B}^0_q(t)\rightarrow \ell^- X)}.
              \label{aslq}
\end{equation}
This quantity is independent of the decay time $t$, and can be expressed as
\begin{equation}
\aslq = \left|\frac{\Gamma^q_{12}}{M^q_{12}} \right| \sin \phi_{12}^q =
\frac{\Delta \Gamma_q}{\Delta M_q} \tan \phi_{12}^q.
\label{phiq}
\end{equation}
For a much more detailed discussion of this topic, we refer the reader to Ref.~\cite{lenz}.

In experimental measurements, the muon is much easier to identify than any other lepton. Therefore
all experimental results on the semileptonic charge asymmetry are obtained with $\ell=\mu$ in
Eq. (\ref{aslq}). The SM predicts values of $\asld$ and $\asls$ which are not detectable with the
current experimental precision \cite{LN}:
\begin{equation}
\asld |_{\rm SM}= -(4.1 \pm 0.6) \times 10^{-4}, 
\quad 
\asls |_{\rm SM} = (1.9 \pm 0.3) \times 10^{-5}.
\end{equation}
Additional contributions to CP violation via
loop diagrams appear in some extensions of the SM \cite{Randall,Hewett,Hou,Soni,Buras,Buras1}
and can result in these asymmetries
within experimental reach.

The D\O\ experiment performed several measurements of the semileptonic $\Bd$ and $\Bs$ charge asymmetry.
The polarities of the toroidal and solenoidal magnetic fields of D\O\ detector 
were regularly reversed so that the four solenoid-toroid polarity
combinations were exposed to approximately the same integrated luminosity. 
This feature is especially important in the measurements of the charge asymmetry,
because the reversal of magnets polarities allows for a cancellation of first order
effects related with the instrumental asymmetry and the reduction of the corresponding systematic uncertainty.

One of the D\O\ results \cite{asl-d0} consists in measuring the like-sign dimuon charge asymmetry $\aslb$. 
Assuming that this asymmetry is produced by CP violation in the mixing of the $\Bd$ and $\Bs$ mesons, it can
be expressed as
\begin{equation}
\aslb = C_d \asld + C_s \asls,
\end{equation}
where the coefficients $C_d$ and $C_s$ depend on the mean mixing probabilities $\chi_d$ and $\chi_s$
and the production rates of the $\Bd$ and $\Bs$ mesons. Using the integrated luminosity of 9.1 fb$^-1$, the 
D\O\ experiment obtained
\begin{equation}
\aslb = [-0.787 \pm 0.172 \mbox{(stat)} \pm 0.093 \mbox{(syst)}] \%. 
\end{equation}
This result differs by 3.9 standard deviation from the SM prediction \cite{LN}:
\begin{equation}
\aslb |_{\rm SM}=  (-2.3 \pm 0.4) \times 10^{-4}.
\end{equation}
From the study of the impact parameter
dependence of the asymmetry, the D\O\ experiment extracted separate values of $\asld$ and $\asls$
\begin{eqnarray}
\asld & = & (-0.12 \pm 0.52) \%, \nonumber \\
\asls & = & (-1.81 \pm 1.06) \%.
\end{eqnarray}
The correlation $\rho_{ds}$ between these two quantities is
\begin{equation}
\rho_{ds} = -0.799.
\end{equation}

The D\O\ experiment also performed separate measurements of the asymmetries $\asld$ and $\asls$ using
the semileptonic decays $\Bd \to \mu^+ \nu D^- X$, $\Bd \to \mu^+ \mu D^{*-} X$ \cite{asld-d0}, 
and $\Bs \to \mu^+ \nu D_s^- X$ \cite{asls-d0}, respectively. They obtained the following values:
\begin{eqnarray}
\asld = [+0.68 \pm 0.45 \mbox{(stat)} \pm 0.14 \mbox{(syst)}] \%, \\
\asls = [-1.08 \pm 0.72 \mbox{(stat)} \pm 0.17 \mbox{(syst)}] \%.
\end{eqnarray}

Recently, the LHCb collaboration has performed a similar measurement \cite{asls-lhcb} 
of the asymmetry $\asls$ using the decays
$\Bs \to \mu^+ \nu D_s^- X$ and has obtained the most precise value to date : 
\begin{equation}
\asls = [-0.24 \pm 0.54 \mbox{(stat)} \pm 0.33 \mbox{(syst)}] \%. 
\end{equation}
All these results are consistent with one another, 
although the LHCb measurement does not confirm the significant
deviation from the SM observed by the D\O\ experiment.
The $\asld$ asymmetry has also been measured at B-factories with a very good accuracy~\cite{asld-hfag} : 
\begin{equation}
\asld = (0.02 \pm 0.31) \%
\end{equation}
Putting everything together, the deviation from the SM is significantly reduced to 2.4 standard deviations~\cite{asld-hfag}. However, the current size of the  
experimental uncertainties still allows for possible NP contributions.

\boldmath
\subsection{CP Violation in $B^0_s\to J/\psi \phi$}\label{CPV-BsJpsiphi}
\unboldmath
The most prominent $\Bs$-meson decay to explore CP violation is the $B^0_s\to J/\psi \phi$ 
channel. It is the $\Bs$ counterpart of the $B^0_d\to J/\psi K_{\rm S}$ decay, which was
in the main focus of the $B$ factories in the previous decade and has allowed the BaBar
and Belle collaborations to establish CP violation in the $\Bd$ system. 
In the SM, the CP-violating asymmetry is proportional to $\sin 2 \beta$, where
$\beta = \mbox{arg}(-V_{cd}^{}V_{cb}^*/V_{td}^{}V_{tb}^*)$ denotes the usual angle of the
CKM unitarity triangle.  

In the case of $B^0_s\to J/\psi \phi$, CP-violating effects allow us to probe the
CP-violating $B^0_s$--$B^0_s$ mixing phase $\phi_s$, which was introduced in
(\ref{phis-def}) and takes the tiny SM value given in (\ref{phis-SM}). 
Since the final state involves two vector mesons which can be present in final state
configurations $f\in\{0,\parallel,\perp\}$ \cite{rosner}, we have to deal with a mixture of 
CP-even and CP-odd eigenstates. For the extraction of $\phi_s$, these CP eigenstates
have to be disentangled, which can be accomplished with the help of a a time-dependent 
angular analysis of the $J/\psi\to\mu^+\mu^-$ and $\phi\to K^+K^-$ decay products 
\cite{DDLR,DDF}.

\begin{figure}[tbp] 
  \hspace*{1.9truecm} 
 \includegraphics[width=5.2truecm]{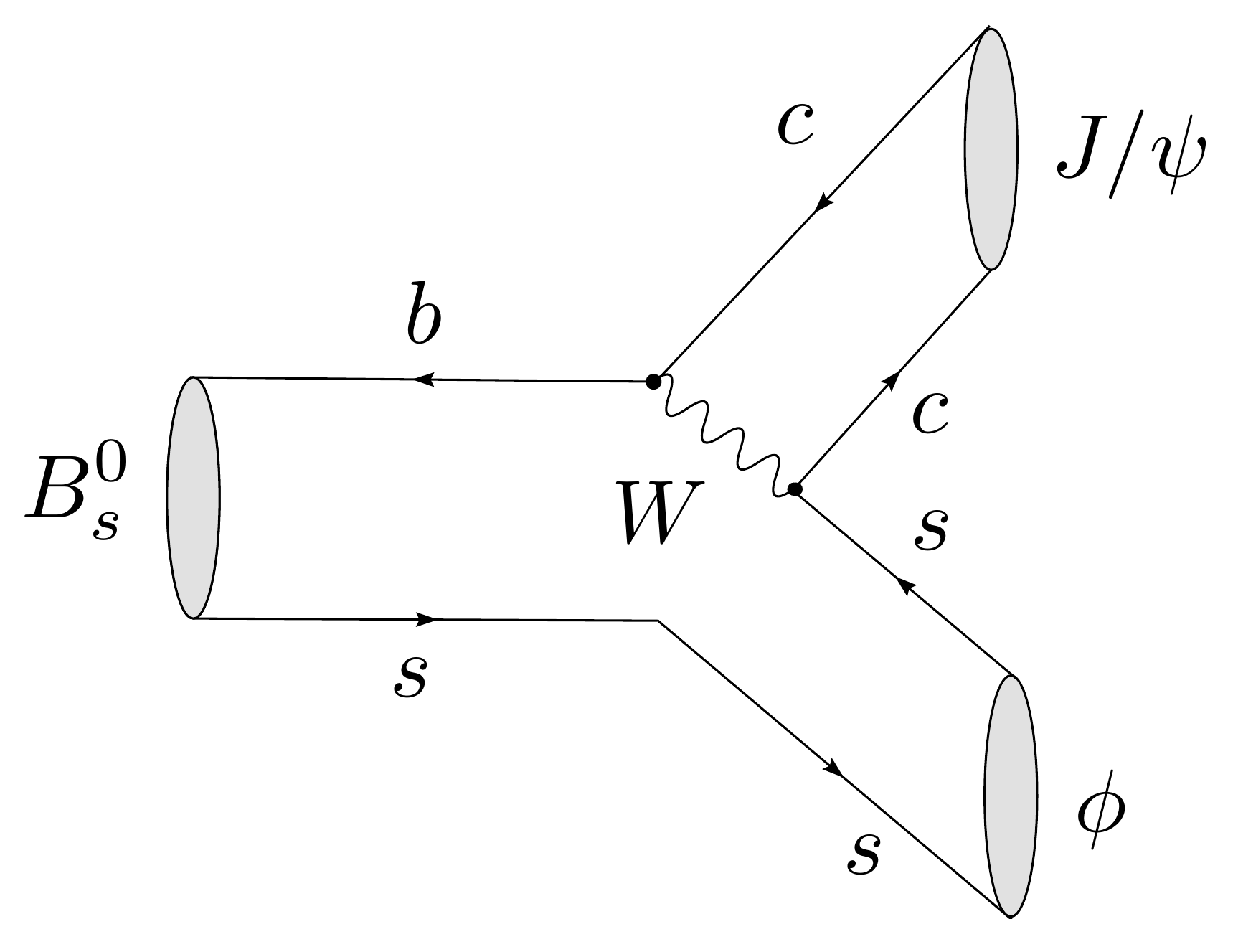}
   \includegraphics[width=5.2truecm]{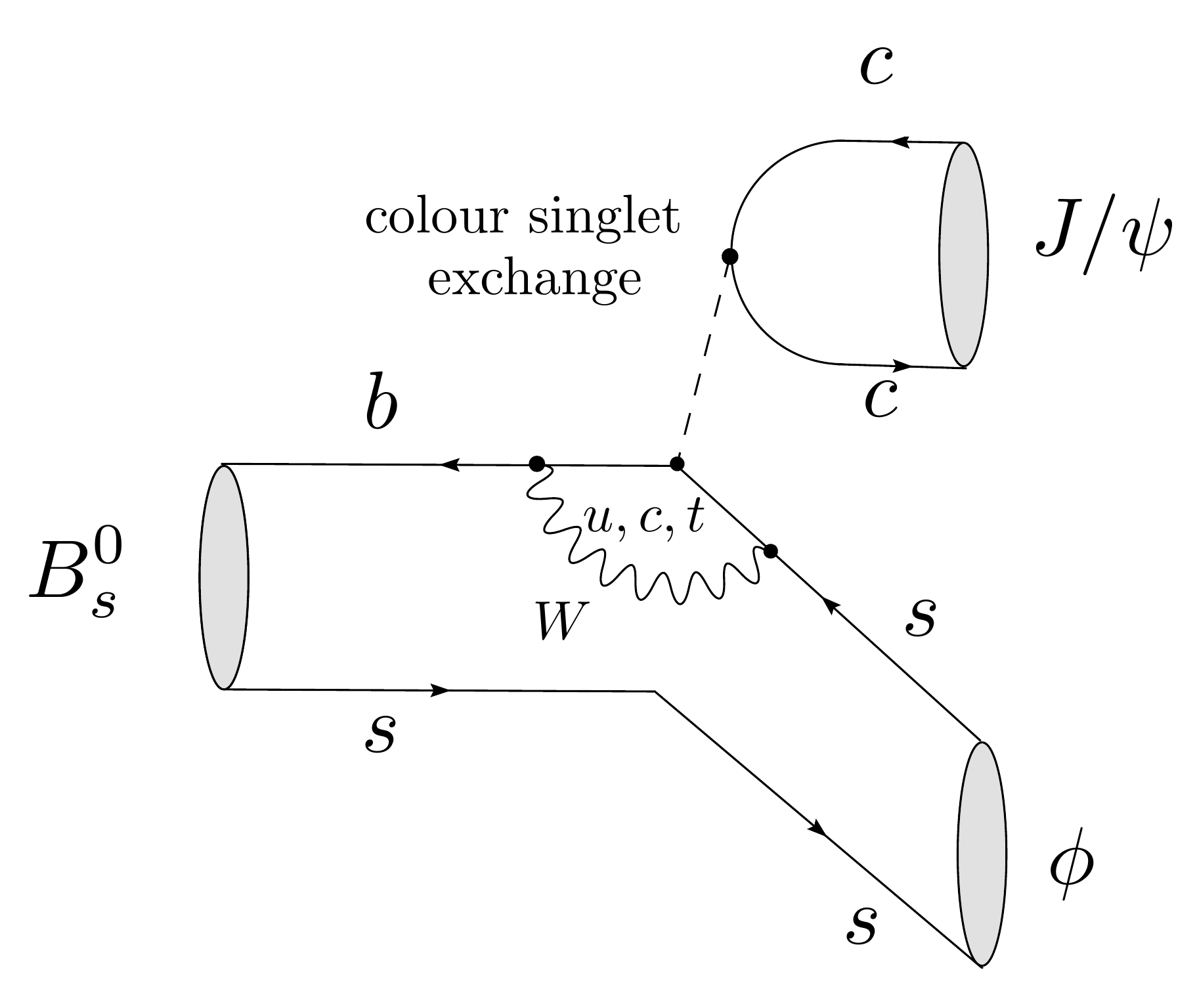} 
   \caption{Decay topologies contributing  to the $B^0_s\to J/\psi \phi$ 
   decay in the SM.}\label{fig:BsJpsiphi}
\end{figure}

In the SM, the $B^0_s\to J/\psi \phi$  decay receives contributions from color-suppressed tree
and penguin topologies, as illustrated in Fig.~\ref{fig:BsJpsiphi}. For a given final-state 
configuration $f$, the corresponding transition amplitude can be written as
follows \cite{FFM}:
\begin{equation}\label{BsJpsiphi-ampl}
A(B^0_s\to (J/\psi \phi)_f)=\left(1-\lambda^2/2\right)
 {\cal A}_f'\left[1+ \epsilon\, a_f' e^{i\theta_f'}{e^{i\gamma}}\right],
\end{equation}
where the following CP-conserving parameters enter:
\begin{equation}\label{hadr}
{\cal A}_f'\equiv \lambda^2 A \left[A_{{\rm T},f}^{(c)'}+A_{{\rm P},f}^{(c)'}-
A_{{\rm P},f}^{(t)'}\right], \quad a_f' e^{i\theta_f'}\equiv R_b
\left[\frac{A_{{\rm P},f}^{(u)'}-A_{{\rm P},f}^{(t)'}}{A_{{\rm T},f}^{(c)'}+
A_{{\rm P},f}^{(c)'}-A_{{\rm P},f}^{(t)'}}\right].
\end{equation}
Here $A_{{\rm T},f}^{(c)'}$ is the color-suppressed tree contribution
and the $A_{{\rm P},f}^{(q)'}$  denote penguin topologies with internal $q$ quarks shown
in Fig.~\ref{fig:BsJpsiphi}. The primes are a reminder that we are dealing with 
a $\bar b\to \bar cc \bar s$  transition. Moreover,
the decay amplitude involves the CKM factors 
\begin{equation}
A\equiv \frac{1}{\lambda^2}|V_{cb}|\sim 0.8, \quad 
R_b\equiv \left(1-\frac{\lambda^2}{2}\right)\frac{1}{\lambda}
\left|\frac{V_{ub}}{V_{cb}}\right|\sim0.5, \quad 
\epsilon\equiv\frac{\lambda^2}{1-\lambda^2}=0.053.
\end{equation}
The  parameters in (\ref{hadr}) suffer from large hadronic uncertainties, in particular
the $a'e^{i\theta'}$, which is a measure of the ratio of the tree to penguin 
contributions. However, as the latter quantity is doubly Cabibbo-suppressed in 
(\ref{BsJpsiphi-ampl}) by the tiny $\epsilon$ parameter,  it is usually neglected. 

The angular analysis allows to construct CP asymmetries in analogy to 
(\ref{CP-asym}) for the final-state configurations $f\in\{0,\parallel,\perp\}$, where
the CP-violating observables can be written as follows \cite{FFM}:
\begin{equation}
C(B_s\to (J/\psi \phi)_f)=
-\frac{2\epsilon a_f'\sin\theta_f'\sin\gamma}{1+2\epsilon a_f'\cos\theta_f'\cos\gamma+
\epsilon^2a_f'^2}
\end{equation}
\begin{equation}\label{S-expr-0}
\frac{S(B_s\to (J/\psi \phi)_f)}{\sqrt{1-C(B_s\to (J/\psi \phi)_f)^2}}
=\sin(\phi_s+\Delta\phi_s^f).
\end{equation}
Here the $\Delta\phi_s^f$ denotes a hadronic phase shift, which is given by
\begin{equation}\label{tanDels}
\tan \Delta\phi_s^f =\frac{2 \epsilon a_f'\cos\theta_f' \sin\gamma+\epsilon^2a_f'^2
\sin2\gamma}{1+ 2 \epsilon a_f'\cos\theta_f' \cos\gamma+\epsilon^2a_f'^2\cos2\gamma}.
\end{equation}
Using data for direct CP violation in $B_d\to J/\psi K_{\rm S}$, the correction due to the
square root in (\ref{S-expr-0}) is tiny, so that this expression can be simplified as
\begin{equation}\label{S-Bspsiphi}
S(B_s\to (J/\psi \phi)_f)=  \sin(\phi_s+\Delta\phi_s^f).
\end{equation}

In the literature, it is usually assumed that
$\Delta\phi_s^f=0$. Making this assumption, HFAG has compiled the 
most recent average  
$\phi_s=-(0.74^{+5.2}_{-4.8})^\circ$ \cite{HFAG}, which is fully consistent with the
SM value in (\ref{phis-SM}). Once the experimental precision improves further, 
even a small phase shift $\Delta \phi_s^f$ 
at the $1^\circ$ level may have a significant impact on the extraction of $\phi_s$ from
(\ref{S-Bspsiphi}) and the resolution of possible CP-violating NP contributions to
$B^0_s$--$\bar B^0_s$ mixing.

A channel to probe these penguin contributions is offered by $B^0_s\to J/\psi \bar K^{*0}$, with
a SM decay amplitude of the structure
\begin{equation}\label{BsJpsiKast}
A(B_s^0\to (J/\psi \bar K^{*0})_f)=-\lambda\,{\cal A}_f\left[1-a_f e^{i\theta_f} e^{i\gamma}\right].
\end{equation}
The key feature is that here the $a_f e^{i\theta_f}$ term is not suppressed by the 
tiny $\epsilon$ parameter. Neglecting penguin annihilation ($PA$) and exchange 
topologies $(E)$, which can be constrained by the upper bound on BR$(B_d\to J/\psi\phi)$ as
 $|E+PA|/|T|\lsim 0.1$, and using the $SU(3)$ flavor symmetry, 
we get the relations $a_f=a_f'$ and $\theta_f=\theta_f'$, allowing us to get a handle on
the penguin shift $\Delta\phi_s^f$  \cite{FFM}. 

Since $B^0_s\to J/\psi \bar K^{*0}$ is a flavor-specific decay and does not exhibit 
mixing-induced CP violation, the implementation of this method has to use 
measurements of untagged and direct CP-violating observables, and an 
angular analysis is required to disentangle the final-state configurations $f$.

The experimental measurement of  CP violation in $\Bs \to J/\psi \phi$ decays has been pioneered by the
CDF and D\O\ collaborations. Both these experiments reported their final study of this channel with the full statistics.
The CDF collaboration \cite{cdf-jpsi} reconstructs about 11000 such decays using the integrated
luminosity 9.6 fb$^{-1}$. 
The obtained confidence regions for the quantity $\beta_s \equiv -\phi_s/2$ is
\begin{eqnarray}
\beta_s & \in & [-\pi/2, -1.51] \cup [-0.06, 0.30] \cup [1.26, \pi/2] ~ \mbox{(68\% C.L.)}, \\
\beta_s & \in & [-\pi/2, -1.36] \cup [-0.21, 0.53] \cup [1.04, \pi/2] ~ \mbox{(95\% C.L.).}
\end{eqnarray}
Assuming the SM value for the CP-violating phase $\beta_s$, the CDF collaboration measured
\begin{eqnarray}
\tau_s & = & [1.528 \pm 0.019 ~\mbox{(stat)} \pm 0.009 ~\mbox{(syst)} ]~\mbox{ps}, \\
\Delta \Gamma_s & = & [0.068 \pm 0.026 ~\mbox{(stat)} \pm 0.009 ~\mbox{(syst)} ]~\mbox{ps}^{-1}.
\end{eqnarray}

A similar analysis 
by the D\O\ collaboration \cite{d0-jpsi}
is based on 6500 signal events collected using the integrated luminosity 8 fb$^{-1}$. 
The result is consistent with the SM prediction:
\begin{eqnarray}
\tau_s & = & [1.443_{-0.035}^{+0.038}]~\mbox{ps}, \nonumber \\
\Delta \Gamma_s & = & [0.163^{+0.065}_{-0.064} ]~\mbox{ps}^{-1}, \nonumber \\
\phi_s & = & -0.55^{+0.38}_{-0.36}.
\end{eqnarray}

\begin{figure}
\centerline
{\includegraphics[height=6cm]{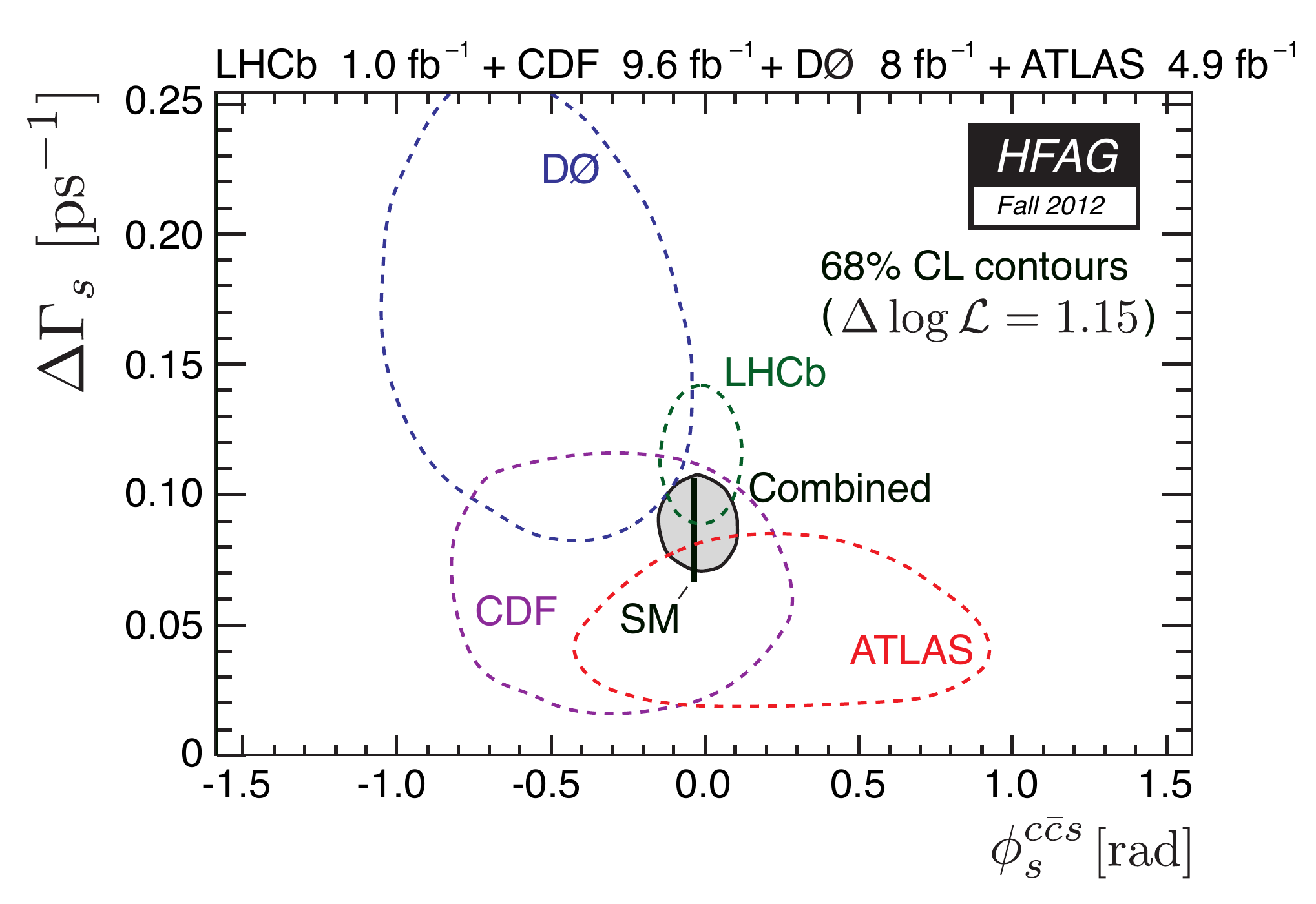}}
\vspace*{-0.5truecm}
\caption{Constraints of all measurements of CP violation in the $\Bs \to J/\psi \phi$ decay  in the 
$\phi_s$--$\Delta \Gamma_s$  plane (from Ref.~\cite{asld-hfag}).}
\label{phis-comb}
\end{figure}

Recently the LHCb experiment reported \cite{jpsi-lhcb} the most precise 
analysis of such measurement. 
Using a data sample of 0.37 fb$^{-1}$, they obtained
\begin{eqnarray}
\Gamma_s & = & [0.657 \pm 0.009 ~\mbox{(stat)} \pm 0.008 ~\mbox{(syst)} ]~\mbox{ps}^{-1}, \\
\Delta \Gamma_s & = & [0.123 \pm 0.029 ~\mbox{(stat)} \pm 0.011 ~\mbox{(syst)} ]~\mbox{ps}^{-1}, \\
\phi_s & = & 0.15  \pm 0.18 ~\mbox{(stat)} \pm 0.06 ~\mbox{(syst)}.
\end{eqnarray}
Finally, the ATLAS experiment also performed a study  of this final state \cite{jpsi-atlas}, 
reporting the following results:
\begin{eqnarray}
\Gamma_s & = & [0.677 \pm 0.007 ~\mbox{(stat)} \pm 0.004 ~\mbox{(syst)} ]~\mbox{ps}^{-1}, \\
\Delta \Gamma_s & = & [0.053 \pm 0.021 ~\mbox{(stat)} \pm 0.010 ~\mbox{(syst)} ]~\mbox{ps}^{-1}, \\
\phi_s & = & 0.22  \pm 0.41 ~\mbox{(stat)} \pm 0.10 ~\mbox{(syst)}.
\end{eqnarray}

The combination of all measurements of CP violation in $\Bs \to J/\psi \phi$ 
 is shown in Fig.\ \ref{phis-comb} taken from~\cite{asld-hfag}.
Since the time-dependent differential decay rates describing the decay $\BsToJpsiPhi$ are invariant under the transformation $(\phi_s, \Delta \Gamma_s) \leftrightarrow  (\pi - \phi_s, - \Delta \Gamma_s)$, together with an appropriate transformation for the strong phases, two solutions are allowed. This ambiguity can be resolved  using the decay $\Bs \to J/\psi K^+ K^-$~\cite{bib:phis_ambiguity_method}. The total decay amplitude is a coherent sum of a slowly varying S-wave (either due to the $f_0(980)$ or a non-resonant contribution) and a P-wave varying rapidly in the $\phi (1020)$ mass region.  
By measuring this phase difference as a function of the $K^+ K^-$ invariant mass, the LHCb collaboration has been able to resolve this ambiguity~\cite{bib:phis_ambiguity_lhcb}: the sign of $ \Delta \Gamma_s$ is determined to be positive, as predicted in the SM.

The $B^0_s\to J/\psi \bar K^{*0}$ decay was observed by CDF \cite{CDF-BsJpsiKS}
and LHCb \cite{LHCb-BsKast}. The most recent LHCb branching ratio 
$(4.4^{+0.5}_{-0.4}\pm0.8) \times 10^{-5}$ agrees well with the prediction 
$(4.6\pm0.4) \times 10^{-5}$ obtained from the BR$(\Bd\to J/\psi \rho^0)$ by means
of the $SU(3)$ flavor symmetry \cite{FFM}, and the polarization fractions agree well with 
those of $\Bd\to J/\psi K^{*0}$. 

The experimental sensitivity for the extraction of $\phi_s$ from $\Bs\to J/\psi \phi$ 
at the LHCb upgrade (50\,$\mbox{fb}^{-1}$) is expected as 
$\Delta\phi_s|_{\rm exp}\sim 0.008 = 0.46 ^\circ$ \cite{LHCb-Strat}. In view of this impressive
precision on the one hand and $\Delta\phi_d=-(1.28\pm0.74)^\circ$ following from the current 
data for $\Bd \to J/\psi \pi, J/\psi K$ decays with a dynamics similar to $\Bs\to J/\psi \phi$
on the other hand \cite{RF-CKM-2012-pen}, it will be crucial to get a handle on the penguin 
effects at the LHCb upgrade as they may mimic NP effects.

\boldmath
\subsection{CP Violation in $B^0_s\to J/\psi f_0(980)$}\label{CPV-BsJpsif0}
\unboldmath
Another $\Bs$-meson decay which has recently entered the stage
is  $B^0_s\to J/\psi f_0(980)$. This channel
was observed by the LHCb \cite{LHCb-f0}, Belle \cite{Belle-f0}, CDF \cite{CDF-f0} and
D\O\ \cite{D0-f0} collaborations. The dominant
decay mode proceeds via $f_0\to\pi^+\pi^-$, with a branching ratio about four times smaller than
that of $B_s^0\to J/\psi \phi$ with $\phi\to K^+ K^-$. On the other hand, since the
$f_0\equiv f_0(980)$ is a scalar $J^{PC}=0^{++}$ state, the final state is CP-odd so that
no angular analysis is required in order to disentangle CP eigenstates as in the case of
the $B^0_s\to J/\psi \phi$ decay. Consequently, the analysis is simplified considerably
and offers an interesting alternative for the determination of $\phi_s$ \cite{SZ,CDeFW}.

The impact of hadronic uncertainties on the extraction of $\phi_s$ from CP violation in
$B^0_s\to J/\psi f_0$ was studied in detail in \cite{FKR}, and for the
$B_{s,d}\to J/\psi \eta^{(\prime)}$ system in \cite{FKR-eta}. The general formalism is
analogous to the discussion in Subsection~\ref{CPV-BsJpsiphi}:
\begin{equation}
A(B^0_s\to J/\psi f_0)\propto
\left [1+\epsilon b e^{i\vartheta} e^{i\gamma}  \right].
\end{equation}
The key feature is again that  the hadronic penguin parameters is doubly Cabibbo-suppressed
by the tiny $\epsilon$. The mixing-induced CP asymmetry can be written as
\begin{equation}\label{Sf0}
S(B^0_s\to J/\psi f_0)=\sqrt{1-C(B^0_s\to J/\psi f_0)^2}\sin(\phi_s+\Delta\tilde\phi_s),
\end{equation}
where $\Delta\tilde\phi_s$ is given by an expression analogous to (\ref{tanDels}).
However, in contrast to the $\Bd\to J/\psi K_{\rm S}$ and $\Bs\to J/\psi \phi$ decays,
the $\Bs\to J/\psi f_0$ channel suffers from the fact that the hadronic structure of the
$f_0(980)$ is poorly known: popular benchmark scenarios are the quark--antiquark
and tetraquark pictures. In the latter case, a peculiar decay topology arises at the tree level
that does not have a counterpart in the quark--antiquark description \cite{FKR}.

The parameter $b$ depends on the hadronic composition of the $f_0$ and is therefore
essentially unknown. Making the conservative assumption $0\leq b\leq0.5$ (where the upper
bound of $0.5$ is related to the $R_b\sim0.5$ factor in (\ref{hadr})) and
$0^\circ\leq\vartheta\leq360^\circ$ yields $\Delta\tilde\phi_s\in [-2.9^\circ, 2.8^\circ]$.
This range translates into the SM range
\begin{equation}\label{f0-SM}
\left.S(B_s \to J/\psi f_0)\right|_{\rm SM} \in [ -0.086, -0.012],
\end{equation}
while the naive value with $\Delta\tilde\phi_s=0^\circ$
reads  $(\sin\phi_s)|_{\rm SM}=-0.036\pm 0.002$ \cite{FKR}.

As we have noted in Subsection~\ref{sec:Bslifetimes}, effective $\Bs$ decay lifetimes
offer an interesting alternative for the extraction of $\phi_s$ and $\Delta\Gamma_s$.
This is also the case for the $B^0_s\to J/\psi f_0$ channel \cite{FK,RK}, where the
situation corresponding to the current data is shown in Fig.~\ref{fig:lifetimes}.
The \Bs\ effective lifetime in the $J/\psi f_0(980)$ final state is currently known with a precision of the order of 3\% (see Section \ref{sec:Bslifetimes}).
The corresponding contour in the $\phi_s$--$\Delta\Gamma_s$ plane is very robust with
respect to the hadronic penguin uncertainties, thereby nicely complementing the
analysis of (\ref{Sf0}). A future measurement of the effective lifetime $\tau_{J/\psi f_0}$
with $1\%$ uncertainty would be most interesting.

A study of CP violation in the $\Bs \to J/\psi f_0(980)$ decay was done by the LHCb collaboration \cite{jpsif0-lhcb}, with the result
\begin{equation}
\phi_s =  -0.44  \pm 0.44 ~\mbox{(stat)} \pm 0.02 ~\mbox{(syst)}.
\end{equation}
In addition, LHCb studied  $\Bs \to J/\psi \pi^+ \pi^-$ decays \cite{jpsipipi-lhcb}, which
includes both the $f_0(980)$ and non-resonant final state. The obtained value of $\phi_s$ is more precise
\begin{equation}
\phi_s =  -0.019^{+0.173}_{-0.174} ~\mbox{(stat)} ^{+0.004}_{-0.003} ~\mbox{(syst)}.
\end{equation}

In view of the large errors in the value of $\phi_s$,
the hadronic corrections discussed above are not yet relevant
in this analysis. However, once the experimental result enters the SM range in (\ref{f0-SM}),
the hadronic phase shift $\Delta\tilde\phi_s$ has to be controlled in order to match the
theoretical and experimental precisions.

In order to obtain insights into these effects, it would be interesting to compare the
separate measurements of $\phi_s$ from $B^0_s\to J/\psi \phi$ and $B^0_s\to J/\psi f_0$
with each other as the hadronic penguin effects have a different impact on these
determinations. With the LHCb upgrade project, the foreseen experimental
uncertainties with 50~\invfb,
are equal to $0.46^\circ$ and $0.80^\circ$ for the
$B^0_s\to J/\psi \phi$ and $B^0_s\to J/\psi f_0$ decays, respectively~\cite{LHCb-Strat}.
In particular, in the high-precision era of the LHCb upgrade, these measurements
should not be averaged in a naive way, neglecting the hadronic corrections.

Another possibility  to probe the penguin effects directly  is offered by
the $B^0_d\to J/\psi f_0$ channel. Estimates have shown that its branching ratio
with $f_0\to\pi^+\pi^-$ could be as large as ${\cal O}(10^{-6})$ \cite{FKR}.
The translation of the corresponding penguin parameters into their counterparts
entering the $B_s \to J/\psi f_0$ mode depends unfortunately also on assumptions
about the hadronic structure of the $f_0(980)$. However, a better picture of this
still unsettled hadronic scalar state may be available once these challenging measurements
can be performed in practice.

\boldmath
\subsection{CP Violation in $B^0_s\to J/\psi K_{\rm S}$}\label{CPV-BsJpsiKS}
\unboldmath
The decay $B^0_s\to J/\psi K_{\rm S}$ originates from $\bar b\to \bar c c\bar d$ 
quark-level processes and is related to $B^0_d\to J/\psi K_{\rm S}$ through the 
$U$-spin flavor symmetry of strong interactions, which relates down and strange
quarks to each other in a manner similar to the $SU(2)$ isospin symmetry connecting
the up and down quarks \cite{RF-BsJpsiK}. In the SM, the decay amplitude of this
channel can be written as 
\begin{equation}\label{BsJpsiK}
A(B_s^0\to J/\psi\, K_{\rm S})=-\lambda\,{\cal A}\left(1-a e^{i\theta}e^{i\gamma}\right),
\end{equation}
which has a structure similar to (\ref{BsJpsiKast}). On the other hand, we have
\begin{equation}\label{BJpsiK-ampl}
A(B_d^0\to J/\psi\, K_{\rm S})=\left(1-\lambda^2/2\right){\cal A'}
\left(1+\epsilon a'e^{i\theta'}e^{i\gamma}\right),
\end{equation}
where the penguin parameter $a'e^{i\theta'}$ enters in a doubly Cabibbo-suppressed way. 
The $U$-spin symmetry implies the relation
\begin{equation}\label{U-spin-1}
a'=a, \quad \theta'=\theta.
\end{equation}

As was pointed out in Ref.~\cite{RF-BsJpsiK}, the information offered by the ratio
of the $B_{s,d} \to J/\psi K_{\rm S}$ branching ratios
and the direct and mixing-induced CP asymmetries 
of the $B_s\to J/\psi K_{\rm S}$ channel 
can be converted into the angle $\gamma$ and the penguin parameters $a$, $\theta$ by means
of the $U$-spin symmetry. 

In 1999, the $\gamma$ determination 
appeared the most interesting aspect of this strategy \cite{RF-BsJpsiK}. A feasibility study was
preformed in Ref.~\cite{DeBFK}. It showed that the extraction of $\gamma$ looks feasible 
for the LHCb upgrade era but will probably not be competitive with other methods for 
the extraction of $\gamma$. On the other hand, if $\gamma$ is used as an input, the 
penguin parameters $a$ and $\theta$ can be extracted in a theoretically clean way from 
the CP-violating $B^0_s\to J/\psi K_{\rm S}$ asymmetries. Using (\ref{U-spin-1}), the 
penguin parameters affecting the extraction of the angle $\beta$ of the unitarity triangle 
from the CP violation in $B^0_d\to J/\psi K_{\rm S}$ can then be determined. 

The extraction of $a$ and $\theta$ will be the key application of the $B^0_s\to J/\psi K_{\rm S}$
channel. Since the dynamics is similar to that of the $B^0_s\to J/\psi \bar K^{*0}, J/\psi \phi$ system,
valuable insights into the size of the penguin uncertainties affecting the extraction of $\phi_s$,
as discussed in Subsection~\ref{CPV-BsJpsiphi}, can be obtained. 

The $B_s\to J/\psi K_{\rm S}$ channel has been observed by CDF \cite{CDF-BsJpsiKS}
and LHCb \cite{LHCb-BsJpsiKS}, but so far only measurements of the branching ratio 
are available, where the subtleties related to the sizable $B_s$ decay width difference 
$\Delta\Gamma_s$ discussed in Subsection~\ref{sec:untagged} have to be taken
into account. A test of the $SU(3)$ flavor symmetry is provided by the following
ration \cite{DeBFK,LHCb-BsJpsiKS}:
\begin{equation}
\Xi_{SU(3)}\equiv
\frac{\Phi^d_{J/\psi \pi^0}}{\Phi^s_{J/\psi K_{\rm S}}} \frac{\tau_{B_d}}{\tau_{B_s}}
\left[
\frac{\mbox{BR}(B^0_s\to J/\psi \bar K^0)_{\rm theo}}{2\mbox{BR}(B^0_d\to J/\psi\pi^0)_{\rm theo}}
\right]=0.93\pm0.15,
\end{equation}
where the $\Phi$ and $\tau_{B_q}$ denote phase-space factors and $B_q$ lifetimes, respectively,
and the ``theoretical" branching ratios refer to the definition in (\ref{BR-theo}). In this
expression, tiny penguin annihilation and exchange topologies (which can be constrained
through experimental data) have been neglected. In the $SU(3)$ limit, we have then
$\Xi_{SU(3)}=1$, which is consistent with the numerical value following from the current
branching ratio measurement.

We look forward to future measurements of the effective $B_s\to J/\psi K_{\rm S}$ lifetime
and the CP asymmetries of this channel. 

\boldmath
\subsection{CP Violation in $B^0_s\to K^+K^-$}\label{CPV-BsKK}
\begin{figure} 
\centerline{
 \includegraphics[width=4.6truecm]{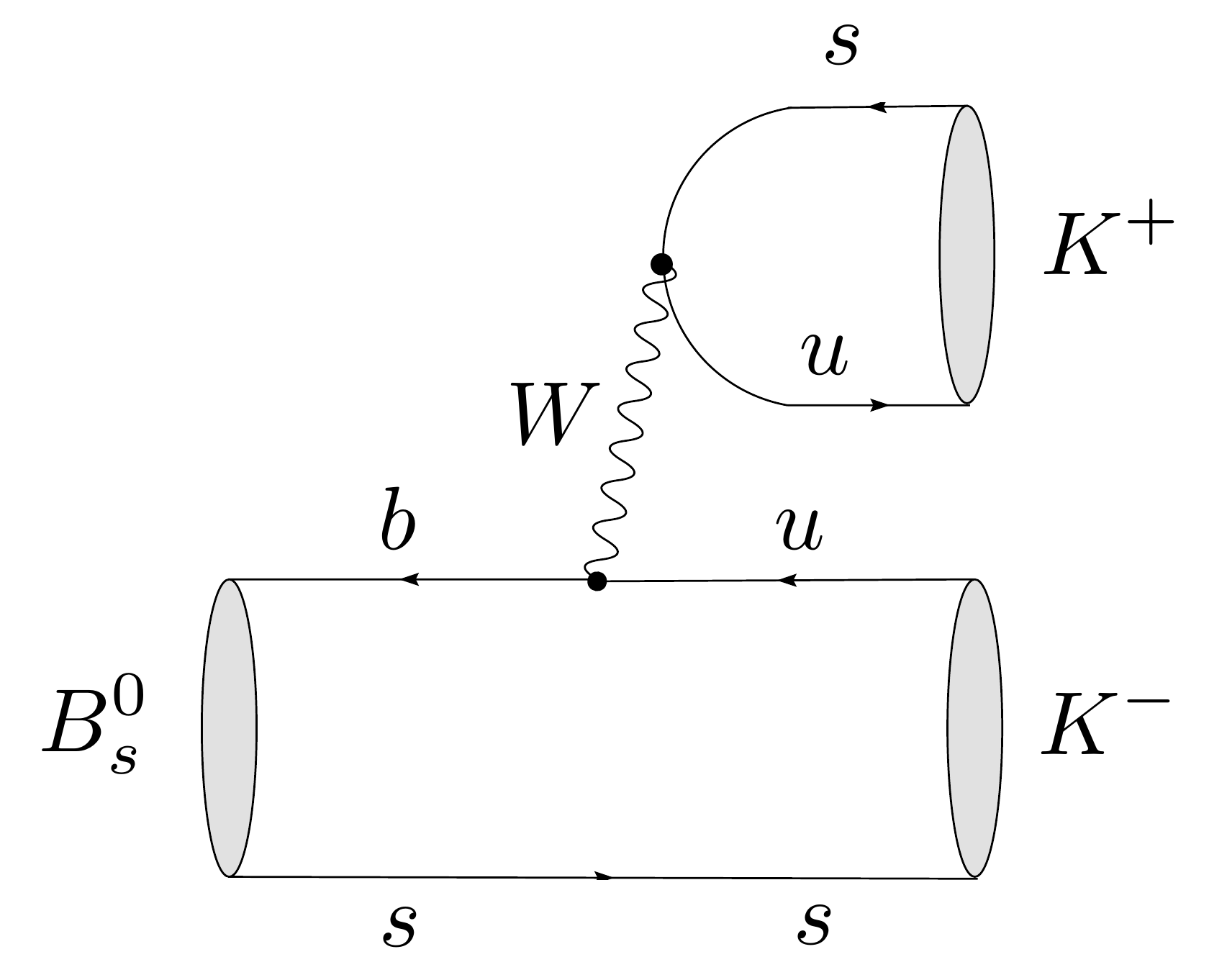}
 \hspace*{0.5truecm}
 \includegraphics[width=5.0truecm]{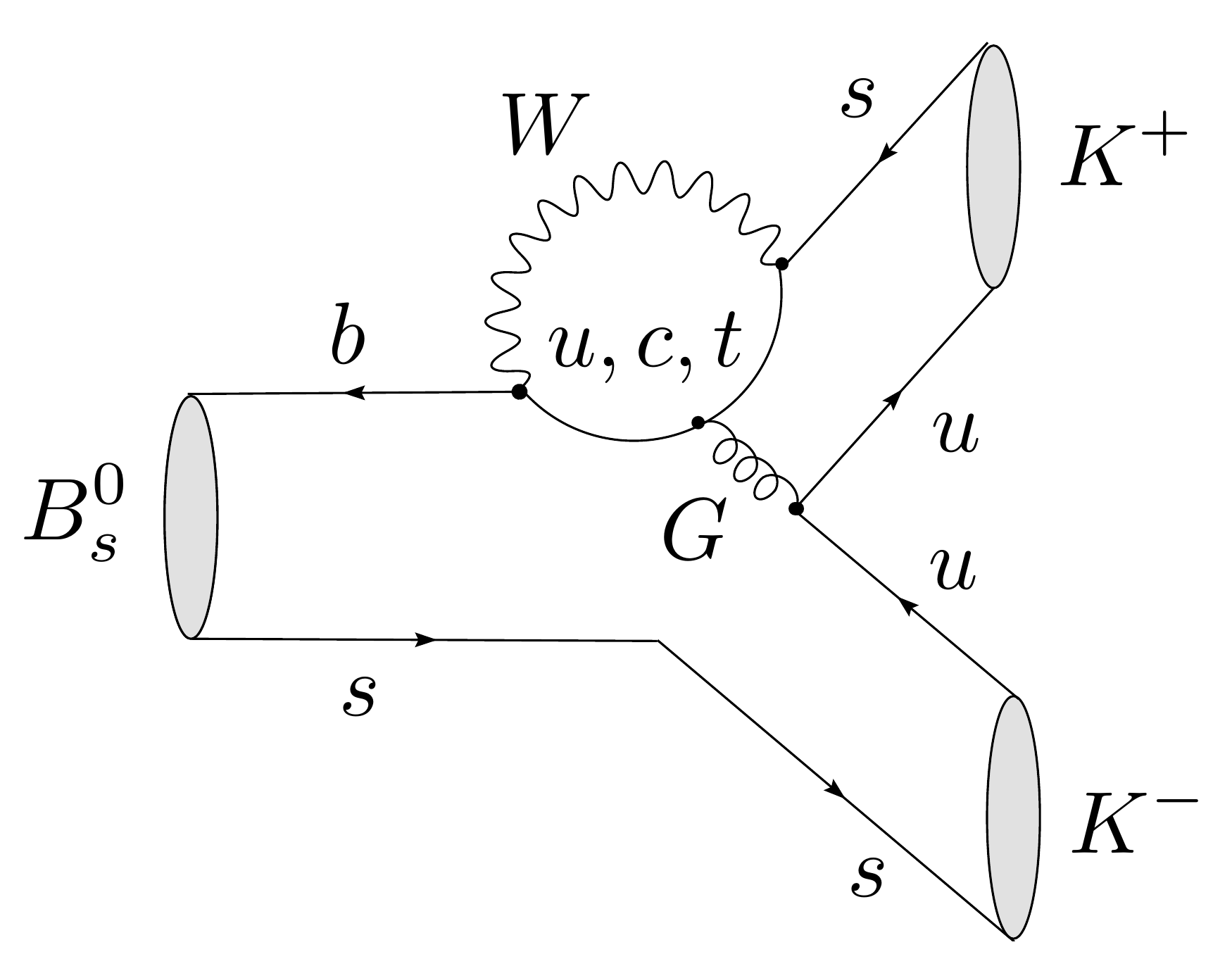}  
 }
 \vspace*{-0.5truecm}
\caption{Illustration of the tree (left) and penguin (right) diagrams 
contributing to the $B^0_s\to K^+K^-$ decay.}\label{fig:BsKK-diag}
\end{figure}

\unboldmath
The decay $B^0_s\to K^+K^-$ originates from $\bar b \to \bar s u \bar u$ quark-level
transitions and receives contributions from tree and penguin topologies,
as illustrated in Fig.~\ref{fig:BsKK-diag}. In the SM, the corresponding decay
amplitude can be written as follows \cite{RF-BsKK}:
\begin{equation}\label{BsKK-ampl}
A(B_s^0\to K^+K^-)=e^{i\gamma}\lambda\,{\cal C}'\left[1+\frac{1}{\epsilon}
d'e^{i\theta'}e^{-i\gamma}\right],
\end{equation}
where ${\cal C'}$ and $d' e^{i\theta'}$ are CP-conserving parameters. While the
former is governed by the tree topology, the latter is a measure of the ratio of the
penguin to tree amplitudes. Interestingly, thanks to the peculiar CKM structure
of (\ref{BsKK-ampl}), the $B^0_s\to K^+K^-$ decay is dominated by the
penguin topologies.

Looking at the diagrams in Fig.~\ref{fig:BsKK-diag}, we observe that we get the
topologies for the $B^0_d\to\pi^+\pi^-$ decay by interchanging the roles of all
down and strange quarks. Consequently, the two decays are related to each
other through the $U$-spin symmetry of strong interactions, in analogy to the
$B_{s,d}\to J/\psi K_{\rm S}$ system discussed in Section~\ref{CPV-BsJpsiKS}. 
The corresponding decay amplitude is given in the SM as follows:
\begin{equation}\label{Bdpipi-ampl}
A(B_d^0\to\pi^+\pi^-)=e^{i\gamma}\left(1-\frac{\lambda^2}{2}\right){\cal C}
\left[1-d\,e^{i\theta}e^{-i\gamma}\right],
\end{equation}
where ${\cal C}$ and $d\,e^{i\theta}$ are the counterparts of the primed quantities in 
(\ref{BsKK-ampl}). The $B^0_d\to \pi^+\pi^-$ channel is dominated by the tree contributions. 

If we apply the $U$-spin symmetry, we obtain the relations \cite{RF-BsKK}
\begin{equation}\label{U-spin-rel-2}
d'=d, \quad \theta'=\theta.
\end{equation}
As the $B^0_{d,s}$--$\bar B^0_{d,s}$ mixing phases are known, the direct and 
mixing-induced CP asymmetries of the $B_d\to\pi^+\pi^-$ and $B_s\to K^+K^-$
decays allow the determination of theoretically clean contours in the
$\gamma$--$d$ and $\gamma$--$d'$ planes, respectively. Using the first
relation in (\ref{U-spin-rel-2}), $\gamma$ and the hadronic parameters $d$, $\theta$ and
$\theta'$ can be determined \cite{RF-BsKK}, where the strong phases $\theta$
and $\theta'$ offer an internal test of the $U$-spin symmetry.  

In Refs.~\cite{RF-Bhh,FK-BsKK}, detailed discussions of this strategy can be found. 
The $B^0_s\to K^+K^-$ decay has been observed by the CDF \cite{CDF-BsKK}, Belle 
\cite{Belle-BsKK} and LHCb \cite{bs-kk-lhcb-1} collaborations. Using the branching
ratio information, non-perturbative QCD sum rule calculations of the form-factor ratio
entering $|{\cal C}'/{\cal C}|$ \cite{DuMe}, and measurements of CP violation in
$B_d\to\pi^+\pi^-$, $B_d\to\pi^\mp K^\pm$, the following result for $\gamma$ 
was determined in Ref.~\cite{FK-BsKK}:
\begin{equation}\label{gam-BsKK}
\gamma=(68.3^{+4.9}_{-5.7}|_{\rm input}\mbox{}^{+5.0}_{-3.7}|_\xi
\mbox{}^{+0.1}_{-0.2}|_{\Delta\theta})^\circ,
\end{equation}
where the first error is due to the uncertainties of the input quantities, and the latter
errors describe $U$-spin-breaking effects parametrized as $\xi\equiv d'/d = 1 \pm 0.15$
and $\Delta\theta\equiv\theta'-\theta = \pm 20^\circ$. The result for $\gamma$ in 
(\ref{gam-BsKK})  is in excellent agreement with the fits of the unitarity triangle. 

The usefulness of the effective $B_s\to K^+K^-$ lifetime has already been addressed in
Section~\ref{sec:Bslifetimes} and Fig.~\ref{fig:lifetimes}. For a more detailed discussion,
the reader is referred to Ref.~\cite{FK}. 

A variant of the $B_s\to K^+K^-$, $B_d\to \pi^+\pi^-$ strategy for the extraction of $\gamma$
discussed above has recently been discussed in \cite{CFMS}. The experimental prospects
of the exploration of the $B_s\to K^+K^-$, $B_d\to \pi^+\pi^-$ system are promising for
the LHCb experiment \cite{LHCb-Strat}. 
First results have been obtained by the LHCb collaboration using only a fraction of the 2011 data 
(0.69 \invfb)~\cite{bib:CPV-BsToKK-LHCb} 
In the notation of Eq.~(\ref{CP-asym}), these results read as follows:
\begin{eqnarray}
C(B_s \to K^+K^-)  &= &-0.02 \pm 0.18 (\rm stat) \pm 0.04 (\rm syst) \stackrel{{\rm SM}}{=}
0.098\pm0.04, \nonumber\\
S(B_s \to K^+K^-)  &= &0.17 \pm 0.18 (\rm stat) \pm 0.05 (\rm syst)\stackrel{{\rm SM}}{=}
0.215^{+0.060}_{-0.047};
\end{eqnarray}
for comparison, we give also the SM predictions obtained in Ref.~\cite{FK-BsKK}.

\boldmath
\subsection{CP violation in $B^0_s\to \pi^+ K^-$}
\unboldmath
Another interesting decay is $B^0_s\to \pi^+ K^-$, which receives contributions from
penguin and tree topologies and can be combined with $B^0_d\to \pi^- K^+$ to determine the
CKM angle $\gamma$ \cite{RF-Bhh,groro,FK-CKM}. As this channel is flavor-specific, it does not
exhibit mixing-induced CP violation. Using $SU(3)$ flavor-symmetry arguments
yields the relations
\begin{eqnarray}
\lefteqn{{\cal A}_{\rm CP}^{\rm dir}(B^0_s\to \pi^+ K^-)\approx
{\cal A}_{\rm CP}^{\rm dir}(B^0_d\to \pi^+ \pi^-)}\nonumber\\
&&\approx
-\left[\frac{\mbox{BR}(B_d\to\pi^\mp K^\pm)}{BR(B_s\to\pi^\pm K^\mp)}\right]
{\cal A}_{\rm CP}^{\rm dir}(B^0_d\to \pi^- K^+)\label{rel-CPdir}
\end{eqnarray}
between the direct CP asymmetries of the $B^0_s\to \pi^+ K^-$, $B^0_d\to \pi^-K^+$
and $B^0_d\to \pi^+ \pi^-$ channels, arising from the interference between tree and
penguin contributions. The $B^0_d\to \pi^-K^+$ decay has allowed the establishment
of direct CP violation in the $B$-meson system
\cite{CPV_BABAR_BsToKPi,Belle-CPdir,TeV-CPdir},
thereby complementing the measurement of direct CP violation in $K_{\rm L,S}\to\pi\pi$
decays through a non-vanishing value of
the Re$(\varepsilon'/\varepsilon)$ parameter by the  KTeV \cite{KTeV}  and NA48 \cite{NA48}
collaborations.

The $B^0_s\to \pi^+ K^-$ channel has been extensively studied at the Tevatron
\cite{CPV_CDF_BsToKPi} and the Belle experiment  \cite{Belle-CPdir}.

\begin{figure}
{\includegraphics[height=5cm]{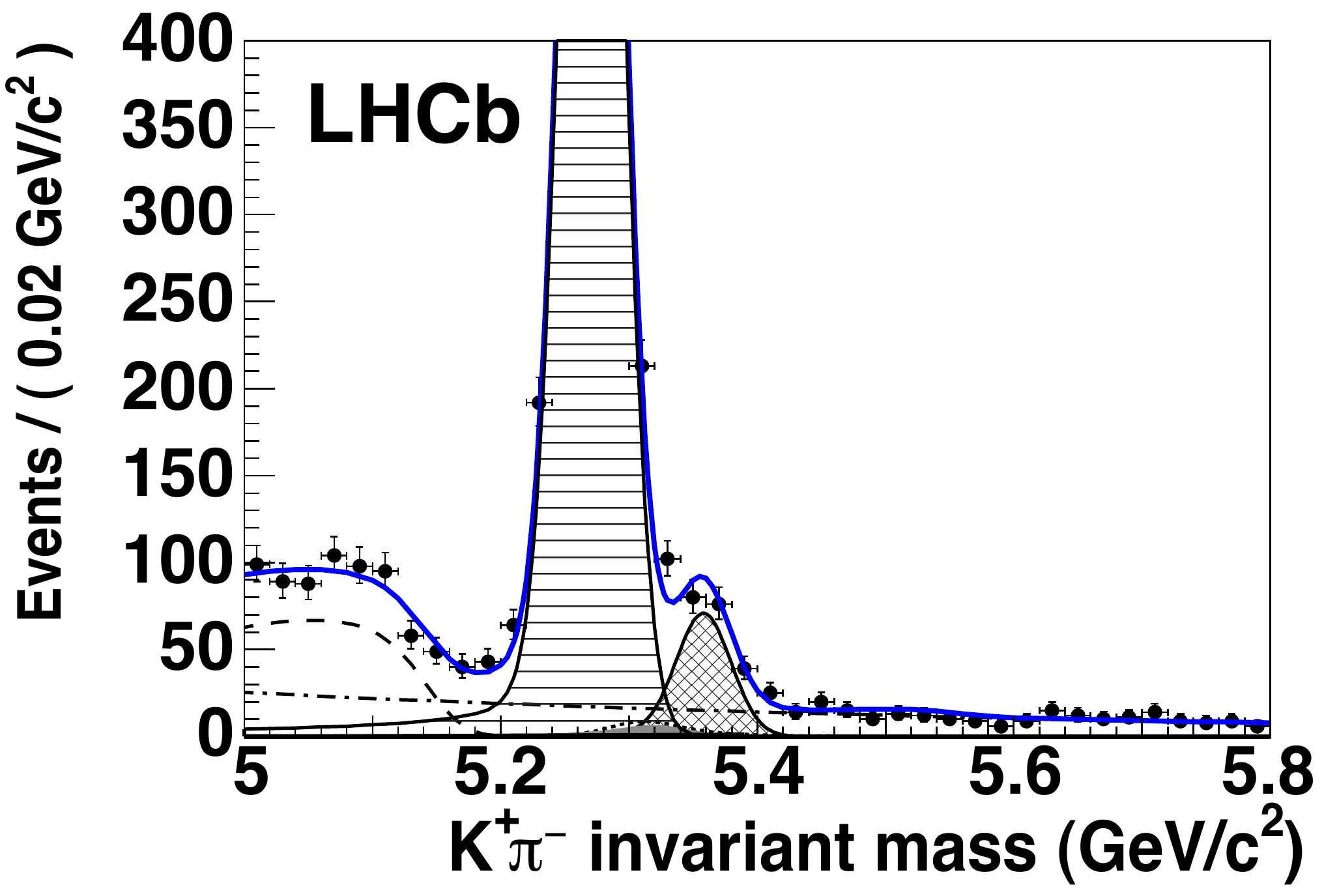}}
\vskip -5cm
\hskip 9cm
{\includegraphics[height=5cm]{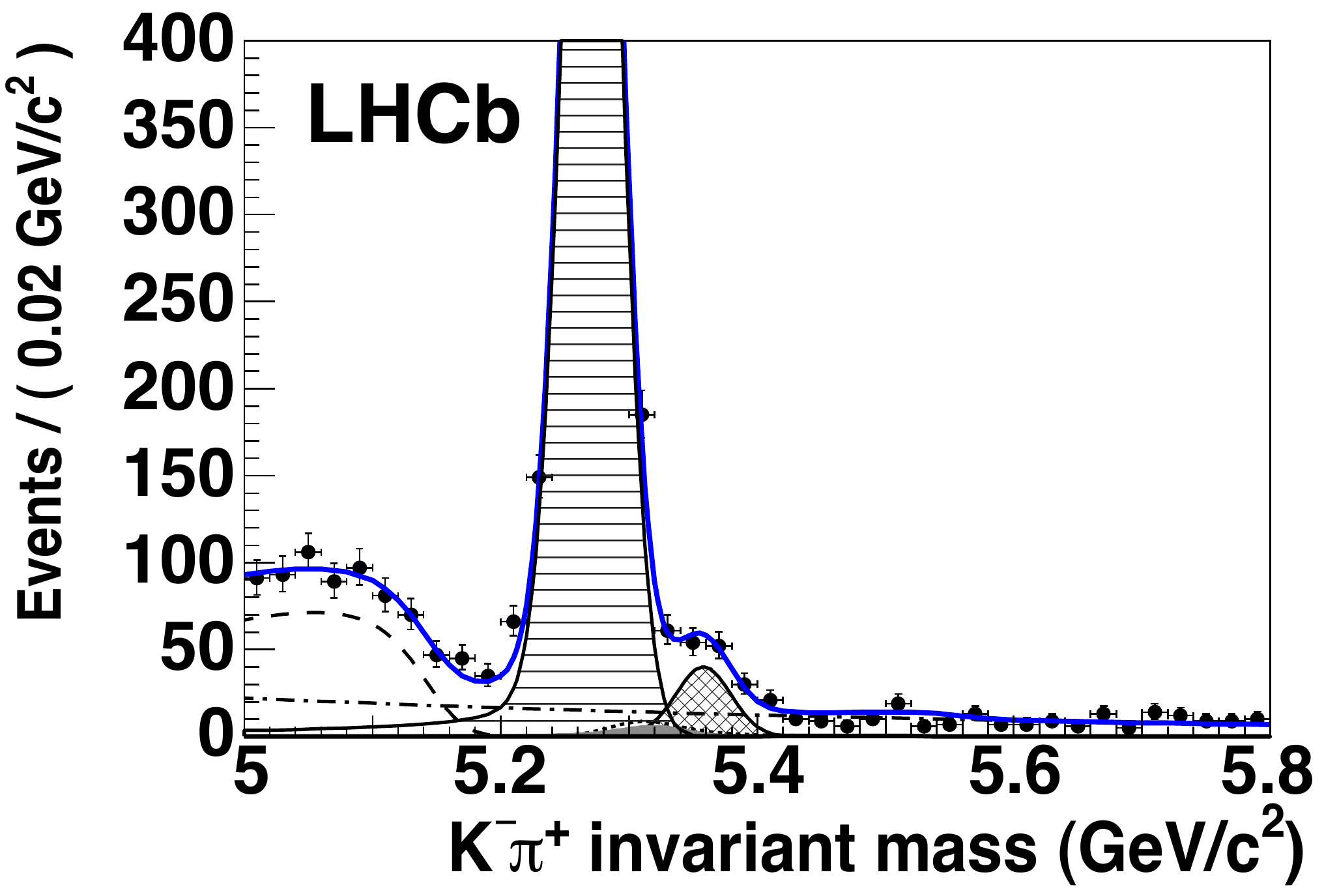}}
\vskip -4.7cm
\hskip 12.8cm
{\includegraphics[height=3.cm]{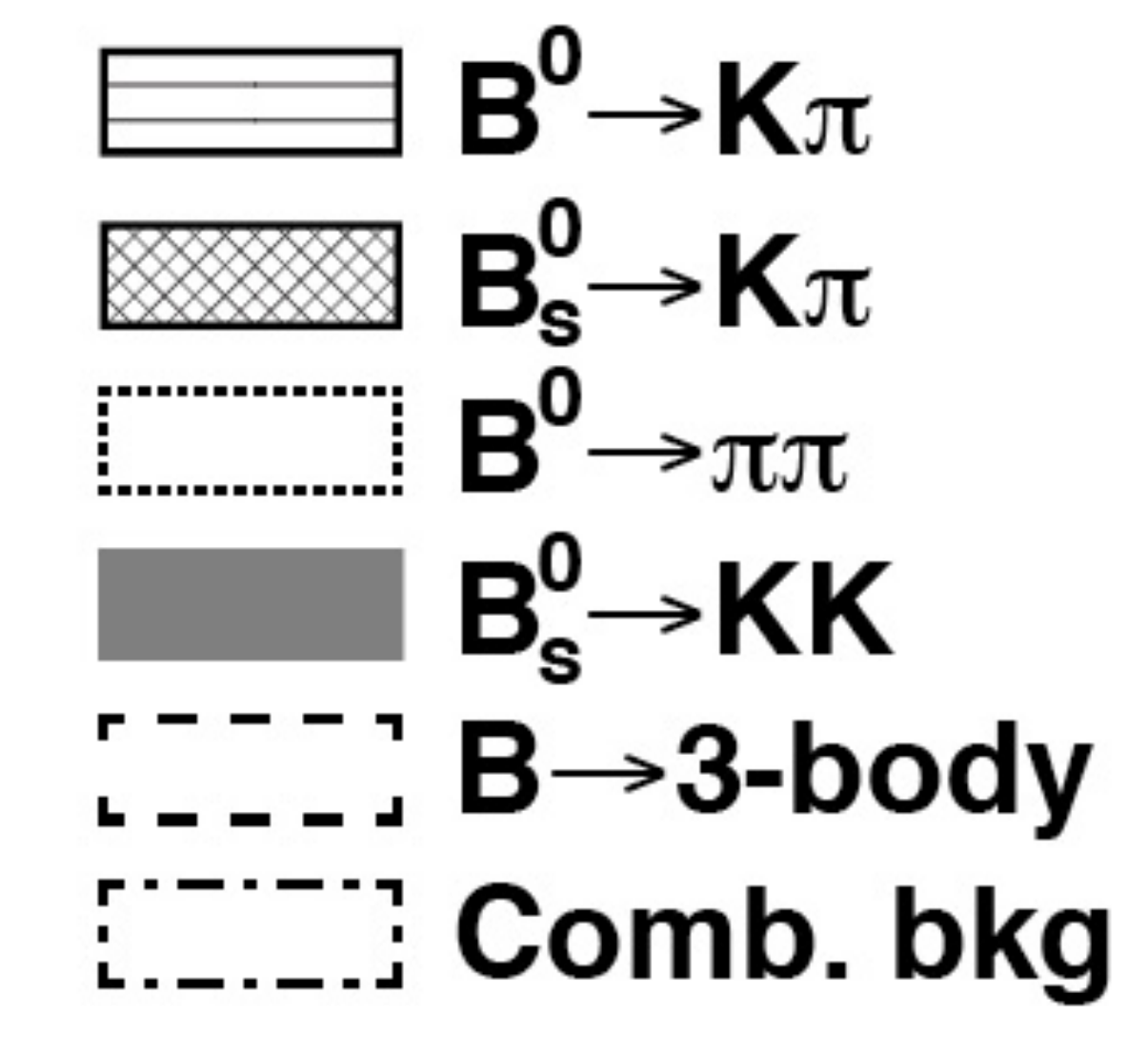}}
\vskip 1cm
\begin{center}
\vspace*{0.8truecm}
\caption{Invariant $K \pi$ mass spectra. The right plot corresponds to the $K^+ \pi^-$ invariant mass while the left plot
corresponds to the $K^- \pi^+$ invariant mass. The results of the unbinned maximum likelihood fits are overlaid. The main components contributing to the fit model are also shown (from 
Ref.~\cite{CPV_LHCb_BsToKPi}).}
\label{fig:CPV_BsToKPi}
\end{center}
\end{figure}

In 2012, the LHCb experiment has provided  the
first evidence of direct CP violation in this channel  \cite{CPV_LHCb_BsToKPi},
confirmed later by the CDF experiment \cite{CPV_CDF_BsToKPi-1}:
\begin{eqnarray}
{\cal A}_{\rm CP}^{\rm dir}(B^0_s\to \pi^+ K^-)
& = & 0.27 \pm 0.08 \mbox{ }(\rm stat) \pm 0.02 \mbox{ }(\rm syst)~\mbox{(LHCb)}, \\
{\cal A}_{\rm CP}^{\rm dir}(B^0_s\to \pi^+ K^-)
& = & 0.22 \pm 0.07 \mbox{ }(\rm stat) \pm 0.02 \mbox{ }(\rm syst)~\mbox{(CDF)},
\end{eqnarray}
which is the first signal of CP violation in the $B_s$-meson system.
The clear asymmetry is already visible from the raw signal yields from Fig.~\ref{fig:CPV_BsToKPi},
and the result is consistent with the relations in (\ref{rel-CPdir}) for the most recent measurements
of CP violation in $B^0_d\to\pi^+\pi^-$.

\boldmath
\subsection{CP Violation in $B_s\to D_s^\pm K^\mp$ Decays}\label{BsDsK}
\begin{figure}[t]
\centerline{
 \includegraphics[width=6.2truecm]{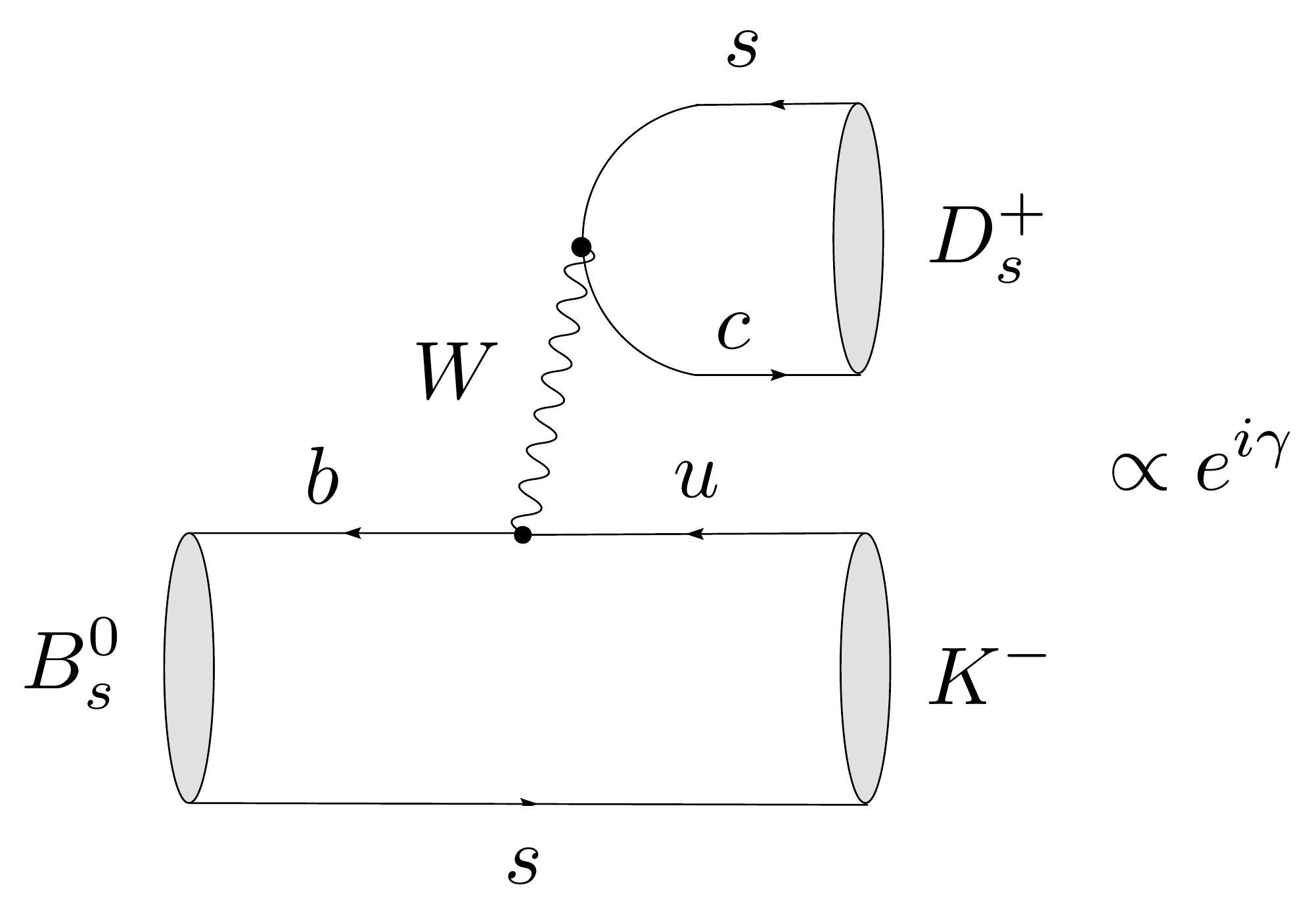}
 \hspace*{0.5truecm}
 \includegraphics[width=6.2truecm]{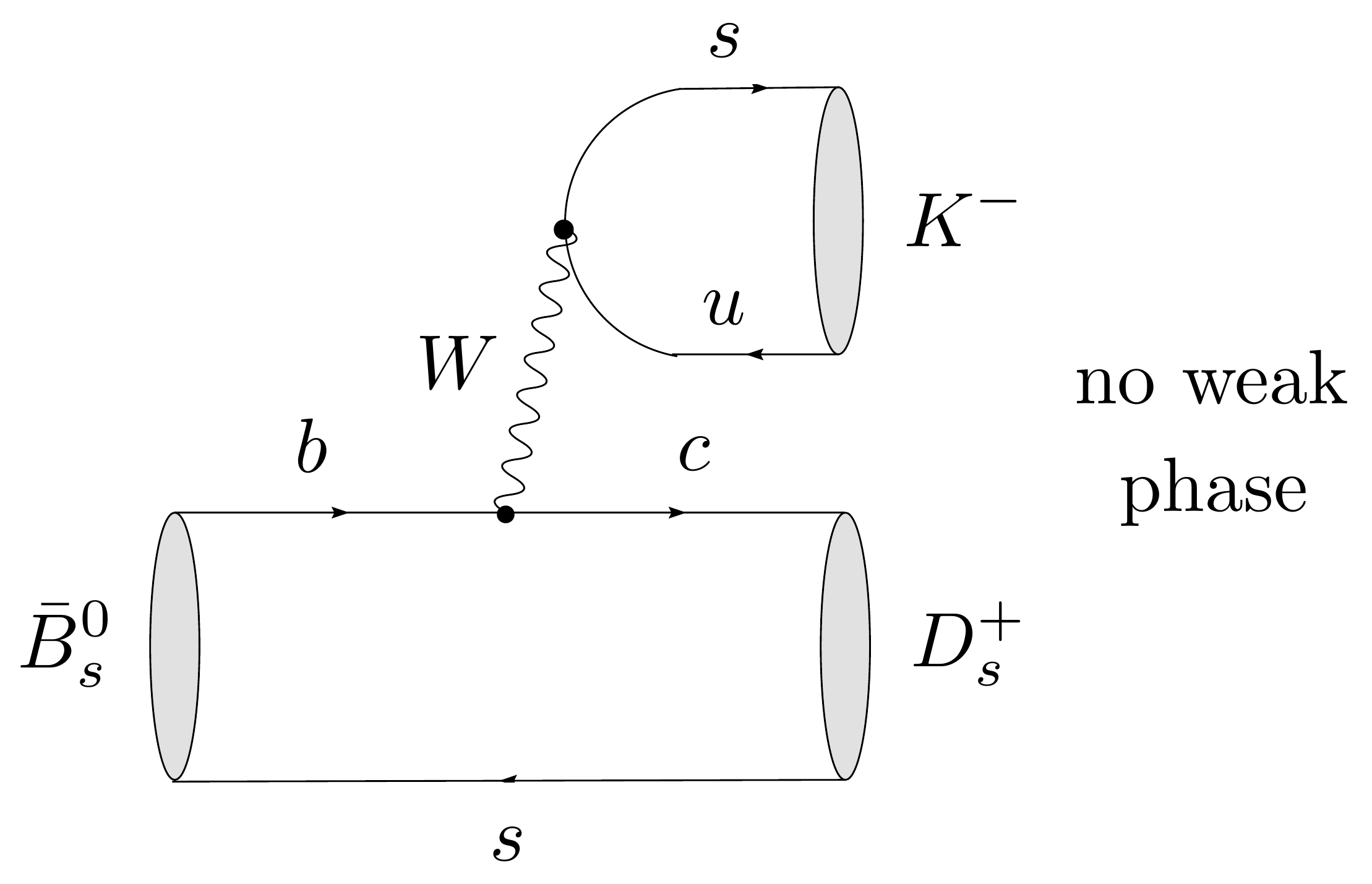}
 }
 \vspace*{-0.3truecm}
\caption{Feynman diagrams contributing to $B^0_s\to D_s^+K^-$
and $\bar B^0_s\to D_s^+ K^-$ decays.}\label{fig:BsDsK}
\end{figure}

\unboldmath
In contrast to the modes discussed in the previous sections,
$B_s\to D_s^\pm K^\mp$ decays receive only
contributions from tree-diagram-like topologies, i.e.\ there are no penguin contributions
present. As can be seen in Fig.~\ref{fig:BsDsK}, both $B^0_s$ and $\bar B^0_s$ mesons
can decay into the $D_s^+ K^-$ final state. Consequently, interference effects between
$B^0_s$--$\bar B^0_s$ mixing and decay processes lead to a time-dependent CP-violating
rate asymmetry, which has the same form as the expression in (\ref{CP-asym}). Moreover,
both decay paths are of the same order $\lambda^3$ in the Wolfenstein expansion \cite{wolf},
thereby leading to large interference effects. The corresponding CP asymmetries
provide sufficient information to determine the phase $\phi_s+\gamma$ in a theoretically
clean way \cite{BsDsK,RF-BsDsK}. The decay width difference $\Delta\Gamma_s$
offers new observables for this method and allows an unambiguous determination of
$\phi_s+\gamma$ \cite{RF-BsDsK}, as studied in
Refs.~\cite{Cavoto:2006um}--\cite{Gligorov:2011id}. For a detailed recent analysis of the
$B_s\to D_s^{(*)\pm} K^\mp$ system in view of the sizable $\Delta\Gamma_s$,
the reader is referred to Ref.~\cite{BsDsK-12}.

Concerning the experimental status, the CDF \cite{CDF-BsDsK}, Belle \cite{Belle-BsDsK}
and LHCb \cite{LHCb-BsDsK} collaborations have reported first measurements of the
$B_s \to D_s^\pm K^\mp$ branching ratio:
\begin{equation}\label{LHCb-CDF-BR}
	\frac{{\rm BR}(B_s\to
	D_s^\pm K^\mp)_{\rm exp}}{{\rm BR}(B_s\to D_s^\pm \pi^\mp)_{\rm exp}} =
	\left\{
	\begin{array}{lcl}	
		0.097 \pm 0.018\:({\rm stat.}) \pm 0.009\:({\rm syst.}) &  &{\rm [CDF],}\\
		0.065^{+0.035}_{-0.029} \:({\rm stat.}) &  &{\rm [Belle],}\\
		0.0646 \pm 0.0043\:({\rm stat.}) \pm 0.0025\:({\rm syst.}) &  &
		{\rm [LHCb];}
	\end{array}\right.
\end{equation}
the errors of the Belle result are dominated by the small $B_s\rightarrow D_s^\pm K^\mp$
data sample. These branching ratios correspond to the ``experimental" time-integrated
branching ratios, as introduced in Eq.~(\ref{BR-exp}).  In Ref.~\cite{BsDsK-12}, it was pointed
out that there is a theoretical lower bound of $0.080 \pm 0.007$ for the ratio in (\ref{LHCb-CDF-BR}).
Using data for $B_d\to D^\pm \pi^\mp$ decays and the $SU(3)$ flavor symmetry results in a
sharper picture, with the following prediction \cite{BsDsK-12}:
\begin{equation}
\left.\frac{{\rm BR}(B_s\to
	D_s^\pm K^\mp)_{\rm exp}}{{\rm BR}(B_s\to D_s^\pm \pi^\mp)_{\rm exp}}\right|_{SU(3)} =
	0.0864^{+0.0087}_{-0.0072}.
\end{equation}

In addition to the ratio in (\ref{LHCb-CDF-BR}), another interesting observable is provided by
the following asymmetry (see also Ref.~\cite{NN}):
\begin{eqnarray}
\lefteqn{\frac{{\rm BR}(B_s\to D_s^{+} K^-)_{\rm exp}-
{\rm BR}(B_s\to D_s^{-} K^+)_{\rm exp}}{{\rm BR}(B_s\to D_s^{+} K^-)_{\rm exp}+
{\rm BR}(B_s\to D_s^{-} K^+)_{\rm exp}}}\label{BR-diff}\\
&&=y_s\left[\frac{{\cal A}_{\Delta\Gamma}(B_s\to D_s^{+} K^-)-
{\cal A}_{\Delta\Gamma}(B_s\to D_s^{-} K^+)}{2+
y_s\{{\cal A}_{\Delta\Gamma}(B_s\to  D_s^{+} K^-)+
{\cal A}_{\Delta\Gamma}(B_s\to D_s^{-} K^+)\}}
\right]\stackrel{SU(3)}{=}-0.027^{+0.052}_{-0.019},\nonumber
\end{eqnarray}
where we give also the theoretical prediction following from the $SU(3)$ flavor symmetry
and the $B_d\to D^\pm \pi^\mp$ data. An experimental non-vanishing value of
(\ref{BR-diff}) would establish a difference between the
${\cal A}_{\Delta\Gamma}(B_s\to D_s^{+} K^-)$ and
${\cal A}_{\Delta\Gamma}(B_s\to D_s^{-} K^+)$ observables. The corresponding
effective lifetimes defined in analogy to (\ref{taueff}) offer also useful information
\cite{BsDsK-12 ,NN}.

Using 1~\invfb of $pp$ collision data recorded in 2011 at a center of mass energy of $\sqrt s = 7$ TeV,
LHCb has reported the first measurement of the time-dependent CP asymmetries of \BsToDsK\ decays~\cite{LHCb-BsDsK-CPV}. In the notation of Eq.~(\ref{CP-asym}), the results read as
\begin{eqnarray}
C(B_s\to D_s^+K^-) = -C(B_s \to D_s^-K^+) &= & - 1.01 \pm 0.50 ({\rm stat}) \pm 0.23
({\rm syst}), \nonumber\\
S(B_s \to D_s^+K^-)	 	 &= &-1.25 \pm 0.56 ({\rm stat}) \pm 0.24 ({\rm syst}), \nonumber \\
S(B_s \to D_s^-K^+ )	& = &-0.08 \pm 0.68 ({\rm stat}) \pm 0.28 ({\rm syst}), \nonumber\\
{\cal A}_{\Delta\Gamma}(B_s \to D_s^+K^-)\	& =& -1.33\pm 0.60 ({\rm stat}) \pm 0.26 ({\rm syst}),
\nonumber\\
{\cal A}_{\Delta\Gamma}(B_s \to D_s^-K^+)& =& -0.81 \pm 0.56 ({\rm stat}) \pm 0.26
({\rm syst}).
\end{eqnarray}

Since $\phi_s$ is known from analyses of the $B_s\to J/\psi \phi$ and
$B_s\to J/\psi f_0$ decays as we have discussed in Subsections~\ref{CPV-BsJpsiphi}
and \ref{CPV-BsJpsif0}, the phase $\phi_s+\gamma$ which
can be extracted in the future from these measurements can be straightforwardly converted
into $\gamma$. This determination 
complements nicely the  well-established time-integrated methods
to extract $\gamma$ from $B^- \ra D^{(\ast)}K^{(\ast)-}$ and $\Bd \ra D^0 K^{ \ast 0}$,
which are also pure tree-level  decays \cite{LHCb-Strat}.

\boldmath
\subsection{Further $B_s$ Decays to Explore CP Violation}\label{CPV-other}
\unboldmath
The $B_s$-meson system offers various other decays with an interesting physics potential
for the exploration of CP violation. Since a detailed presentation goes beyond the scope
of this review, let us just briefly list promising channels. 

The decay $B^0_s\to D_s^{+}D_s^{-}$ originates from $\bar b\to \bar c c \bar s$ quark-level
processes, and receives contributions from a tree topology and doubly Cabibbo-suppressed
penguin amplitudes. It offers yet another determination of the $B^0_s$--$\bar B^0_s$
mixing phase $\phi_s$ and can be combined with the $B^0_d\to D_d^{+}D_d^{-}$ channel
through the $U$-spin symmetry to extract the CKM angle $\gamma$ and the relevant
penguin parameters \cite{RF-BsJpsiK,RF-BsDsDs}. Performing an angular analysis, 
information about $\phi_s$ can also be extracted from the $B^0_s\to  D_s^{*+}D_s^{*-}$ 
channel. A first step towards this analysis is the measurement of the branching fraction of 
the $B^0_s\to D_s^{+}D_s^{-}$ decay mode. It has been measured relative to the 
$\Bd\to D_s^{+}D^{-}$ channel  by the LHCb collaboration with 1 \invfb ~\cite{bib:LHCb-DD}: 
\begin{equation}
\frac{{\rm BR}(\Bs \to D_s^{+}D_s^{-} )}{{\rm BR}(\Bd \to D_s^{+}D^{-} )} = 0.508  \pm 0.026 ({\rm stat}) \pm 0.043 ({\rm syst}) 
\end{equation}
in good agreement with the current world average \cite{pdg-2012} but with an higher precision. 

The $B^0_s\to K^{*0}\bar K^{*0}$ mode is caused by $\bar b \to \bar d d \bar s$ transitions and
receives therefore only contributions from penguin topologies.  It is hence of interesting to test
the SM description of CP violation. The $B^0_s\to K^{*0}\bar K^{*0}$ decay can be related to 
the $B^0_d\to K^{*0}\bar K^{*0}$ channel by means of the $U$-spin symmetry, thereby 
allowing the extraction of $\phi_s$ and $\gamma$ from the observables of the 
time-dependent angular distribution \cite{RF-ang}. The $B^0_s\to K^{*0}\bar K^{*0}$ mode 
has received increasing interest in the context with the determination of 
$\phi_s$ \cite{DGMV,DGMV-2,FG,BDIL}, and the LHCb collaboration has recently reported 
the first observation of this channel \cite{LHCbKstarKstar}.

The final $B_s$ decay of our discussion of CP violation is the $B^0_s\to \phi\phi$ channel,
which is caused by $\bar b\to \bar s s \bar s$ quark-level processes. It is again a pure 
penguin decay, which offers a sensitive probe to new sources of CP violation; for a detailed
recent theoretical analysis see Ref.~\cite{DDL-1}. It will be
exciting to measure its time-dependent CP-violating rate asymmetry at the LHCb experiment. 
So far, only time-integrated measurements of angular 
distributions are available~\cite{bib:CDF-BsToPhiPhi,bib:LHCb-BsToPhiPhi}.
Such measurements offer interesting analyses of triple product asymmetries \cite{DDL-2,GR-trip}.

\boldmath
\section{Rare Decays of $\Bs$ Mesons}
\unboldmath
\label{rare_decays}
\boldmath
\subsection{The $\Bs \rightarrow \mu^+ \mu^-$ Decay}
\begin{figure}
   \centering
   \hspace*{-1.0truecm}\includegraphics[width=5.0truecm]{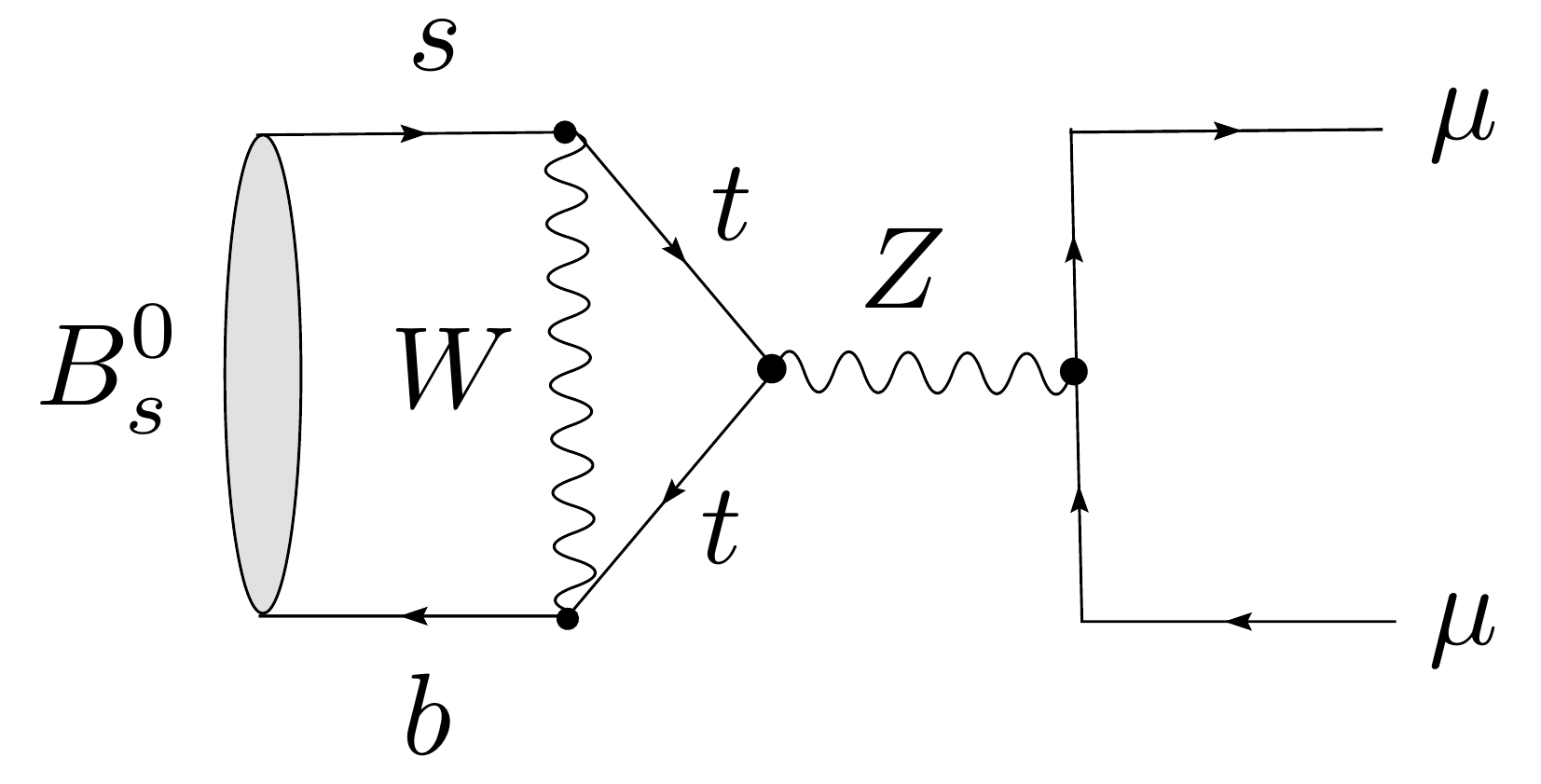}
    \hspace*{0.6truecm}\includegraphics[width=4.5truecm]{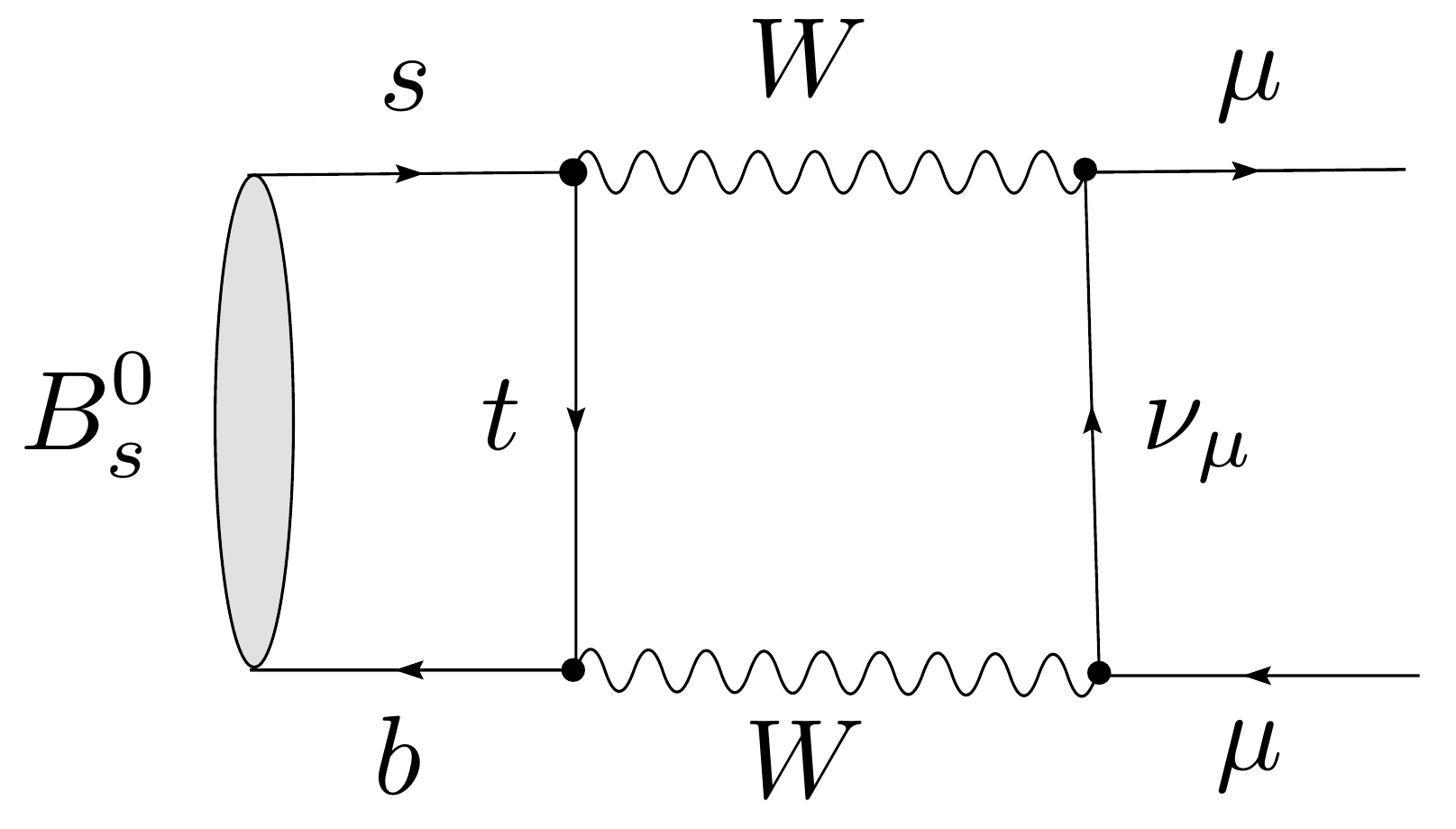}
  \caption{Decay topologies contributing to the $B^0_s\to\mu^+\mu^-$ decay in the SM.}
   \label{fig:Bsmumu-topol}
\end{figure}

\unboldmath
In the SM, the rare decay $B_s\to\mu^+\mu^-$ originates from loop contributions, as
illustrated in Fig.~\ref{fig:Bsmumu-topol}.
Moreover, as only leptons are present in the final state, the hadronic sector is simply
described by the non-perturbative $B_s$ decay constant $f_{B_s}$.
The $B_s\to\mu^+\mu^-$ channel is one of the cleanest rare $B$ decays
and therefore offers a powerful probe to search for NP effects which may enter
through new particles running in the loops or even through new flavor-changing
neutral current contributions at the tree level (see \cite{Buras1} and references therein).

The most recent theoretical update of the $B_s\to\mu^+\mu^-$ branching ratio
arising in the SM is given as follows \cite{BGGI}:
\begin{equation}\label{BRs-SM}
\mbox{BR}(B_s\to \mu^+\mu^-)_{\rm SM}=(3.23\pm0.27)\times 10^{-9},
\end{equation}
where the error is dominated by the lattice QCD value of $B_s$ decay constant
$f_{B_s}$. The extremely small branching ratio makes the experimental search and
analysis of this rare decay very challenging.

For the following discussion, it is useful to have a closer look at the theoretical
description of the $\bar B^0_s\to\mu^+\mu^-$ channel. The starting point is an
appropriate low-energy effective Hamiltonian \cite{BBL}. Using the same notation
as in  \cite{APS}, it can be written as follows:
\begin{equation}\label{Heff}
{\cal H}_{\rm eff}=-\frac{G_{\rm F}}{\sqrt{2}\pi} \alpha V_{ts}^\ast V_{tb}
\bigl[C_{10} O_{10} + C_{S} O_S + C_P O_P
 + C_{10}' O_{10}' + C_{S}' O_S' + C_P' O_P' \bigr].
\end{equation}
Here $G_{\rm F}$ and $\alpha$ are the Fermi and QED fine-structure constants,
respectively, and the $V_{qq'}$ are CKM matrix elements. The short-distance physics
is encoded in the Wilson coefficients $C_i$,  $C_i'$ of the four-fermion operators
\begin{equation}
O_{10}=(\bar s \gamma_\mu P_L b) (\bar\ell\gamma^\mu \gamma_5\ell), \quad
O_S=m_b (\bar s P_R b)(\bar \ell \ell), \quad
O_P=m_b (\bar s P_R b)(\bar \ell \gamma_5 \ell),
\end{equation}
where $P_{L,R}\equiv(1\mp\gamma_5)/2$, $m_b$ is the $b$-quark mass, and the $O'_i$
are obtained from the $O_i$ through the replacements $P_L \leftrightarrow P_R$.
All matrix elements can be expressed in terms of the $B_s$-meson decay constant
$f_{B_s}$.

In the SM, (\ref{Heff}) simplifies considerably. Then we have to deal only with $O_{10}$
and its real Wilson coefficient $C_{10}^{\rm SM}$, which governs the SM prediction in
(\ref{BRs-SM}). Concerning the search for NP effects,
the $\bar B^0_s\to \mu^+\mu^-$ decay has the outstanding feature to offer sensitivity
to the (pseudo-)scalar lepton densities entering the
$O_{(P)S}$ and $O_{(P)S}'$ operators, which is particularly relevant for models with
extended Higgs sectors. The Wilson coefficients of these operators are still
largely unconstrained by the current data (see, for instance, \cite{APS}).

\begin{figure}
\begin{center}
{\includegraphics[height=7cm]{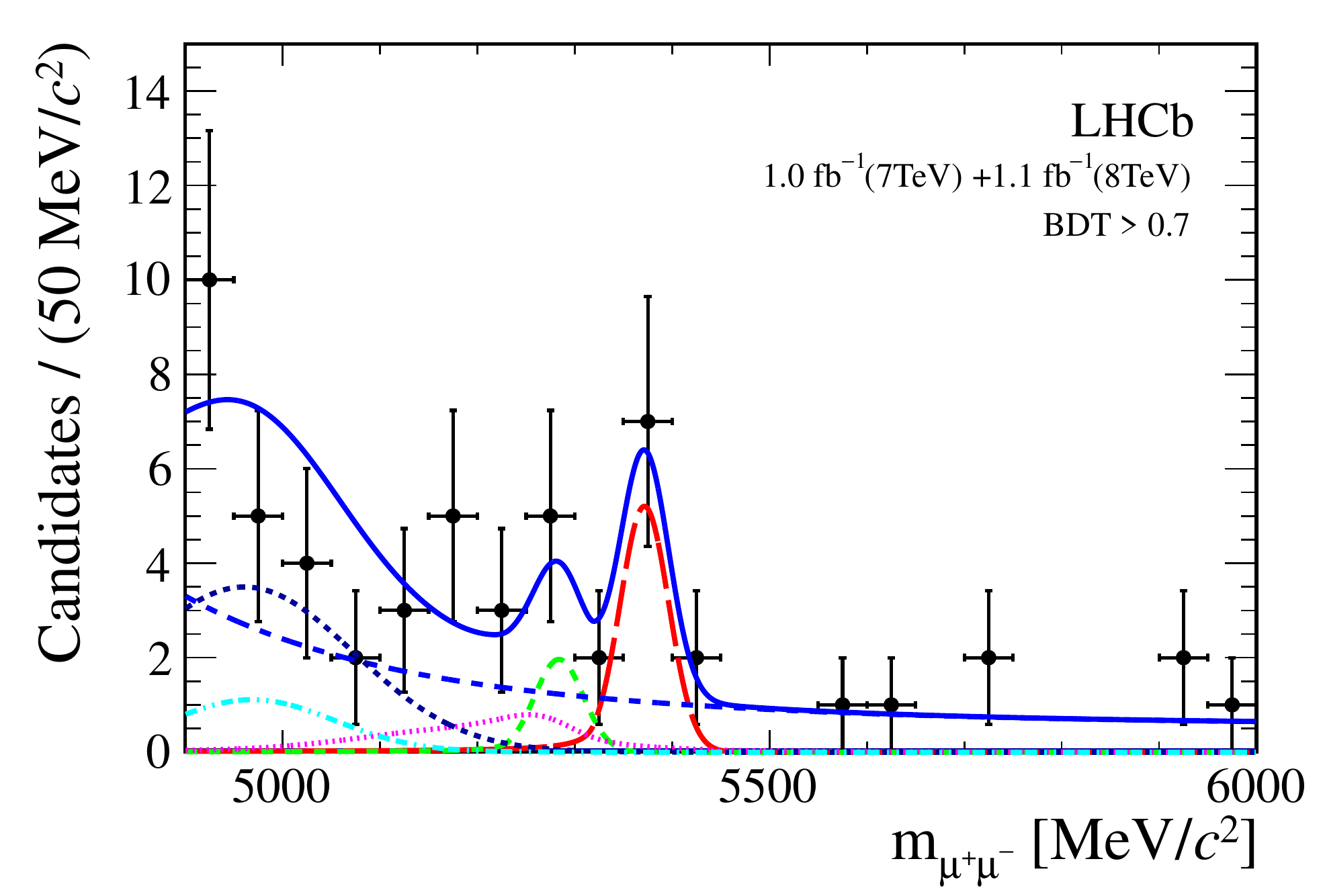}}
\caption{Invariant mass distribution of the selected \BsToMuMu candidates (black dots) in a signal enriched region \cite{LHCb-Bsmumu}. The result of the fit is overlaid (blue solid line) and the different component detailed: \BsToMuMu (red long dashed),  $\Bd \rightarrow \mu^+ \mu^-$ (green medium dashed), $B^0_{(s)} \rightarrow h^+ h^{'-}$ (pink dotted), $\Bd \rightarrow \pi^- \mu^+ \nu_{\mu}$ (black short dashed) and $B^{0(+)} \rightarrow \pi^{0(+)} \mu^+ \mu^-$  (light blue dot dashed), and the combinatorial background (blue medium dashed).}
\label{fig:Rare_BsToMuMu_LHCb}
\end{center}
\end{figure}

The first step for the experimental exploration of the $B_s\to\mu^+\mu^-$ decay
is the measurement of the branching ratio. Since it is experimentally very challenging
to measure the muon helicity, we consider the untagged combination
\begin{equation}\label{rate-no-lam}
\langle\Gamma({B}_s(t)\to \mu^+\mu^-)\rangle \equiv \sum_{\lambda={\rm L,R}}
\left[\Gamma({B}_s^0(t)\to \mu^+_\lambda \mu^-_\lambda) +
\Gamma(\bar B^0_s(t)\to \mu^+\mu^-)\right].
\end{equation}
Ignoring the time information in this rate, we obtain the time-integrated ``experimental
branching" ratio as defined in Eq.~(\ref{BR-exp}).

The search for $B_s\to\mu^+\mu^-$ at the Tevatron experiments
has finally reached the region of about ten times the SM value (\ref{BRs-SM}),
where D\O\ and CDF
obtain the upper bounds $5.1 \times 10^{-8}$ \cite{Rare_BsToMuMu_D0}
and $4.0 \timesÊ10^{-8}$ \cite{Rare_BsToMuMu_CDF} for the branching ratio
at 95 \% C.L., respectively.
The search has been continued by the LHC experiments
\cite{Rare_BsToMuMu_CMS,Rare_BsToMuMu_ATLAS,Rare_BsToMuMu_LHCb_limit},
reaching a combined limit of BR$(\BsToMuMu) < 4.2 \mbox{ }Ê10^{-9} $ (95 \% C.L.), which is only 20\% larger than the SM prediction.

In November 2012, the LHCb collaboration
has reported the first  evidence for $B^0_s\to \mu^+\mu^-$
at the $3.5\sigma$ level, with the following branching ratio \cite{LHCb-Bsmumu}:
\begin{equation}\label{LHCb-BR}
\mbox{BR}(B_s\to \mu^+\mu^-)=(3.2^{+1.5}_{-1.2})\times10^{-9}.
\end{equation}
The invariant mass distribution of the
\BsToMuMu candidates in a signal enriched region is
shown in Fig. \ref{fig:Rare_BsToMuMu_LHCb}.

While the experimental upper bounds and the LHCb result in (\ref{LHCb-BR}) refer to
(\ref{rate-no-lam}) and
the time-integrated ``experimental" branching ratio  (\ref{BR-exp}),  the SM prediction in
Eq.~(\ref{LHCb-BR})  refers to the ``theoretical" branching ratio defined in (\ref{BR-theo}).
As was pointed out in Ref.~\cite{Bsmumu-paper}, the conversion between these
two branching ratio concepts is given by the following expression:
\begin{equation}\label{BRratio}
        {\rm BR}(B_s \to \mu^+\mu^-)_{\rm theo}
        =   \left[\frac{1-y_s^2}{1 + {\cal A}_{\Delta\Gamma}\, y_s}\right]
        {\rm BR}(B_s \to \mu^+\mu^-)_{\rm exp},
\end{equation}
where it is essential that the observable
\begin{equation}\label{ADG-lam-mu}
{\cal A}_{\Delta\Gamma}
=\frac{|P|^2\cos 2\varphi_P-|S|^2\cos 2\varphi_S}{|P|^2+|S|^2}
\end{equation}
does not depend on the muon helicity (note also that $f_{B_s}$ cancels).
Here the combinations of Wilson coefficients
\begin{equation}\label{P-expr}
P\equiv |P|e^{i\varphi_P}\equiv \frac{C_{10}-C_{10}'}{C_{10}^{\rm SM}}+\frac{M_{B_s} ^2}{2 m_\mu}
\left(\frac{m_b}{m_b+m_s}\right)\left(\frac{C_P-C_P'}{C_{10}^{\rm SM}}\right)
\end{equation}
and
\begin{equation}\label{S-expr}
S\equiv  |S|e^{i\varphi_S} \equiv \sqrt{1-4\frac{m_\mu^2}{M_{B_s}^2}}
\frac{M_{B_s} ^2}{2 m_\mu}\left(\frac{m_b}{m_b+m_s}\right)
\left(\frac{C_S-C_S'}{C_{10}^{\rm SM}}\right)
\end{equation}
with their CP-violating phases $\varphi_{P,S}$ have been introduced in such a way
that $P=1$ and $S=0$ in the SM. In (\ref{ADG-lam-mu}), the NP contribution
to the $B^0_s$--$\bar B^0_s$ mixing phase (\ref{phis-def})
was neglected. This
effect can straightforwardly be included through $2\varphi_{P,S}\to 2\varphi_{P,S}-\phi_s^{\rm NP}$.
However, the LHCb data for CP violation in $B_s\to J/\psi \phi, J/\psi f_0(980)$
already constrain $\phi_s^{\rm NP}$ to the few-degree level (see Subsections
\ref{CPV-BsJpsiphi} and \ref{CPV-BsJpsif0}), whereas the $\varphi_{P,S}$ are
still essentially unconstrained.

As can be seen in (\ref{ADG-lam-mu}), since NP may enter through the Wilson coefficients,
the ${\cal A}_{\Delta\Gamma}$ observable is currently unknown. On the other
hand, the SM gives the theoretically clean prediction of ${\cal A}_{\Delta\Gamma}^{\rm SM}=+1$.
Using (\ref{BRratio}),  we hence rescale the theoretical SM branching ratio in
(\ref{BRs-SM}) by a factor of $1/(1-y_s)$, yielding
\begin{equation}
\mbox{BR}(B_s\to \mu^+\mu^-)_{\rm SM}|_{y_s}=(3.54\pm0.30)\times 10^{-9}
\end{equation}
for the value of $y_s$ in (\ref{ys-def}). This is the SM reference value for the
comparison with the experimental branching ratio (\ref{LHCb-BR}).

Once the currently emerging $B_s\to\mu^+\mu^-$ signal has been well established and
more data become available, also the decay-time information for the untagged data sample
can be included, which will allow the measurement of the effective lifetime
$B_s\to\mu^+\mu^-$ lifetime, which is defined in analogy to (\ref{taueff})
\cite{BR-paper,Bsmumu-paper}. The effective lifetime allows
the conversion of the experimental $B_s\to\mu^+\mu^-$ branching ratio into its theoretical
counterpart through an expression that is analogous to (\ref{BRratioT}). Moreover, also the
observable
\begin{equation}
 {\cal A}_{\Delta\Gamma}  = \frac{1}{y_s}\left[\frac{(1-y_s^2)\tau_{\mu^+\mu^-}-(1+
 y_s^2)\tau_{B_s}}{2\tau_{B_s}-(1-y_s^2)\tau_{\mu^+\mu^-}}\right]
\end{equation}
can be extracted from the data.

These measurements offer are exciting new aspects for the exploration of the
$B_s\to\mu^+\mu^-$ decay at the high-luminosity upgrade of the LHC. An extrapolation
from current measurements of the effective $B_s\to J/\psi\,f_0(980)$ and
$B_s\to K^+K^-$ lifetimes by the CDF and LHCb collaborations to
$\tau_{\mu^+\mu^-}$ indicates that a precision of $5\%$ or better may be feasible
\cite{Bsmumu-paper}. Detailed experimental studies are strongly encouraged.

The $\Delta\Gamma_s$ effects propagate also into the NP constraints that can be obtained
from the comparison of the experimental $B_s\to\mu^+\mu^-$ branching ratio with the SM,
where it is useful to introduce the following ratio \cite{Bsmumu-paper}:
\begin{equation}\label{Rmumu-def}
R\equiv
\frac{\mbox{BR}(B_s \to \mu^+\mu^-)_{\rm exp}}{\mbox{BR}(B_s \to \mu^+\mu^-)_{\rm SM}}=
\left[\frac{1+y_s\cos2\varphi_P}{1-y_s^2}  \right] |P|^2+
\left[\frac{1-y_s\cos2\varphi_S}{1-y_s^2}  \right] |S|^2.
\end{equation}
Using (\ref{BRs-SM}) and (\ref{LHCb-BR}) yields $R=1.0^{+0.5}_{-0.4}$, where the errors have
been added in quadrature.

The $R$ ratio can be converted into ellipses in the $|P|$--$|S|$ plane which depend on the
CP-violating phases $\varphi_{P,S}$. Since the latter quantities are unknown, $R$ fixes
actually a circular band with the upper bounds $|P|, |S|\leq\sqrt{(1+y_s)R}$. As the
experimental information on $R$ does not allow us to separate the $S$ and $P$ contributions,
still significant NP contributions may be hiding in the $B_s\to\mu^+\mu^-$ channel.

This situation can be resolved by measuring the effective lifetime $\tau_{\mu^+\mu^-}$ and the associated ${\cal A}_{\Delta\Gamma}$ observable, as illustrated in
the figures shown in \cite{Bsmumu-paper}.

In the most recent analyses of the constraints on NP parameter space that are implied
by the experimental upper bound on the $B_s\to\mu^+\mu^-$ branching ratio for various
extensions of the SM, authors have now started to take the effect of $\Delta\Gamma_s$
into account (see, for instance, \cite{BDeFG}--\cite{BG-2}
and the papers in \cite{WCSY}--\cite{Nan}).

\boldmath
\subsection{The $\Bs \to \phi \gamma$ Decay}
\unboldmath
Another interesting rare $\Bs$ decay 
is the $\Bs \to \phi \gamma$ channel, which arises in the SM from penguin topologies. 
It is the $B_s$ counterpart of the $\Bd \to K^{*0}\gamma$ mode and originates from 
$\bar b\to \bar s\gamma$ quark-level processes. The data for $\Bd \to K^{*0}\gamma$ and
the inclusive $B\to X_s\gamma$ mode are in agreement with the SM within the errors and
have put strong constraints on NP models.

The SM prediction reads BR$(\Bs \to \phi \gamma) = (4.3 \pm 1.4) \times 10^{-5}$ \cite{bsg-theory},
where the uncertainty is due to the non-perturbative QCD effects which are encoded in the
corresponding form factors. A more precise theoretical prediction is given for the
ratio $\mbox{BR}(\Bd \to K^{*0} \gamma) / \mbox{BR}(\Bs \to \phi \gamma) = 1.0 \pm 0.2$ \cite{bsg-theory}.

The decay $\Bs \to \phi \gamma$ was first measured by the Belle collaboration \cite{bsg-belle}:
\begin{equation}
\mbox{BR}(\Bs \to \phi \gamma) = \left[5.7 ^{+1.8}_{-1.5}\mbox{(stat)} ^{+1.2}_{-1.1}\mbox{(syst)}\right] \times 10^{-5}.
\end{equation}
Recently, the LHCb collaboration published \cite{bsg-lhcb} the measurement of the ratio 
\begin{equation}
\frac{\mbox{BR}(\Bd \to K^{*0} \gamma)} {\mbox{BR}(\Bs \to \phi \gamma)} =
1.12 \pm 0.08 \mbox{(stat)} ^{+0.06}_{-0.04}\mbox{(syst)} ^{+0.09}_{-0.08}\mbox{(frag)}.
\end{equation}
The last uncertainty is due to the ratio $f_s / f_d$ of fragmentation fractions 
discussed in Subsection~\ref{sec:Bsprod}. 
This result was obtained using 0.37 \invfb\ of $p p$ collisions at LHC. 
Using this measurement and the world average value \cite{pdg-2012} of the $\mbox{BR}(\Bd \to K^{*0} \gamma)$,
the LHCb collaboration has obtained the currently most precise measurement of 
$\mbox{BR}(\Bs \to \phi \gamma)$, which is given by
\begin{equation}
\mbox{BR}(\Bs \to \phi \gamma) = (3.9 \pm 0.5) \times 10^{-5}.
\end{equation}
The obtained experimental results are consistent with each other and with the SM prediction.

An interesting aspect of the $\Bs \to \phi \gamma$ channel is that $\Delta\Gamma_s$ offers
an observable, which allows to measure the photon polarization and is sensitive to 
right-handed currents appearing in scenarios of physics beyond the SM \cite{MXZ}. 
This feature distinguishes $\Bs \to \phi \gamma$ from $\Bd \to K^{*0} \gamma$ as 
$\Delta\Gamma_d$ is negligibly small.

\boldmath
\subsection{The $\Bs \to \phi \mu^+\mu^-$ Decay}
\unboldmath
Another interesting rare decay is $\Bs \to \phi \mu^+\mu^-$, which is the
$B_s$ counterpart of the well-known $B^0_d\to K^{*0}\mu^+\mu^-$ channel. For a
detailed study of the theoretical aspects of the $\Bs \to \phi \mu^+\mu^-$ mode, 
we refer the reader to Ref.~\cite{BHP}. 

The CDF collaboration performed an extensive study of decays originating from  
$b \to s \mu^+ \mu^-$ quark-level processes with different hadrons in the initial and 
final states. The analysis is based on the statistics
corresponding to 9.6 fb$^{-1}$ of $p \bar p$ collisions. This result \cite{b-to-s-cdf} 
was still unpublished at the time of preparing this review. In each case the branching fraction of the decay
$H_b \to h \mu^+ \mu^-$ is normalized to the well identified decay $H_b \to J/\psi h$ with $J/\psi \to \mu^+ \mu^-$.
Such a normalization significantly reduces the systematic uncertainty of the measurements.
The following result for the decay $\Bs \to \phi \mu^+ \mu^-$ is obtained:
\begin{eqnarray}
\mbox{Br}(\Bs \to \phi \mu^+ \mu^-) & = & [1.17 \pm 0.18 \mbox{(stat)} \pm 0.37 \mbox{(syst)}]\times 10^{-6}.
\end{eqnarray}
These measurements are continued by the LHCb experiment which has recently obtained with 1 \invfb \cite{b-to-s-lhcb}: 
\begin{equation}
\mbox{Br}(\Bs \to \phi \mu^+ \mu^-)  =  [0.78 \pm 0.10 \mbox{(stat)} \pm 0.06 \mbox{(syst)} \pm 0.28 \mbox{(BR)}]\times 10^{-6}
\end{equation}
where the last uncertainty comes from the knowledge of the $\BsToJpsiPhi$ branching fraction. 
In future, a better precision, could reveal possible NP contributions. In particular angular analyses, similar to those performed for the $\Bd \to K^{*0} \mu^+ \mu^- $ decay, will be extremely interesting. 

\section{Conclusions and Outlook}\label{sec:concl}
\label{conclusions}
The $B_s$-meson system plays a key role in the testing of the quark-flavor sector of the
SM. After pioneering work at the Tevatron and measurements at the Belle experiment, 
the exploration of weak decays of $B_s$ mesons has now fully shifted to the LHC. In 2012, 
we have seen two particularly exciting developments in the exploration of the $B_s$ system: 
the common efforts of the Tevatron and LHC experiments has established a non-vanishing decay width difference $\Delta\Gamma_s$,
and the first evidence for the rare decay $B^0_s\to\mu^+\mu^-$ has been reported at
the $3.5\sigma$ level by the LHCb experiment, 
thereby complementing the previous constraints from the CDF, D\O\,
ATLAS and CMS collaborations. Both measurements are in accordance with the SM 
although the branching ratio of $B^0_s\to\mu^+\mu^-$ has still a large error. It will be
very interesting to monitor the future evolution of the experimental picture.

Concerning the theoretical aspects of these results, the sizable value of 
$\Delta\Gamma_s$ leads to subtleties 
in the interpretation of time-integrated $B_s$ rates in terms of branching ratios but provides also 
new observables which can be accessed through effective $B_s$ decay lifetimes. The emerging
signal for the $B^0_s\to\mu^+\mu^-$ decay has entered many analyses of specific NP models, 
in particular supersymmetric scenarios, where strong constrains on the corresponding parameter
space emerge. The effective lifetime of $B_s\to\mu^+\mu^-$ offers a new, theoretically clean
observable for the search for NP that is complementary to the branching ratio. Detailed studies
for the high-luminosity upgrade of the LHC are strongly encouraged. 

In the exploration of CP violation in the $B_s$-meson system, we have seen more precise
experimental analyses of the benchmark decays $B^0_s\to J/\psi \phi$ and 
$B^0_s\to J/\psi f_0$ in 2012. The resulting picture of smallish CP violation -- with $\phi_s$ in 
the few degree regime -- is again consistent with the SM description of CP violation through the
Kobayashi--Maskawa 
mechanism. In view of this development, penguin contributions to the corresponding
decay amplitudes have to be controlled. Since these topologies enter in a doubly 
Cabibbo-suppressed way they are usually neglected. However, once the experimental precision
increases further, in particular at the LHCb upgrade, these effects may lead to fake NP signals
and it will be crucial to match the experimental with the theoretical precision. Thanks to
their non-perturbative nature, experimental control channels have to be used to probe the
importance of the penguin contributions. In this context, the $B^0_s\to J/\psi K_{\rm S}$ and
$B^0_s\to J/\psi \bar K^{*0}$ decays play key roles, where first measurements of branching
ratios and angular observables are already available. Another highlight of the exploration 
of CP violation is the $B^0_s\to \pi^+ K^-$ channel, where LHCb and CDF established a direct
CP asymmetry in 2012, which is again in accordance with the SM and  $SU(3)$ relations
to the direct CP asymmetries of the $B^0_d\to\pi^-K^+$ and $B^0_d\to\pi^+\pi^-$ modes. 
Important first steps in the measurement
of CP violation in the $B_s\to K^+K^-$ and $B_s\to D_s^\pm K^\mp$  channels could also
be made by LHCb. 

We look forward to many more exciting results in the exploration of rare decays and CP
violation in the $B_s$ system!


\begin{thebibliography}{99}
%
%
%
\bibitem{detector-aleph}
D. Decamps {\it et al.} [ALEPH Collaboration],   {\it Nucl. Instrum. Methods in Phys. Res.\/} A {\bf 294}, 121 (1990). 
\bibitem{detector-delphi}
P. Aarnio {\it et al.} [DELPHI Collaboration],  {\it Nucl. Instrum. Methods in Phys. Res.\/} A {\bf 303}, 233 (1991). 
\bibitem{detector-opal}
K. Ahmet {\it et al.} [OPAL Collaboration],  {\it Nucl. Instrum. Methods in Phys. Res.\/} A {\bf 305}, 275 (1991). 
\bibitem{detector-cdf}
A. Abulencia {\it et al.} [CDF Collaboration], {\it J. Phys. \/}  G {\bf 34}, 2457 (2007). \\
A. Sill {\it et al.}, {\it Nucl. Instrum. Methods in Phys. Res.\/} A {\bf 447}, 1 (2000).
A. Affolder {\it et al.} [CDF Collaboration], {\it Nucl. Instrum. Methods in Phys. Res.\/} A {\bf 526}, 249 (2004). \\
C. S. Hill [on behalf of the CDF Collaboration], {\it Nucl. Instrum. Methods in Phys. Res. A\/} {\bf 530}, 1 (2004). 

\bibitem{muon-cdf}
G. Ascoli {\it et al.}, {\it Nucl. Instrum. Methods in Phys. Res.\/} A {\bf 268}, 33 (1988). \\
T. Dorigo {\it et al.}, {\it Nucl. Instrum. Methods in Phys. Res.\/} A {\bf 461}, 560 (2001).

\bibitem{trigger-cdf}
E. J. Thomson {\it et al.}, {\it IEEE Trans. Nucl. Sci.\/} {\bf 49}, 1063 (2002).
J. A. Adelman {\it et al.} [CDF Collaboration], {\it Nucl. Instrum. Methods in Phys. Res.\/} A {\bf 572}, 361 (2007). \\
A. Abulencia {\it et al.}) [CDF Collaboration], {\it Phys. Rev. Lett.\/} {\bf 98}, 122002 (2007).

\bibitem{detector-d0}
V.M. Abazov {\it et al.} [D0 Collaboration], {\it Nucl. Instrum. Methods in Phys. Res.\/} A {\bf 565}, 463 (2006). \\
S.N. Ahmed {\it et al.}, {\it Nucl. Instrum. Methods in Phys. Res.\/} A {\bf 634}, 8 (2011); \\
R. Angstadt {\it et al.}, {\it Nucl. Instrum. Methods in Phys. Res.\/} A {\bf 622}, 298 (2010).

\bibitem{muon-d0}
V.M. Abazov {\it et al.}, {\it Nucl. Instrum. Methods in Phys. Res.\/}  A {\bf 552}, 372 (2005).

\bibitem{LHCb-detector} 
A.A. Alves~Jr. {\it et al.} [LHCb collaboration],  
{\it JINST} 3 (2008) S08005.

  \bibitem{cab}N.~Cabibbo,
  {\it Phys.\ Rev.\ Lett.}\  {\bf 10}, 531 (1963).
  
\bibitem{KM}M.~Kobayashi and T.~Maskawa,
  {\it Prog.\ Theor.\ Phys.}\  {\bf 49}, 652 (1973).

\bibitem{BuGi}A.~J.~Buras and J.~Girrbach,
  {\it Acta Phys.\ Polon.}\ B {\bf 43}, 1427 (2012)
  [arXiv:1204.5064 [hep-ph]].
  
  \bibitem{asl-d0}
V. Abazov {\it et al.} [D0 Collaboration], {\it Phys. Rev. D\/} {\bf 84}, 052007 (2011).

\bibitem{asld-d0}
V. Abazov {\it et al.} [D0 Collaboration], {\it Phys. Rev. D\/} {\bf 86}, 072009 (2012).

\bibitem{asls-d0}
V. Abazov {\it et al.} [D0 Collaboration], arXiv:1207.1769 [hep-ex] (2012).

\bibitem{asls-lhcb}
LHCb collaboration, Conference report LHCb-CONF-2012-022 (2012).

\bibitem{cdf-jpsi} T. Aaltonen {\it et al.} [CDF Collaboration], {\it Phys. Rev. Lett.\/} {\bf 109}, 171802 (2012).


\bibitem{d0-jpsi} V.M. Abazov {\it et al.} [D0 Collaboration], {\it Phys. Rev. D\/} {\bf 85}, 032006 (2012).

\bibitem{jpsi-lhcb}
R. Aaij {\it et al.} [LHCb Collaboration], {\it Phys. Rev. Lett.\/} {\bf 108}, 101803 (2012).

\bibitem{jpsi-atlas}
G. Aad {\it et al.} [ATLAS Collaboration], arXiv: 1208.0572 [hep-ex] (2012).

\bibitem{jpsif0-lhcb}
R. Aaij {\it et al.} [LHCb Collaboration], {\it Phys. Lett. B\/} {\bf 707}, 497 (2012).

\bibitem{jpsipipi-lhcb}
R. Aaij {\it et al.} [LHCb Collaboration], {\it Phys. Lett. B\/} {\bf 713}, 378 (2012).

  \bibitem{RF-BsJpsiK}R.~Fleischer,
  {\it Eur.\ Phys.\ J.}\ C {\bf 10}, 299 (1999)
  [hep-ph/9903455].

\bibitem{CPS}M.~Ciuchini, M.~Pierini and L.~Silvestrini,
  {\it Phys.\ Rev.\ Lett.}\  {\bf 95}, 221804 (2005) [hep-ph/0507290];
  arXiv:1102.0392 [hep-ph].

\bibitem{FFJM}S.~Faller, R.~Fleischer, M.~Jung  and T.~Mannel,
  {\i Phys.\ Rev.}\ D {\bf 79}, 014030 (2009)
  [arXiv:0809.0842 [hep-ph]].
  
\bibitem{FFM}S.~Faller, R.~Fleischer and T.~Mannel,
  {\it Phys.\ Rev.}\ D {\bf 79}, 014005 (2009)
  [arXiv:0810.4248 [hep-ph]].
    
\bibitem{DeBFK}K.~De Bruyn, R.~Fleischer and P.~Koppenburg,
  {\it Eur.\ Phys.\ J.}\ C {\bf 70}, 1025 (2010)
  [arXiv:1010.0089 [hep-ph]].
  
\bibitem{FKR}R.~Fleischer, R.~Knegjens and G.~Ricciardi,
  {\it Eur.\ Phys.\ J.}\ C {\bf 71}, 1832 (2011)
  [arXiv:1109.1112 [hep-ph]].
  
\bibitem{FKR-eta}R.~Fleischer, R.~Knegjens and G.~Ricciardi,
  {\it Eur.\ Phys.\ J.}\ C {\bf 71}, 1798 (2011)
  [arXiv:1110.5490 [hep-ph]].
  
\bibitem{MJ}M.~Jung,
  {\it Phys.\ Rev.}\ D {\bf 86}, 053008 (2012)
  [arXiv:1206.2050 [hep-ph]].

\bibitem{Rare_BsToMuMu_D0} 
V. M. Abazov {\it et al.} [D0 Collaboration],   {\it Phys. Lett.}\ B {\bf 693}, 539 (2010)  [arXiv:1006.3469].

\bibitem{Rare_BsToMuMu_CDF} 
T. Aaltonen {\it et al.}  [CDF Collaboration],  {\it Phys. Rev. Lett.}\  {\bf 107}, 191801 (2011) 
[arXiv:1107.2304].

\bibitem{Rare_BsToMuMu_CMS} 
S. Chatrchyan {\it et al.} [CMS Collaboration], {\it JHEP} 04, 033 (2012)  [arXiv:1203.3976].

\bibitem{Rare_BsToMuMu_ATLAS} 
G. Aad {\it et al.} [ATLAS Collaboration],  {\it Phys. Lett.}\ B {\bf 713}, 387 (2012)  [arXiv:1204.0735].

\bibitem{LHCb-Bsmumu}R. Aaij {\it et al.}  [LHCb Collaboration],
  {\it Phys.\ Rev.\ Lett.} {\bf 110}, 021801 (2013).

  \bibitem{bs-kk-lhcb-1}
R. Aaij {\it et al.} [LHCb Collaboration], {\it Phys. Lett.\/} B{\bf 707}, 349 (2012).

\bibitem{bs-kk-lhcb-2}
R. Aaij {\it et al.} [LHCb Collaboration], {\it Phys. Lett.\/} B {\bf 716}, 393 (2012).

\bibitem{bs-f0-cdf}
T. Aaltonen {\it et al.} [CDF Collaboration], {\it Phys. Rev.\/} D {\bf 84}, 052012 (2011).

\bibitem{bs-f0-lhcb}
R. Aaij {\it et al.} [LHCb Collaboration],  {\it Phys. Rev. Lett.}\ {\bf 109},152002 (2012).


\bibitem{BR-paper}K.~De Bruyn, R.~Fleischer, R.~Knegjens, P.~Koppenburg, M.~Merk 
and N.~Tuning,
 {\it  Phys.\ Rev.}\ D {\bf 86}, 014027 (2012) 
  [arXiv:1204.1735 [hep-ph]].
    
 \bibitem{Bsmumu-paper}K.~De Bruyn, R.~Fleischer, R.~Knegjens, P.~Koppenburg, 
 M.~Merk, A.~Pellegrino and N.~Tuning,
  {\it Phys.\ Rev.\ Lett.}\  {\bf 109}, 041801 (2012) 
  [arXiv:1204.1737 [hep-ph]].

 \bibitem{bib:LHCbTrigger} R. Aaij {\it et al.} [LHCb Collaboration], 
 arXiv:1211.3055, submitted to JINST. 
 \bibitem{bib:ATLAS} ATLAS Collaboration, JINST, 3:S08003, 2008.
 \bibitem{bib:CMS} CMS Collaboration, JINST, 3:S08004, 2008.
  \bibitem{bib:Pythia} T. Sjostrand, S. Mrenna, and P. Skands, {\it JHEP} {\bf05}, 026 (2006) [arXiv:hep-ph/0603175].
 \bibitem{LHCb-JpsiProd}
 R. Aaij {\it et al.}, 
 {\it Eur. Phys. J.}\ C 71, 1645 (2011).

\bibitem{FST}R.~Fleischer, N.~Serra and N.~Tuning,
  {\it Phys.\ Rev.}\ D {\bf 82}, 034038 (2010)
  [arXiv:1004.3982 [hep-ph]].
  
  
\bibitem{FST-fact}R.~Fleischer, N.~Serra and N.~Tuning,
  {\it Phys.\ Rev.}\ D {\bf 83}, 014017 (2011) 
  [arXiv:1012.2784 [hep-ph]].
    

\bibitem{LHCb-hadr}R.~Aaij {\it et al.}  [LHCb Collaboration],
  {\it Phys.\ Rev.\ Lett.}\  {\bf 107},  211801 (2011) 
  [arXiv:1106.4435 [hep-ex]].

\bibitem{LHCb-sl}R.~Aaij {\it et al.}  [LHCb Collaboration],
  {\it Phys.\ Rev.}\ D {\bf 85},  032008 (2012)
  [arXiv:1111.2357 [hep-ex]].

\bibitem{FF-lat}J.~A.~Bailey, A.~Bazavov, C.~Bernard, C.~M.~Bouchard, C.~DeTar, D.~Du, A.~X.~El-Khadra and J.~Foley {\it et al.},
  {\it Phys.\ Rev}.\ D {\bf 85},  114502 (2012) 
  [arXiv:1202.6346 [hep-lat]].

\bibitem{LHCb-hadr-update}R.~Aaij {\it et al.}  [LHCb Collaboration], 	arXiv:1301.5286 [hep-ex].

\bibitem{wolf}L.~Wolfenstein,
  {\it Phys.\ Rev.\ Lett.}\  {\bf 51}, 1945 (1983).
  
\bibitem{CKM-fitter}J.~Charles, O.~Deschamps, S.~Descotes-Genon, R.~Itoh, H.~Lacker, A.~Menzel, S.~Monteil and V.~Niess {\it et al.},
  {\it Phys.\ Rev.}\ D {\bf 84}, 033005 (2011)
  [arXiv:1106.4041 [hep-ph]].
\bibitem{bib:UTFit}\begin{verbatim} http://www.utfit.org/UTfit/\end{verbatim}

\bibitem{RF-Phys-Rep}R.~Fleischer,
  {\it Phys.\ Rept.}\  {\bf 370}, 537 (2002)
  [hep-ph/0207108].
  
\bibitem{BaFl}P.~Ball and R.~Fleischer,
  {\it Eur.\ Phys.\ J.}\  C {\bf 48}, 413 (2006) 
  [arXiv:hep-ph/0604249].
  
\bibitem{LPP}Z.~Ligeti, M.~Papucci and G.~Perez,
  {\it Phys.\ Rev.\ Lett.}\  {\bf 97},  101801 (2006)  
  [hep-ph/0604112].
  
\bibitem{RG}J.~L.~Rosner and M.~Gronau,
  PoS BEAUTY {\bf 2011}, 045 (2011)
  [arXiv:1105.1923 [hep-ph]].
  
\bibitem{bur-fla}A.~J.~Buras,
  PoS BEAUTY {\bf 2011}, 008 (2011)
  [arXiv:1106.0998 [hep-ph]].
  
\bibitem{LN}A.~Lenz and U.~Nierste,
  JHEP {\bf 0706},  072  (2007) 
  [arXiv:hep-ph/0612167];
  arXiv:1102.4274 [hep-ph].
  
\bibitem{CFLMT}M.~Ciuchini, E.~Franco, V.~Lubicz, F.~Mescia and C.~Tarantino,
  JHEP {\bf 0308}, 031 (2003) 
  [arXiv:hep-ph/0308029].

\bibitem{lenz}A.~Lenz,
  arXiv:1205.1444 [hep-ph].
  
  \bibitem{Dmd_Argus} 
H. Albrecht {\it et al.}  [ARGUS Collaboration],  {\it Phys. Lett.}\ B {\bf 192}, 245 (1987).

\bibitem{Dms_D0}
V. M. Abazov {\it et al.} [D0 Collaboration],  
{\it Phys. Rev. Lett.}\  {\bf 97}, 021802 (2006).

 \bibitem{Dms_CDF} 
A. Abulencia {\it et al.} [CDF Collaboration],  {\it Phys. Rev. Lett.}\  {\bf 97}, 062003 (2006).

\bibitem{Dms_LHCb} 
 R. Aaij {\it et al.} [LHCb Collaboration],  {\it Phys. Lett.}\ B {\bf 709}, 177 (2012). 
\bibitem{pdg-2012} 
 J. Beringer {\it et al.}  [Particle Data Group], {\it Phys.\ Rev.}\ D {\bf 86}, 010001 (2012). 
 \bibitem{lattice-xi}
J. Laiho, E. Lunghi,and R.S. Van de Water, {\it Phys. Rev.}\ D {\bf 81}, 034503 (2010); \\
the latest results are given at http://www.latticeaverages.org.
  
\bibitem{LHCb-Mor-12}R. Aaij {\it et al.}\ [LHCb Collaboration],  LHCb-CONF-2012-002. 

\bibitem{DFN}I.~Dunietz, R.~Fleischer and U.~Nierste,
  Phys.\ Rev.\ D {\bf 63}, 114015 (2001) [hep-ph/0012219].
  
\bibitem{LHCbKstarKstar}R. Aaij  {\it et al.}\ [LHCb Collaboration],  
  Phys.\ Lett.\ B {\bf 709}, 2 (2012)  [arXiv:1111.4183 [hep-ex]].
  
\bibitem{DGMV}S.~Descotes-Genon, J.~Matias and J.~Virto,
  Phys.\ Rev.\ D {\bf 85}, 034010 (2012) 
  [arXiv:1111.4882 [hep-ph]].

\bibitem{FK}R.~Fleischer and R.~Knegjens,
  Eur.\ Phys.\ J.\ C {\bf 71},   1789 (2011)
  [arXiv:1109.5115 [hep-ph]].
  
\bibitem{RK}R.~Knegjens,
  arXiv:1209.3206 [hep-ph].


\bibitem{LHCb-Strat}R. Aaij {\it et al.}  [LHCb Collaboration] and A. Bharucha {\it et al.},
  arXiv:1208.3355 [hep-ex].
  

\bibitem{grossman}Y.~Grossman,
  Phys.\ Lett.\ B {\bf 380}, 99 (1996)
  [hep-ph/9603244].
  
\bibitem{Randall}
L.~Randall and S.~Su, {\it Nucl. Phys. B \/} {\bf 540}, 37 (1999).

\bibitem{Hewett}
J.L.~Hewett, arXiv:hep-ph/9803370 (1998).

\bibitem{Hou}
G.W.S.~Hou, arXiv:0810.3396 [hep-ph] (2008).

\bibitem{Soni}
A.~Soni {\it et al.}, {\it Phys. Lett. B \/} {\bf 683}, 302 (2010); 
A.~Soni {\it et al.}, {\it Phys. Rev. D \/} {\bf 82}, 033009 (2010) and references therein.

\bibitem{Buras}
M.~Blanke {\it et al.}, J.~High~Energy~Phys. {\bf 12}, 003 (2006); 
W. Altmannshofer, {\it et al.}, Nucl. Phys. B {\bf 830}, 17 (2010).

\bibitem{Buras1}
A. Buras and J. Girrbach, {\it Acta Phys.\ Polon.}\ B {\bf 43}, 1427 (2012)
  [arXiv:1204.5064 [hep-ph]].

\bibitem{asld-hfag}
\begin{verbatim}
http://www.slac.stanford.edu/xorg/hfag/osc/fall_2012/HFAG_Chapter3_oct2012.pdf 
\end{verbatim}
and references therein.


\bibitem{rosner}J. L. Rosner, {\it Phys.\ Rev.}\ D {\bf 42}, 3732 (1990).

\bibitem{DDLR}A.~S.~Dighe, I.~Dunietz, H.~J.~Lipkin and J.~L.~Rosner,
  {\it Phys.\ Lett.}\ B {\bf 369}, 144 (1996)
  [hep-ph/9511363].

\bibitem{DDF}A.~S.~Dighe, I.~Dunietz and R.~Fleischer,
  {\it Eur.\ Phys.\ J.}\ C {\bf 6}, 647 (1999)
  [hep-ph/9804253].

\bibitem{HFAG}Y.~Amhis {\it et al.}  [Heavy Flavor Averaging Group Collaboration],
  arXiv:1207.1158 [hep-ex]; for online updates, see http://www.slac.stanford.edu/xorg/hfag/

\bibitem{bib:phis_ambiguity_method} Y. Xie, P. Clarke, G. Cowan and F. Muheim, 
JHEP {\bf 09}, 074 (2009).

\bibitem{bib:phis_ambiguity_lhcb}R. Aaij {\it et al.}  [LHCb Collaboration],  {\it Phys. Rev. Lett.\/} {\bf 108}, 241801 (2012).

\bibitem{CDF-BsJpsiKS}T.~Aaltonen {\it et al.}  [CDF Collaboration],
  {\it Phys.\ Rev.}\ D {\bf 83}, 052012 (2011)
  [arXiv:1102.1961 [hep-ex]].
  
\bibitem{LHCb-BsKast}R. Aaij {\it et al.}  [LHCb Collaboration],
  {\it Phys.\ Rev.}\ D {\bf 86}, 071102 (2012)
  [arXiv:1208.0738 [hep-ex]].
  
\bibitem{RF-CKM-2012-pen}R.~Fleischer,
  arXiv:1212.2792 [hep-ph].

\bibitem{LHCb-f0}R.~Aaij {\it et al.}  [LHCb Collaboration],
  {\it Phys.\ Lett.}\  B {\bf 698}, 115 (2011) 
  [arXiv:1102.0206 [hep-ex]].
 
\bibitem{Belle-f0}J.~Li {\it et al.}  [Belle Collaboration],
  {\it Phys.\ Rev.\ Lett.}\  {\bf 106}, 121802 (2011) 
  [arXiv:1102.2759 [hep-ex]].
 
\bibitem{CDF-f0}T.~Aaltonen {\it et al.}  [CDF Collaboration],
  {\it Phys.\ Rev.}\ D {\bf 84}, 052012 (2011)
  [arXiv:1106.3682 [hep-ex]].
  
\bibitem{D0-f0}V.~M.~Abazov {\it et al.}  [D0 Collaboration],
  {\it Phys.\ Rev.}\ D {\bf 85}, 011103 (2012)
  [arXiv:1110.4272 [hep-ex]].

\bibitem{SZ}S.~Stone and L.~Zhang,
  {\it Phys.\ Rev.}\  D {\bf 79}, 074024 (2009)  [arXiv:0812.2832 [hep-ph]];
  arXiv:0909.5442 [hep-ex].
  
\bibitem{CDeFW}P.~Colangelo, F.~De Fazio and W.~Wang,
  {\it Phys.\ Rev.}\ D {\bf 83}, 094027 (2011)
  [arXiv:1009.4612 [hep-ph]].
  

\bibitem{LHCb-BsJpsiKS}R. Aaij {\it et al.}  [LHCb Collaboration],
  {\it Phys.\ Lett.}\ B {\bf 713}, 172 (2012)
  [arXiv:1205.0934 [hep-ex]].
  
\bibitem{RF-BsKK}R.~Fleischer,
  {\it Phys.\ Lett.}\ B {\bf 459}, 306 (1999)
  [hep-ph/9903456].
  
\bibitem{RF-Bhh}R.~Fleischer,
  {\it Eur.\ Phys.\ J.}\ C {\bf 52}, 267 (2007)
  [arXiv:0705.1121 [hep-ph]].
  
\bibitem{FK-BsKK}R.~Fleischer and R.~Knegjens,
  {\it Eur.\ Phys.\ J.}\ C {\bf 71}, 1532 (2011)
  [arXiv:1011.1096 [hep-ph]].

\bibitem{CDF-BsKK}A.~Abulencia {\it et al.}  [CDF Collaboration],
  {\it Phys.\ Rev.\ Lett.}~{\bf 97},  211802 (2006);
M.~Morello  [CDF Collaboration],
  {\it Nucl.\ Phys.\ Proc.\ Suppl.}\  {\bf 170}, 39 (2007) 
  [arXiv:hep-ex/0612018].

\bibitem{Belle-BsKK} R.~Louvot  [Belle Collaboration],
  PoS E {\bf PS-HEP2009}, 170 (2009) 
  [arXiv:0909.2160 [hep-ex]].
 
\bibitem{CFMS}M.~Ciuchini, E.~Franco, S.~Mishima and L.~Silvestrini,
  JHEP {\bf 1210}, 029 (2012)
  [arXiv:1205.4948 [hep-ph]].

\bibitem{DuMe}G.~Duplan$\check{\mbox{c}}$i\'c and B.~Meli\'c,
  {\it Phys.\ Rev.}\  D {\bf 78}, 054015 (2008) 
  [arXiv:0805.4170 [hep-ph]].
  
  \bibitem{bib:CPV-BsToKK-LHCb} 
  R. Aaij { \it et al.} [LHCb Collaboration], LHCb-CONF-2012-007
  
 \bibitem{groro}M.~Gronau and J.~L.~Rosner,
  {\it Phys.\ Lett.}\ B {\bf 482}, 71 (2000)
  [hep-ph/0003119].
  
\bibitem{FK-CKM}R.~Fleischer and R.~Knegjens,
  arXiv:1012.0839 [hep-ph].
  

\bibitem{CPV_BABAR_BsToKPi} 
B. Aubert {\it et al.} [BABAR Collaboration],   
{\it Phys. Rev. Lett.}\ {\bf 99}, 021603 (2007).

\bibitem{Belle-CPdir}
S.W. Lin {\it et al.} [Belle Collaboration],  Nature {\bf 452}, 332 (2008).

\bibitem{TeV-CPdir}
T. Aaltonen {\it et al.} [CDF Collaboration], 
{\it Phys. Rev. Lett.}\ {\bf 106}, 181802 (2011).

\bibitem{KTeV}A.~Alavi-Harati {\it et al.}  [KTeV Collaboration],
  {\it Phys.\ Rev.}\ D {\bf 67}, 012005 (2003)
  [Erratum-ibid.\ D {\bf 70}, 079904 (2004)]
  [hep-ex/0208007].
  
\bibitem{NA48}J.~R.~Batley {\it et al.}  [NA48 Collaboration],
  {\it Phys.\ Lett.}\ B {\bf 544}, 97 (2002)
  [hep-ex/0208009].

\bibitem{CPV_CDF_BsToKPi} 
 T. Aaltonen {\it et al.} [CDF Collaboration], {\it Phys. Rev. Lett.}\ {\bf 106}, 181802 (2011).
  
%
\bibitem{CPV_LHCb_BsToKPi} 
 R. Aaij { \it et al.} [LHCb Collaboration], {\it Phys.\ Rev.\ Lett.} {\bf 108}, 201601 (2012) .
 
 \bibitem{CPV_CDF_BsToKPi-1} 
 T. Aaltonen {\it et al.}  [CDF Collaboration], Conference note 10726, \\
 http://www-cdf.fnal.gov/physics/new/bottom/120628.blessed-Bhh9fb/
  
\bibitem{BsDsK}R. Aleksan, I. Dunietz and B. Kayser,
{\it Z.\ Phys.}\ C {\bf 54}, 653 (1992).

\bibitem{RF-BsDsK}R.~Fleischer,
  {\it Nucl.\ Phys}.\ B {\bf 671}, 459 (2003)
  [hep-ph/0304027].

\bibitem{Cavoto:2006um} 
  G.~Cavoto, R.~Fleischer, T.~Gershon, A.~Soni, K.~Abe, J.~Albert, D.~Asner and D.~Atwood {\it et al.},
  hep-ph/0603019.

\bibitem{Gligorov:2008zzb} 
  V.~Gligorov and G.~Wilkinson,
  CERN-LHCB-2008-035.
  
  \bibitem{Gligorov:2011id}
  V.~Gligorov [LHCb Collaboration],
  arXiv:1101.1201 [hep-ex].

\bibitem{BsDsK-12}K.~De Bruyn, R.~Fleischer, R.~Knegjens, M.~Merk, M.~Schiller and N.~Tuning,
  {\it Nucl.\ Phys}.\ B {\bf 868}, 351 (2013) [arXiv:1208.6463 [hep-ph]].

\bibitem{CDF-BsDsK}
  T.~Aaltonen {\it et al.}  [CDF Collaboration],
  {\it Phys.\ Rev.\ Lett.}\  {\bf 103},   191802 (2009).
  [arXiv:0809.0080 [hep-ex]].

\bibitem{Belle-BsDsK}
  R.~Louvot {\it et al.}  [Belle Collaboration],
  {\it Phys.\ Rev.\ Lett.}\  {\bf 102}, 021801 (2009) 
  [arXiv:0809.2526 [hep-ex]].

\bibitem{LHCb-BsDsK}
  R.~Aaij {\it et al.}  [LHCb Collaboration],
  JHEP {\bf 1206}, 115 (2012) 
  [arXiv:1204.1237 [hep-ex]].

\bibitem{NN}S.~Nandi and U.~Nierste,
  {\it Phys.\ Rev.}\ D {\bf 77},  054010 (2008) 
  [arXiv:0801.0143 [hep-ph]].
  
\bibitem{LHCb-BsDsK-CPV}
  R.~Aaij {\it et al.}  [LHCb Collaboration],
  LHCb-CONF-2012-029,\\ http://cds.cern.ch/record/1477943/files/LHCb-CONF-2012-029.pdf

\bibitem{RF-BsDsDs}R.~Fleischer,
  {\it Eur.\ Phys.\ J.}\ C {\bf 51}, 849 (2007)
  [arXiv:0705.4421 [hep-ph]].
  
 \bibitem {bib:LHCb-DD}
  R.~Aaij {\it et al.}  [LHCb Collaboration], 	arXiv:1302.5854 [hep-ex].
  
\bibitem{RF-ang}R.~Fleischer,
  {\it Phys.\ Rev.}\ D {\bf 60}, 073008 (1999)
  [hep-ph/9903540].
  
\bibitem{DGMV-2}S.~Descotes-Genon, J.~Matias and J.~Virto,
  {\it Phys.\ Rev.}\ D {\bf 76}, 074005 (2007)
  [Erratum-ibid.\ D {\bf 84}, 039901 (2011)]
  [arXiv:0705.0477 [hep-ph]].
  
\bibitem{FG}R.~Fleischer and M.~Gronau,
  {\it Phys.\ Lett.}\ B {\bf 660}, 212 (2008)
  [arXiv:0709.4013 [hep-ph]].
  
\bibitem{BDIL}B.~Bhattacharya, A.~Datta, M.~Imbeault and D.~London,
  {\it Phys.\ Lett.}\ B {\bf 717}, 403 (2012)
  [arXiv:1203.3435 [hep-ph]].
  
\bibitem{DDL-1}A.~Datta, M.~Duraisamy and D.~London,
  {\it Phys.\ Rev.}\ D {\bf 86}, 076011 (2012)
  [arXiv:1207.4495 [hep-ph]].
  
    \bibitem {bib:CDF-BsToPhiPhi}
     T.~Aaltonen {\it et al.}  [CDF Collaboration], {\it Phys. Rev. Lett.}\ {\bf 107},  261802 (2011) 
 [arXiv: 1107.4999. [hep-ph]].
    
   \bibitem {bib:LHCb-BsToPhiPhi}
  R.~Aaij {\it et al.}  [LHCb Collaboration], {\it Phys.\ Lett.}\ B {\bf 713}, 369 (2012).
  
\bibitem{DDL-2}A.~Datta, M.~Duraisamy and D.~London,
  {\it Phys.\ Lett.}\ B {\bf 701}, 357 (2011)
  [arXiv:1103.2442 [hep-ph]].
  
 \bibitem{GR-trip}M.~Gronau and J.~L.~Rosner,
  {\it Phys.\ Rev.}\ D {\bf 84}, 096013 (2011)
  [arXiv:1107.1232 [hep-ph]].
    

\bibitem{BGGI}A.~J.~Buras, J.~Girrbach, D.~Guadagnoli and G.~Isidori,
  {\it Eur.\ Phys.\ J.}\ C {\bf 72}, 2172 (2012)
  [arXiv:1208.0934 [hep-ph]].
  
  
\bibitem{BBL}G.~Buchalla, A.~J.~Buras and M.~E.~Lautenbacher,
  {\it Rev.\ Mod.\ Phys.}\  {\bf 68}, 1125 (1996)
  [hep-ph/9512380].

\bibitem{APS}W.~Altmannshofer, P.~Paradisi and D.~M.~Straub,
  JHEP {\bf 1204}, 008 (2012) 
  [arXiv:1111.1257 [hep-ph]].

\bibitem{Rare_BsToMuMu_LHCb_limit} 
R. Aaij et al. [LHCb Collaboration], {\it Phys. Rev. Lett.}\ {\bf 108}, 231801 (2012) [arXiv:1203.4493].
  
\bibitem{BDeFG}A.~J.~Buras, F.~De Fazio and J.~Girrbach,
   JHEP {\bf 1302}, 116 (2013)
  [arXiv:1211.1896 [hep-ph]].

\bibitem{BdeFGC}A.~J.~Buras, F.~De Fazio, J.~Girrbach and M.~V.~Carlucci,
  JHEP {\bf 1302}, 023 (2013)
  [arXiv:1211.1237 [hep-ph]].
  
\bibitem{BG-2}A.~J.~Buras and J.~Girrbach,
   JHEP {\bf 1301}, 007 (2013)
  [arXiv:1206.3878 [hep-ph]].

\bibitem{WCSY}W.~Altmannshofer, M.~Carena, N.~Shah and F.~Yu,
 JHEP {\bf 1301}, 160 (2013)
  [arXiv:1211.1976 [hep-ph]].

\bibitem{KMetal}K.~Kowalska, S.~Munir, L.~Roszkowski, E.~M.~Sessolo, 
S.~Trojanowski and Y.-L.~S.~Tsai,
  arXiv:1211.1693 [hep-ph].

\bibitem{HaiMa}U.~Haisch and F.~Mahmoudi,
  JHEP {\bf 1301}, 061 (2013)
  [arXiv:1210.7806 [hep-ph]].

\bibitem{BCCetal}O.~Buchmueller, R.~Cavanaugh, M.~Citron, A.~De Roeck, M.~J.~Dolan, J.~R.~Ellis, H.~Fl\"acher and S.~Heinemeyer {\it et al.},
  {\it Eur.\ Phys.\ J.}\ C {\bf 72}, 2243 (2012)
  [arXiv:1207.7315 [hep-ph]].
 
 \bibitem{HM}T.~Hurth and F.~Mahmoudi,
 {\it Nucl.\ Phys.}\ B {\bf 865}, 461 (2012)
  [arXiv:1207.0688 [hep-ph]].
  
 \bibitem{AS}W.~Altmannshofer and D.~M.~Straub,
JHEP {\bf 1208}, 121 (2012)
  [arXiv:1206.0273 [hep-ph]].
  
\bibitem{BKMS}D.~Becirevic, N.~Kosnik, F.~Mescia and E.~Schneider,
  {\it Phys.\ Rev.}\ D {\bf 86}, 034034 (2012)
  [arXiv:1205.5811 [hep-ph]].
    
\bibitem{MNO}F.~Mahmoudi, S.~Neshatpour and J.~Orloff,
  JHEP {\bf 1208}, 092 (2012)
  [arXiv:1205.1845 [hep-ph]].
  
 \bibitem{Nan}  T.~Li, D.~V.~Nanopoulos, W.~Wang, X.-C.~Wang and Z.~-H.~Xiong,
  JHEP {\bf 1207}, 190 (2012) 
  [arXiv:1204.5326 [hep-ph]].

\bibitem{bsg-theory}
A. Ali, B. D. Pecjak, and C. Greub, {\it Eur. Phys. J.}\ C {\bf 55}, 577 (2008).

\bibitem{bsg-belle}
J. Wicht {\it et al.} [Belle Collaboration] {\it Phys. Rev. Lett.\/} {\bf 100}, 121801 (2008).

\bibitem{bsg-lhcb}
R. Aaij {\it et al.} [LHCb Collaboration], {\it Phys. Rev.}\ D {\bf 85}, 112013 (2012).

\bibitem{MXZ}F.~Muheim, Y.~Xie and R.~Zwicky,
  {\it Phys.\ Lett.}\ B {\bf 664}, 174 (2008)
  [arXiv:0802.0876 [hep-ph]].

\bibitem{BHP}C.~Bobeth, G.~Hiller and G.~Piranishvili,
  JHEP {\bf 0807}, 106 (2008)
  [arXiv:0805.2525 [hep-ph]].

\bibitem{b-to-s-cdf}
CDF Collab. http://www-cdf.fnal.gov/physics/new/bottom/ \\
120628.blessed-b2smumu\_96/

\bibitem{b-to-s-lhcb}
R. Aaij {\it et al.} [LHCb Collaboration], LHCb-CONF-2012-003



\end{thebibliography}
\end{document}